\documentclass[notitlepage,floatfix,amsmath,amssymb,prf,
  longbibliography]{revtex4-1} 
\usepackage{tabularx}
\usepackage{graphicx}
\usepackage{dcolumn}
\usepackage{color}
\usepackage[normalem]{ulem}
\usepackage{amssymb}
\usepackage[toc,page]{appendix}
\newcommand{\be}{\begin{equation}}
\newcommand{\ee}{\end{equation}}

\usepackage{graphicx}
\usepackage{enumerate}
\newcommand{\RN}[1]{%
  \textup{\uppercase\expandafter{\romannumeral#1}}%
}

\begin{document}
\title{From waves to convection and back again: The phase space of
  stably stratified turbulence}

\author{N.E. Sujovolsky, and P.D. Mininni}
\affiliation{
  Universidad de Buenos Aires, Facultad de Ciencias Exactas y
  Naturales, Departamento de F\'\i sica, \& IFIBA, CONICET, Ciudad
  Universitaria, Buenos Aires 1428, Argentina.}

\begin{abstract}
We show that the phase space of stratified turbulence mainly consists
of two slow invariant manifolds with rich physics, embedded on a larger
basin with fast evolution. A local invariant manifold in the vicinity
of the fluid at equilibrium corresponds to waves, while a global 
invariant manifold corresponds to the onset of local convection. Using
a reduced model derived from the Boussinesq equations, we propose that
waves accumulate energy nonlinearly up to a point such that fluid
elements escape from the local manifold and evolve fast to the global
manifold, where kinetic energy can be more efficiently
dissipated. After this, fluid elements return to the first 
manifold. As the stratification increases, the volume of the first
manifold increases, and the second manifold becomes harder to
access. This explains recent observations of enhanced intermittency
and marginal instability in these flows. The reduced model also allows
us to study structure formation,  alignment of field gradients in the
flow, and to identify balance relations that hold for each fluid
element.
\end{abstract}
\maketitle

\section{Introduction}

% Main question
In {\it The Hobbit} by J.R.R.~Tolkien, Bilbo Baggins leaves a quiet
life in the Shire to go ``there and back again.'' After a long journey
and finding turmoil and adventure in Khazad-d\^um, he returns to the
Shire to find out his family has auctioned off his possessions, but to
also realize that his life has deeply changed. In turbulent flows,
fluid elements explore a complicated phase space, and may or may not
go back to a previous state. Where do fluid elements go in the phase
space of stably stratified turbulence, and how are they changed as
they explore this space? This question has become relevant as recent
observations indicate that fluid elements in stably stratified
turbulence alternate between stable and unstable states, resulting in
the occurence of extreme events, enhanced intermittency, marginal
instability, and critical behavior \cite{Smyth_2013,
  rorai_turbulence_2014, Pearson_2018, feraco_vertical_2018,
  pouquet2019linking, Smyth_2019}. However, properly answering this
question requires studying the flow from a Lagrangian point of view,
while most studies of this problem consider an Eulerian description of
the flow.

% Eulerian description
The correct characterization of stratified turbulence is of
fundamental importance for many environmental processes, as well as
for weather forecasting \cite{bauer2015quiet, bodenschatz2010can}. In
the Eulerian description, a useful way to characterize turbulent flows
is through its energy spectrum and scaling relations. The presence of
gravity provides a preferred direction in the system, which results in
different scaling laws for the spectrum in the directions parallel and
perpendicular to the gravity \cite{waite_stratified_2004,
  riley2008stratified, rorai_stably_2015, pouquet2018scaling, 
  de2019effects}. The anisotropy associated with the stratification
also affects velocity and temperature gradients, resulting in the
generation of pancake-like structures in the flow. These structures
can also be interpreted, from a spectral point of view, as resulting
from a preferential transfer of energy by nonlinear interactions
towards horizontal slow modes \cite{godeferd1994detailed,
  smith_generation_2002, godoy2004vertical,
  marino_large-scale_2014}. The study of stratified turbulence from
the Eulerian point of view has thus led to significant advances which
are useful, e.g., to understand scale-by-scale balances satisfied by
the system, which in turn are relevant for subgrid scale models as the
production of turbulent fluctuations are ultimately controlled by
velocity and buoyancy gradients.

% Mixing
However, important turbulent averages used to estimate mixing, such
as the local shear production, the vertical buoyancy turbulent flux,
and the flux Richardson number, are not obtained from the energy
spectrum but from global balance equations \cite{ivey2008density, 
  pouquet2018scaling}. A correct estimation of mixing is crucial for
weather prediction, as models rely on it to estimate the effect of
turbulence. But averaged values are only relevant if the statistics of
the fields are Gaussian, as averages can fail to correctly capture the
mixing if extreme events and localized structures are present
\cite{rorai_turbulence_2014, Pearson_2018, feraco_vertical_2018}.

% Lagrangian description
In this context, a very useful approach to study mixing and
dissipation in stratified turbulence is the Lagrangian description of
the flow. In homogeneous and isotropic turbulence, whether it is from
a single-particle approach \cite{falkovich2012lagrangian} or from a
multi-particle one \cite{chertkov1999lagrangian}, the Lagrangian
perspective has been key for the understanding of several flow
properties \cite{toschi2009lagrangian, xu2011pirouette}. Both
approaches have been used to study stratified turbulence in
\cite{kimura_diffusion_1996, aartrijk_single-particle_2008,
  sujovolsky2018single, sujovolsky2019vertical, buaria2019single}. 
Vertical dispersion of Lagrangian tracers was found to be strongly
suppressed when compared to the homogeneous and isotropic case
\cite{kimura_diffusion_1996, sujovolsky2019vertical}, while horizontal 
dispersion was found to be enhanced by horizontal winds
\cite{sujovolsky2018single}. Experiments following Lagrangian buoys in
the ocean have also enabled studies of dispersion and dissipation in
stratified flows \cite{d2000lagrangian, lien2006measurement}. More
recently, studies focussed on the Lagrangian evolution of vertical
gradients in stratified turbulence found a non-monotonic enhancement
of extreme events with the level of stratification
\cite{feraco_vertical_2018, sujovolsky2019invariant, 
  pouquet2019linking}.

% Waves
Compared with homogeneous and isotropic turbulence, the Eulerian and
Lagrangian descriptions of stably stratified flows are obfuscated by
the presence of waves: In these flows, buoyancy acts as a restitutive
force allowing for the excitation of internal gravity waves
\cite{davidson_turbulence_2013}. These waves play a crucial role in
the development of the Eulerian spectra \cite{godeferd1994detailed,
  waite_stratified_2004, riley2008stratified} and in the suppression of
vertical transport \cite{kimura_diffusion_1996,
  sujovolsky2019vertical}  discussed above. And while they interact
nonlinearly and contribute to the turbulent transfer of energy to
smaller scales \cite{nazarenko2011wave}, they can be inefficient at
dissipating energy. As a result, under some conditions energy can be
expected to accumulate until wave-like solutions cannot hold anymore,
and the system must search in phase space for other solutions that can
dissipate energy more efficiently \cite{newell2013wave,
  dyachenko2016whitecapping}.

% Restricted models and this paper
This is the main motivation for the present study. What surfaces of
solutions in phase space are explored when internal gravity waves
break, and how does the system fill in the gap between the different
possible states? Using a reduced model for stratified flows introduced
in \cite{sujovolsky2019invariant}, and direct numerical simulations of
stably stratified turbulence, we show that the Boussinesq equations
have two slow invariant manifolds corresponding to two solutions:
waves, and the onset of local convection. Fluid elements stay for long
times in the first region of phase space. When they escape from this
manifold, they travel fast in phase space from one manifold to the
other, where they can dissipate energy more efficiently, to finally
return to the first manifold. Our reduced model is an extension of
restricted Euler models studied in detail in homogeneous and isotropic
turbulence \cite{vieillefosse_local_1982, cantwell_exact_1992,
  chevillard2006lagrangian, meneveau2011lagrangian}. These models
describe the Lagrangian evolution of field gradients using a small
number of ordinary differential equations, and for homogeneous and
isotropic turbulence were able to explain many flow properties such as
the development of intermittency and the origin of flow structures
\cite{chevillard2006lagrangian, meneveau2011lagrangian}. In our case, 
the reduced model allows us to also identify and study: (1) balance
relations that hold for fluid elements as they are advected by the
flow, (2) alignments between the vorticity, the gradient of density
fluctuations, and the strain-rate tensor, and (3) correlations between
different terms in the potential vorticity.

% Index
The structure of the paper is as follows. In
Sec.~\ref{sec:simulations} we introduce the Boussinesq equations, 
and we describe the direct numerical simulations used to study the
evolution of fluid elements in phase space and to compare their
evolution with the predictions from the reduced model.
In Sec.~\ref{sec:reduced} we derive in detail the restricted Euler
model for stratified turbulence originally presented in
\cite{sujovolsky2019invariant}. In Sec.~\ref{sec:phase} we study its
fixed points and invariant manifolds, we compare with the results from
the numerical simulations of the full Boussinesq equations, and show
the evolution of fluid elements in the phase space defined by the
model. Section \ref{sec:balance} discusses implications of the model
for subgrid modeling of stably stratified turbulence and, in
particular, for the turbulent production of buoyancy
gradients. Implications of the results for the alignment of field
gradients are presented in Sec.~\ref{sec:alignment}. Section
\ref{sec:PV} discusses further implications of the model for the
evolution of the potential vorticity. Finally,
Sec.~\ref{sec:conclusions} presents our conclusions.

\section{The Boussinesq equations and direct numerical
  simulations \label{sec:simulations}}

\subsection{The equations for a stably stratified flow}

In this work we will consider the Lagrangian evolution of velocity and
density gradients of individual fluid elements, as they evolve under
the Eulerian dynamics of an incompressible stably stratified turbulent
flow. To describe the flow Eulerian dynamics we work under the
Boussinesq approximation, which describes perturbations to a linear
background density profile (which in our case is stable). On top of
this background profile, density (or ``buoyancy'') fluctuations are
represented by $\theta$. The scalar field $\theta$ has units of
velocity by defining it as $\theta = g\rho/(N \rho_0)$, where $g$ is
the gravitational acceleration, $\rho$ is the actual density
fluctuation, $N$ is the Brunt-V\"{a}is\"{a}l\"{a} frequency
(associated to the frequency of internal gravity waves, to the linear
background density profile, and a measure of how strong stratification
is), and $\rho_0$ is the mean background density (i.e., averaged over
all space). For this scalar field and for a velocity field $\bf{u}$,
the incompressible Boussinesq equations can then be written as
\begin{eqnarray}
\label{eq:n-s_strat}
\frac{\partial {\bf u}}{\partial t} +{\bf u}\cdot{\boldsymbol \nabla}{\bf u}
  &=& -{\boldsymbol \nabla}p - N \theta {\hat z} +\nu \nabla^{2}
  {\bf u} + {\bf f}, \\ 
\label{eq:theta}
\frac{\partial \theta}{\partial t}+ {\bf u}\cdot{\boldsymbol \nabla} \theta 
  &=& N {\bf u} \cdot {\hat z}  + \kappa \nabla^{2} \theta, \\
\label{eq:incomp}
\nabla \cdot {\bf u}&=&0,
\end{eqnarray}
where $p$ is the correction to the hydrostatic pressure, $\nu$ the
kinematic viscosity, $\kappa$ the diffusivity, and ${\bf f}$ an
external mechanical forcing. Equations (\ref{eq:n-s_strat}) and
(\ref{eq:theta}) have three dimensionless parameters that control the
dynamics of the system, the Reynolds, Froude, and Prandtl numbers 
respectively defined as 
\begin{equation}
  \mathrm{Re}=\frac{UL}{\nu}, \,\,\,\,\,\,\,\,
  \mathrm{Fr}=\frac{U}{NL}, \,\,\,\,\,\,\,\,
  \mathrm{Pr}=\frac{\nu}{\kappa} ,
\end{equation}
where $U$ and $L$ are the characteristic velocity and length of the
flow. While $\mathrm{Re}$ measures the strength of the nonlinear term
compared with that of the viscous term in the momentum equation,
$\mathrm{Fr}$ compares the strength of the nonlinear term with the
buoyancy term; $\mathrm{Pr}$ just corresponds to the ratio of
diffusivities.  The strength of the turbulence is also often
characterized using the buoyancy Reynolds number, defined as
\begin{equation}
  \mathrm{Rb}= \mathrm{Re} \, \mathrm{Fr^2} .
\end{equation}

The Boussinesq equations have a well known fixed point for
${\bf u}=\theta=0$. Linearizing Eqs.~(\ref{eq:n-s_strat}) and
(\ref{eq:theta}) in the vicinity of this solution and neglecting
viscosity and diffusivity ($\nu = \kappa =0$), internal gravity waves
are found with dispersion relation \cite{davidson_turbulence_2013}
\begin{equation}
  \omega = N \frac{k_{\perp}}{k} ,
  \label{eq:disp}
\end{equation}
where $\omega$ is the wave frequency, $k=|{\bf k}|$ is the wave
number, ${\bf k}$ the wave vector, and $k_{\perp}$ is the wave number
associated to the components of ${\bf k}$ perpendicular to
gravity. Note that $\omega \le N$, and the Brunt-V\"{a}is\"{a}l\"{a}
frequency is thus the maximum possible frequency of internal gravity
waves.

These waves play a crucial role in the dynamics of stably stratified
flows, and it is known that fluid elements are often found in
wave-like states in the vicinity of ${\bf u}=\theta=0$. This results
in low vertical diffusion and mixing \cite{kimura_diffusion_1996,
  aartrijk_single-particle_2008, sujovolsky2019vertical}. However,
local instabilities also play a key role in these flows. Important
parameters to measure the vertical stability of the flow are given by
the Richardson numbers, which have multiple definitions in the
literature. Here we will consider the gradient Richardson number
\begin{equation}
\mathrm{Ri}_{g} =
  \frac{N(N-\partial_{z}\theta)}{|\partial_{z}{\bf u}_{\perp}|^{2}} ,
\label{eq:rig}
\end{equation}
where ${\bf u}_{\perp}$ is the horizontal flow velocity
\cite{rosenberg_evidence_2015}. This number provides us with a
pointwise estimation of the flow stability. When $\mathrm{Ri}_{g} \leq
1/4$ the flow  can undergo zig-zag and shear instabilities
\cite{billant_theoretical_2000}, while for $\mathrm{Ri}_{g}\leq0$ the
local density fluctuation can overcome the background density gradient
(controlled by $N$) and local convection can develop, significantly
increasing the vertical mixing in the vicinity of that point
\cite{sujovolsky2019vertical, mashayek}. Another way to look at this
latter case is the following: When $\partial_{z}\theta \geq N$ the
local buoyancy steepness is larger in absolute value than that of the
background, breaking down the vertical stability and allowing for a
convective instability to take place.

%%%%%%%%%%%%%%%%%%%%%%%%%%%%%%%%%%%%%%%%%% 
\begin{table}
\begin{ruledtabular}
\begin{tabular*}{\textwidth}{ c c c c c c}
    $N$ & $\textrm{Fr}$ & $\textrm{Re}$ & $\textrm{Rb}$ & $L_B/\eta$
  & $L_{oz}/\eta$ \\
\hline
    $4$  & $0.05$ & $10000$ & $25$ & $9.6$ & $14.4$ \\
    $8$  & $0.03$ & $14000$ & $13$ & $6.0$ & $5.6$ \\
    $12$ & $0.02$ & $15000$ &  $4$ & $4.0$ & $2.8$ \\
\end{tabular*}
\end{ruledtabular}
\caption{Characteristic parameters of the numerical simulations: $N$
  is the Brunt-V\"{a}is\"{a}l\"{a} frequency, $\textrm{Fr}$ is the
  Froude number (typical geophysical values are
  $\textrm{Fr}\approx 10^{-2}$), $\textrm{Re}$ is the Reynolds number, 
  $\textrm{Rb}$ is the buoyancy Reynolds number, and $L_B/\eta$ and
  $L_{oz}/\eta$ are respectively the buoyancy scale $L_b$ and the
  Ozmidov scale $L_{oz}$ normalized by the Kolmogorov dissipation
  scale $\eta$.}
\label{tab:param}
\end{table}
%%%%%%%%%%%%%%%%%%%%%%%%%%%%%%%%%%%%%%%%%% 

\subsection{Numerical solutions of the Boussinesq equations and
  Lagrangian particles integration}

We performed direct numerical simulations of
Eqs.~(\ref{eq:n-s_strat})-(\ref{eq:incomp}) in the regime of developed
turbulence, in a three-dimensional (3D) periodic domain with aspect
ratio $L_x$:$L_y$:$L_z=4$:4:1 (where $L_x$, $L_y$, and $L_z$ are the
lengths of the domain in each direction, and with $L_x=2\pi$ in
dimensionless units), and with spatial resolution of
$768\times768\times192$ grid points. We used a parallel
pseudo-spectral fully-dealiased method to compute spatial derivatives 
and nonlinear terms in the equations, and a second order Runge-Kutta
scheme for time integration \cite{mininni2011hybrid}. A Taylor-Green
forcing was used to sustain the turbulence, a forcing that has been
used in previous studies of stably stratified flows (see, e.g.,
\cite{riley_dynamics_2003, sujovolsky2018generation}), and which is 
given by
\begin{equation}
  {\bf f} = \mathrm{F}_{0} \left[ \sin(x)\cos(y)\cos\left(
      \frac{L_x}{L_z}z\right) \hat{x} -\cos(x)\sin(y)\cos\left(
      \frac{L_x}{L_z}z\right) \hat{y} \right] ,
\label{eq:TG}
\end{equation}
where $L_x/L_z=4$ and $F_0$ is the forcing amplitude (which was chosen
to have a r.m.s.~flow velocity $U \approx 1$ in the turbulent steady
state). The forcing generates counter-rotating large scale vortices
separated by planes of strong horizontal shear. In the analysis that
follows, we also verified that other domain aspect ratios, other
spatial resolutions, and other possible forcing schemes (see, e.g.,
simulations in \cite{sujovolsky2019invariant}) give qualitatively
similar results.

The viscosity and diffusivity ($\nu = \kappa$, and thus
$\mathrm{Pr}=1$) were chosen so that all the relevant flow scales were
properly resolved \cite{waite_stratified_2004}. This includes the
buoyancy scale $L_B=2\pi/k_{B}$ (the scale associated to the typical
height of the strata), the Ozmidov scale $L_{oz}=2\pi/k_{oz}$ (the
scale at which the flow starts recovering isotropy), and the
Kolmogorov dissipation scale $\eta=2\pi/k_\eta$, where all the
corresponding wave numbers are respectively defined as
\begin{equation}
  k_B = \frac{N}{U} , \,\,\,\,\,\,\,\,
  k_{oz}=\left( \frac{N^3}{\epsilon} \right)^{1/2} , \,\,\,\,\,\,\,\,
  k_\eta = \left( \frac{\epsilon}{\nu^3} \right)^{1/4} ,
\end{equation}
and where $\epsilon$ is the energy injection rate. The Kolmogorov
dissipation scale was always slightly larger than the minimum resolved
spatial scale. Three values of the Brunt-V\"{a}is\"{a}l\"{a} frequency
(and thus of the stratification) were considered, with $N=4$, $8$, and
$12$ in dimensionless units. Resulting values of Fr, Re, Rb, and of
the ratios $L_B/\eta$ and $L_{oz}/\eta$ for the three simulations, are
given in table \ref{tab:param}.

To study the evolution of velocity and density gradients as fluid
elements are advected by the fluid, in each simulation we tracked
${\cal O}(10^6)$ Lagrangian particles for over 10 large-scale turnover
times. Each tracer satisfies the ordinary differential equation
\begin{equation}
\frac{d {\bf x}}{d t} = {\bf u}({\bf x}, t) ,
\end{equation}
where ${\bf x}$ is the particle position. This equation was integrated
using a Runge-Kutta method to evolve in time, and a 3D cubic spline
interpolation to estimate the fluid velocity ${\bf u}({\bf x}, t)$ at
the position of the particles \cite{yeung1988algorithm}. The same
method was used to estimate velocity and temperature gradients at
particles' positions, respectively denoted as
${\boldsymbol \nabla} {\bf u}({\bf x}, t)$ and
${\boldsymbol \nabla} \theta({\bf x}, t)$. These quantities were
stored for each particle with very high time cadence. Then, finite
differences and averages over multiple nearby particles (to decrease
fluctuations) were used to obtain time derivatives of
${\boldsymbol \nabla} {\bf u}({\bf x}, t)$ and
${\boldsymbol \nabla} \theta({\bf x}, t)$, to study their evolution in
phase space.

\section{A reduced model for the Lagrangian evolution of field
  gradients \label{sec:reduced}}

%%%%%%%%%%%%%%%%%%%%%%%%%%%%%%%%%%%%%%%%%% 
\begin{figure}
\includegraphics[width=8.5cm]{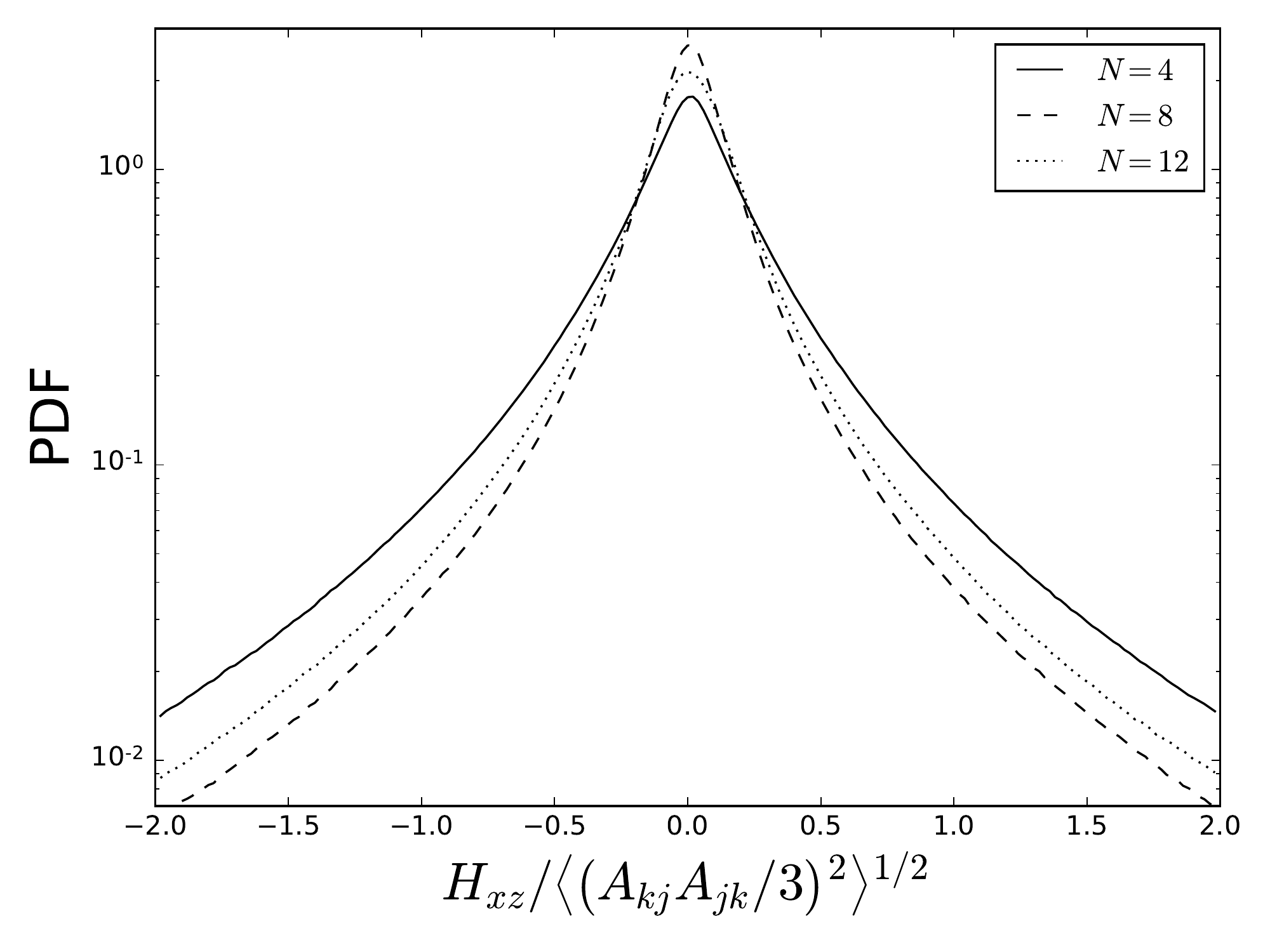}
\includegraphics[width=8.5cm]{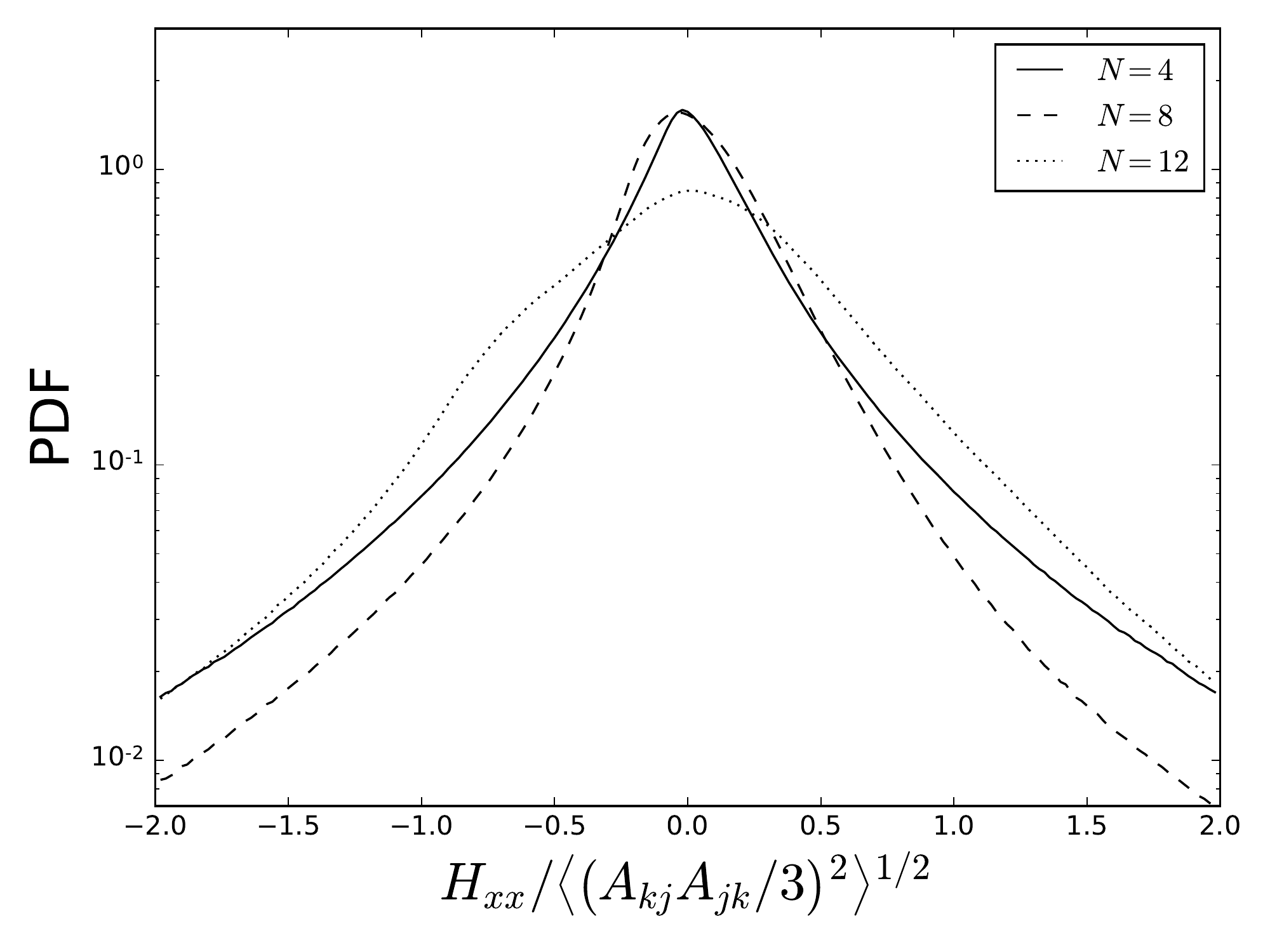}
\caption{Probability density functions (PDFs) of the pressure Hessian
  components $H_{xz}$ ({\it left}), and $H_{xx}$ ({\it right}), in
  both cases normalized by a r.m.s.~typical value of one-third of the
  trace of the nonlinear term in Eq.~(\ref{eq:AijH0}), 
  $\left< (A_{kj}A_{jk}/3)^{2}\right>^{1/2}$, and for the three
  simulations with different Brunt-V\"{a}is\"{a}l\"{a} frequencies
  $N$. Note most fluid elements have relatively small pressure Hessian
  components, although the PDFs also display fat tails.}
\label{f:Hij}
\end{figure}
%%%%%%%%%%%%%%%%%%%%%%%%%%%%%%%%%%%%%%%%%%

From Eqs.~(\ref{eq:n-s_strat})-(\ref{eq:incomp}) a closed model for
the Lagrangian evolution of velocity and buoyancy gradients was 
presented in \cite{sujovolsky2019invariant}. In this section we
introduce a detailed derivation of this model, including intermediate
equations that, albeit not closed, are in many cases exact, and useful
to interpret the physical implications of the reduced model. While
actual stably stratified fluids have external forcing and dissipation,
for practical reasons in this section we neglect both and we consider
the ideal unforced case ($\nu = \kappa = {\bf f} = 0$). As a result,
we can expect the reduced model to give a good approximation to field
gradients dynamics for short times, when the effects of the forcing
and of the dissipation are small compared with other linear and
nonlinear terms in the Boussinesq equations. In spite of this, we will
see that the model gives useful insights into the dynamics of the full
system even when long times are considered.

We start by computing spatial derivatives of Eqs.~(\ref{eq:n-s_strat})
and (\ref{eq:theta}). We write field gradients using index notation,
and define $A_{ij}=\partial_{j}u_{i}$ and 
$\theta_{j}=\partial_{j}\theta$ (for $i$, $j=\{x,y,z\}$). With this
notation, we obtain  
\begin{equation}
\dfrac{DA_{ij}}{Dt} + A_{kj}A_{ik} = - \dfrac{\partial^{2}p}{\partial
  x_{i}\partial x_{j}} - N \theta_{j} \delta_{iz},
\label{eq:Aij} 
\end{equation}
\begin{equation}
\dfrac{D\theta_{j}}{Dt} + A_{kj} \theta_{k} = N A_{zj},
\label{eq:tita_j}
\end{equation}
where $\delta_{ij}$ is the Kronecker delta, and
$D/Dt=(\partial/\partial t + {\bf u}\cdot{\boldsymbol \nabla})$ is the
material derivative. Equation (\ref{eq:Aij}) is the usual Lagrangian
evolution equation for the velocity gradient tensor $A_{ij}$, with the
extra term $-N \theta_{j} \delta_{iz}$ that accounts for the (linear)
creation or destruction of gradients of the vertical velocity by the
buoyancy gradients $\theta_{j}$. Equation (\ref{eq:tita_j}) is the
Lagrangian evolution equation for buoyancy gradients. Here, the
nonlinear term $A_{kj} \theta_{k}$ represents the turbulent production
or destruction of buoyancy gradients by strain and vorticity
\cite{gulitski_velocity_2007-1}, while the linear term $N A_{zj}$
corresponds to the (linear) creation or destruction of these gradients
by gradients of the vertical velocity.

We can remove some of the derivatives of the pressure in
Eq.~(\ref{eq:Aij}) by using the incompressibility condition 
${\boldsymbol \nabla} \cdot {\bf u} = A_{ii}=0$, which for the trace
of Eq.~(\ref{eq:Aij}) implies
\begin{equation}
A_{kl}A_{lk} = - \dfrac{\partial^{2}p}{\partial x_{l}\partial x_{l}} -
  N \theta_{z} .
\label{eq:inc}
\end{equation}
The remaining spatial derivatives of the pressure can be written using
the pressure Hessian, defined as
\begin{equation}
H_{ij} = -\left(  \dfrac{\partial^{2}p}{\partial x_{i}\partial x_{j}}
  - \dfrac{\delta_{ij}}{3}  \dfrac{\partial^{2} p}{\partial
    x_{k}\partial x_{k}}  \right) .
\label{eq:hes}
\end{equation}
Using Eqs.~(\ref{eq:inc}) and (\ref{eq:hes}), Eq.~(\ref{eq:Aij}) can
be finally written as
\begin{equation}
\dfrac{DA_{ij}}{Dt} + A_{kj}A_{ik} -\dfrac{\delta_{ij}}{3}
  A_{kl}A_{lk} =  H_{ij} -  N \theta_{j} \delta_{iz} + N \theta_{z}
  \dfrac{\delta_{ij}}{3}.
\label{eq:AijH0}
\end{equation}
This equation, together with Eq.~(\ref{eq:tita_j}), provides a set of
equations (albeit not closed) for the evolution of all components of
${\boldsymbol \nabla}{\bf u}$ and ${\boldsymbol \nabla} \theta$ along
the trajectories of the fluid elements.

To close this set of equations we use an approximation commonly made
in restricted Euler models of homogeneous and isotropic turbulence 
\cite{cantwell_exact_1992, chevillard2006lagrangian,
  meneveau2011lagrangian}, and we assume that the pressure Hessian can
be neglected. In other words, we assume that $H_{ij}\approx 0$ in
Eq.~(\ref{eq:AijH0}). While for the homogeneous and isotropic case
this condition is not well satisfied, in the stratified case the
pressure Hessian becomes relatively smaller as stratification is 
increased. To illustrate this, in Fig.~\ref{f:Hij} we show probability
density functions (PDFs) of pointwise values of $H_{xz}$ and $H_{xx}$
in the direct numerical simulations in table \ref{tab:param},
normalized by the r.m.s.~value of one-third of the trace of the
nonlinear term in Eq.~(\ref{eq:AijH0}), namely
$\left<(A_{kj}A_{jk}/3)^{2}\right>^{1/2}$. For all simulations, more  
than 80\% of fluid elements have normalized components of the pressure
Hessian smaller than $0.2$ (with an increasing percentage of these
fluid elements as $N$ increases). Note also that for $N=12$ an
asymmetry develops in the probability density functions of $H_{xx}$,
associated with the increasing contribution of the buoyancy terms in
Eq.~(\ref{eq:AijH0}). For sufficiently large $N$ it is better to
compare the amplitude of the diagonal terms of $H_{ij}$ against these
terms instead. But in any case, proper estimation of the pressure
Hessian is an open problem that goes beyond this work (see, e.g.,
discussions in \cite{meneveau2011lagrangian}), and in spite of its
apparent smaller relevance here, just as in homogeneous and isotropic
turbulence we need to neglect it in order to obtain a closed set of
equations.

%%%%%%%%%%%%%%%%%%%%%%%%%%%%%%%%%%%%%%%%%% 
\begin{figure}
\includegraphics[width=5.9cm]{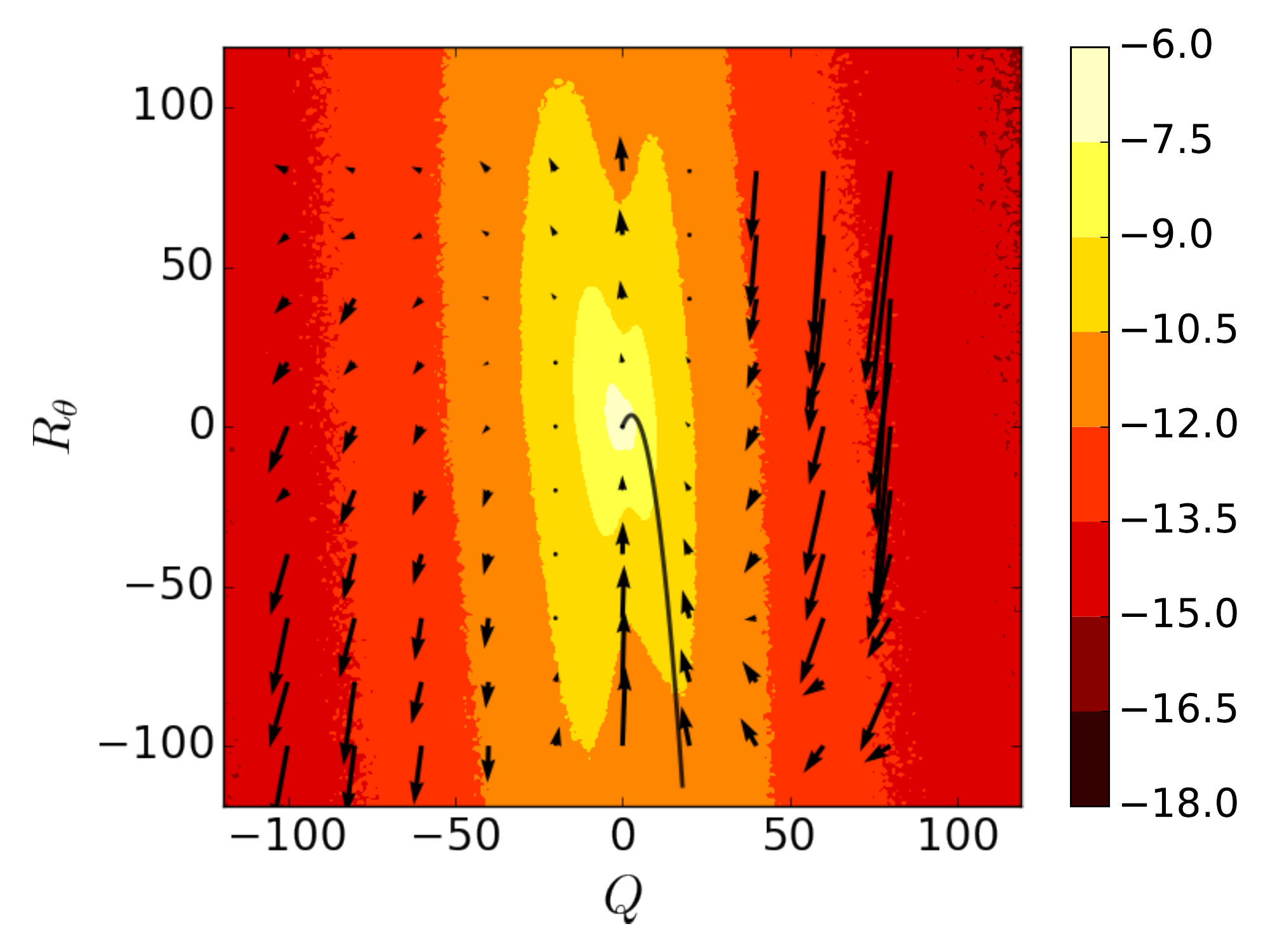}
\includegraphics[width=5.9cm]{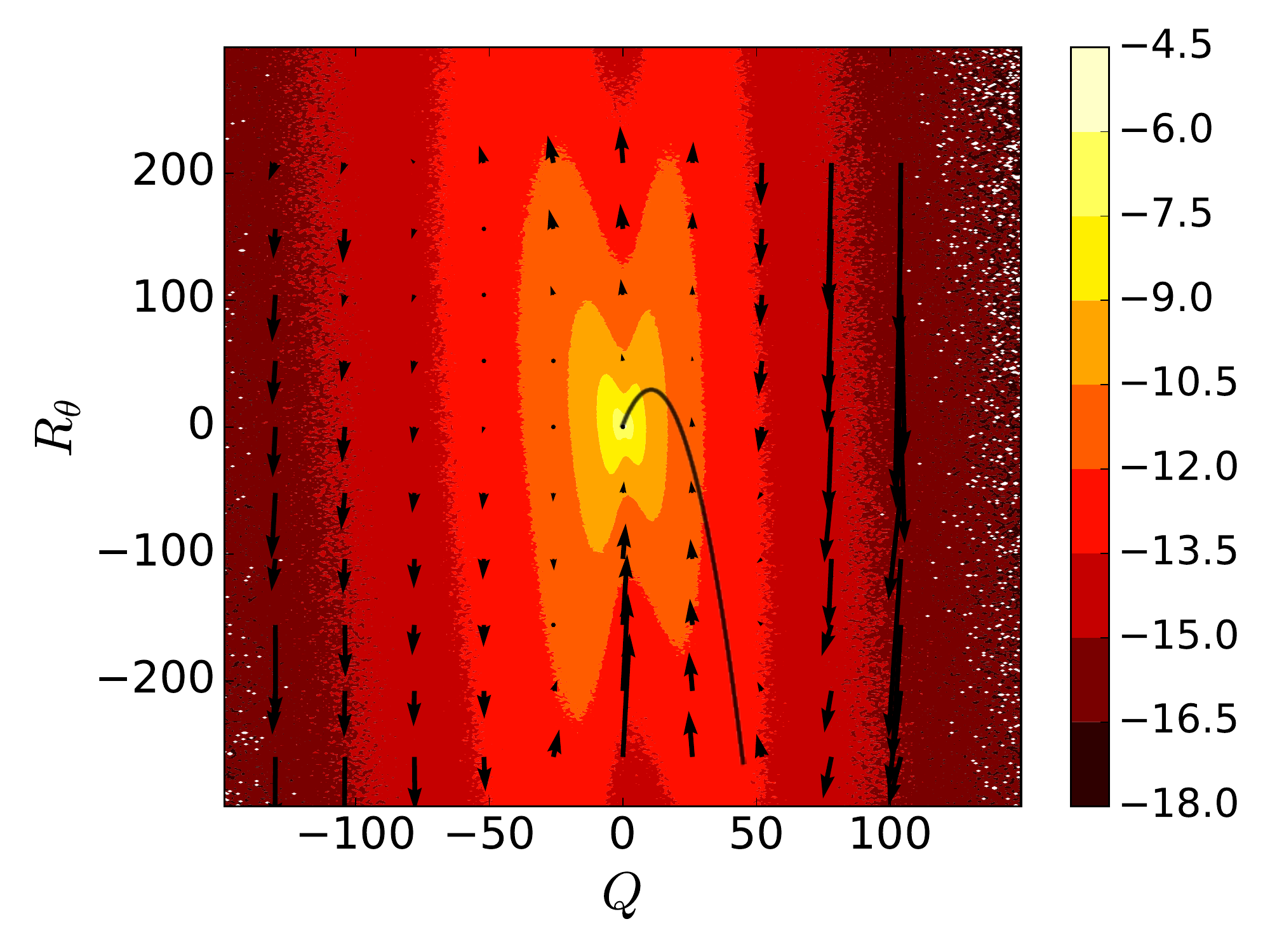}
\includegraphics[width=5.9cm]{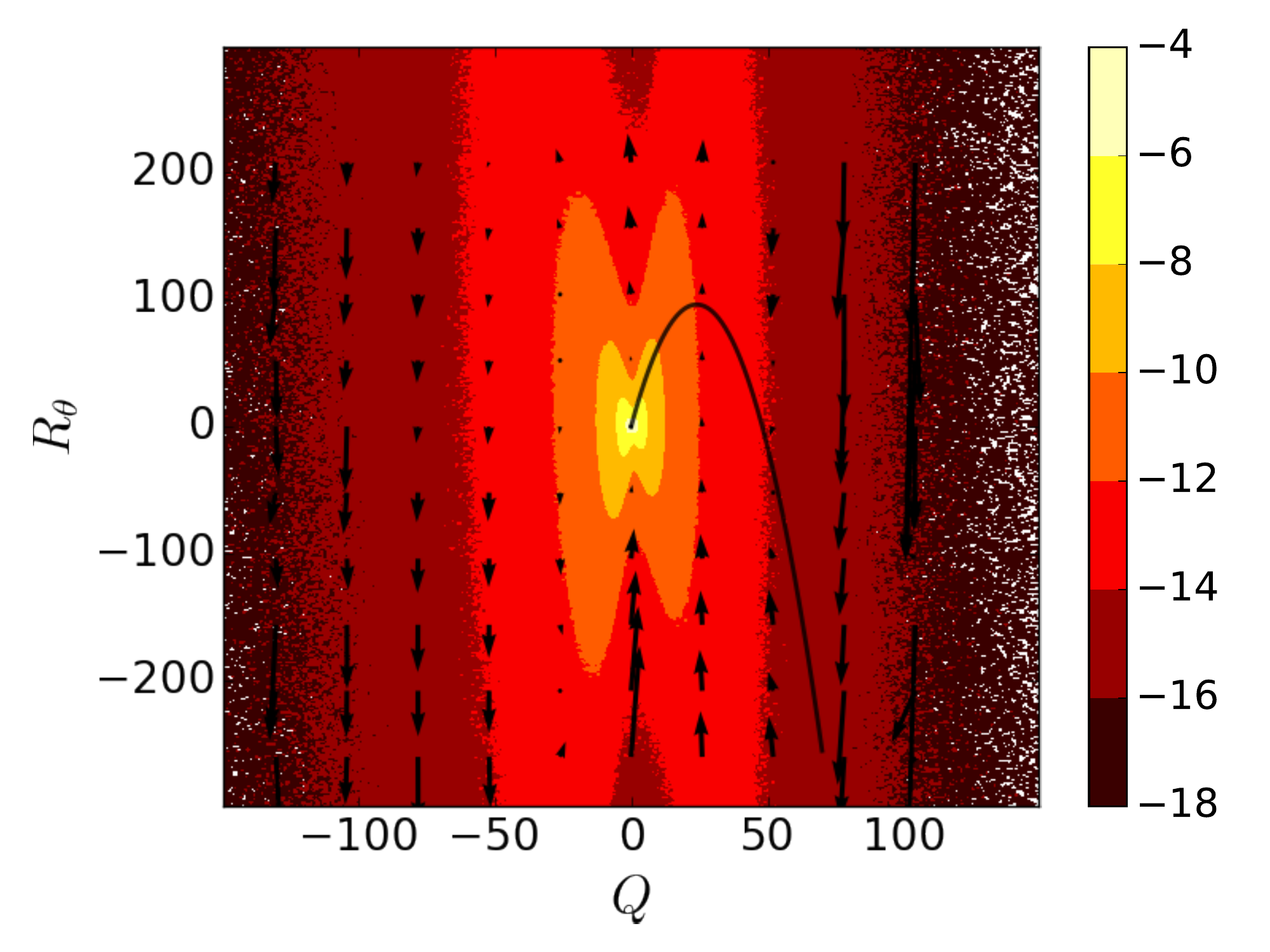}
\caption{Joint probability density functions of $Q$ and $R_{\theta}$
  for $N=4$, $8$, and $12$ (from left to right). Colors represent the
  probability density of finding fluid elements with the corresponding
  values of $Q$ and $R_{\theta}$ in the numerical simulations of the
  Boussinesq equations (color tables here an in the following figures
  are in log scale), while the arrows indicate the direction and speed
  in which fluid elements evolve on the average. The relation
  $R_{\theta}= (2N^{2}Q-6Q^{2}) / (3N)$, associated with fixed point
  $\RN{1}$, is shown as reference with a solid line.}
\label{f:RQtheta}
\end{figure}
%%%%%%%%%%%%%%%%%%%%%%%%%%%%%%%%%%%%%%%%%%

We also want to reduce the information in $A_{ij}$ and $\theta_j$ to
the smallest posible number of scalar quantities that result in an
autonomous system. In the isotropic and homogeneous case, two scalar
quantities (proportional to the traces of ${\bf A}^2$ and of
${\bf A}^3$, with ${\bf A}$ the velocity gradient tensor, and thus
invariant under the group of rotations and reflections) suffice. But
here, as we have more field gradients and as the stratification
introduces anisotropy, the system cannot be closed with just these two
variables. However, using the fact that the system has a preferred
direction and that the equations are axisymmetric around this
direction, we can define
\begin{equation}
\begin{split}
&Q=-A_{ij}A_{ji}/2, \\ 
&R=-A_{ij}A_{jk}A_{ki}/3, \\
&R_{\theta} = \theta_{i}A_{ij}A_{jz}, \\
&B= A_{zi}A_{iz},\\
&T= \theta_{i}A_{iz}, \\
&A=A_{zz}, \\
&S=\theta_{z}.
\end{split}
\label{eq:variables}
\end{equation}

We now want to find Lagrangian evolution equations for these seven
scalars. The equations for the evolution of $Q$ and $R$ are well known
for homogeneous and isotropic turbulence (for a derivation of the
reduced Euler model in this case, see e.g.,
\cite{cantwell_exact_1992}).  To derive an evolution equation for $Q$ 
we evaluate Eq.~(\ref{eq:AijH0}) in $A_{nj}$ and multiply it by
$A_{in}$. Using the derivative product rule we obtain
\begin{equation}
\dfrac{D(A_{in}A_{nj})}{Dt} + 2 A_{ik}A_{kn}A_{nj} - \dfrac{2}{3}A_{ij}
A_{kl}A_{lk} = \dfrac{2}{3} A_{ij}N\theta_{z} - N \theta_{n}
A_{nj}\delta_{iz}-N\theta_{j}A_{iz} ,
\label{eq:AinAnj}
\end{equation}  
where $H_{ij}$ was neglected. Setting $i=j$ in this equation we obtain
$D_{t}{Q}=-3R+NT$ (where $D_{t}$ is shorthand for $D/Dt$). To obtain
an equation for $R$ we multiply Eq.~(\ref{eq:AinAnj}) by the velocity
gradient tensor again, to obtain
 \begin{equation}
\dfrac{D(A_{in}A_{nk}A_{kj})}{Dt} + 3 A_{im}A_{mn}A_{nk}A_{kj} -
  (A_{in}A_{nj})(A_{kl}A_{lk}) = N\theta_{z}A_{ik}A_{kj} - N\theta_{n}
  A_{nj}\delta_{iz}-N\theta_{j}A_{iz}.
\label{eq:AinAnkAkj}
\end{equation}
The trace of this equation results in an equation for the evolution of
$R$, but in order to do so we need to reduce the fourth-order term
$A_{im}A_{mn}A_{nk}A_{ki}$ into second-order terms. To this purpose we
use the Cayley-Hamilton theorem \cite{vieillefosse_local_1982,
  cantwell_exact_1992, meneveau_lagrangian_2011}, which states that
any second-rank tensor $V_{ij}$ satisfies the relation  
\begin{equation}
V_{im}V_{mn}V_{nk}+P_{V} V_{il}V_{lk}+Q_{V} V_{ik}+R_{V}
  \delta_{ik}=0, 
\label{eq:CH}
\end{equation}  
where $P_{V}=V_{ii}$, $Q_{V}=-V_{kl}V_{lk}/2$, and
$R_V=-V_{im}V_{mn}V_{ni}/3$. Taking $V_{ij}=A_{ij}$, then for
incompressible flows $P=0$, $Q_{V}=Q$, and $R_{V}=R$, and
Eq.~(\ref{eq:CH}) reduces to the relation
\begin{equation}
A_{im}A_{mn}A_{nj}=-QA_{ij}-R \delta_{ij}.
\label{eq:CHA}
\end{equation}
Then, using the Cayley-Hamilton theorem, the fourth-order term in 
Eq.~(\ref{eq:AinAnkAkj}) can be written as
$A_{im}A_{mn}A_{nk}A_{ki} = - QA_{ik}A_{ki}-R
  \delta_{ik}A_{ki}=2Q^{2}$. By these means the equation for the
evolution of $R$ reduces to $D_{t}{R}=2Q^{2}/3+2NSQ/3+NR_{\theta}$. 
An evolution equation for $B=A_{zi}A_{iz}$ follows in the same manner
as for $Q$, by taking $i=j=z$ in Eq.~(\ref{eq:AinAnj}), and by
reducing the third-order term $A_{zk}A_{kn}A_{nz}=-QA_{zz}-R$ using
again the Cayley-Hamilton theorem. Then, the equation for $B$ results
$D_{t}{B}=2QA/3+2R-NAS/3-NT$. The equation for
$T=\theta_{k}A_{kz}$ requires using both the $k$ component of 
Eq.~(\ref{eq:tita_j}) multiplied by $A_{kz}$, and the $kz$ component
of Eq.~(\ref{eq:AijH0}) multiplied by $\theta_k$, to obtain
\begin{equation}
\dfrac{D(\theta_{k}A_{kz})}{Dt} + 2 \theta_{m}A_{mk}A_{kz} -
\dfrac{1}{3}\theta_{z} A_{kl}A_{lk}  = N A_{zk}A_{kz} - \dfrac{2}{3}N
\theta_{z}^{2},
\label{eq:T}
\end{equation}
which reduces to $D_{t}{T}=-2R_{\theta} - 2SQ/3 + NB -2NS^{2}/3$. 
To derive an equation for $R_{\theta}=\theta_{k}A_{kl}A_{lz}$ we use the
$k$ component of Eq.~(\ref{eq:tita_j}) and the $kz$ component of
Eq.~(\ref{eq:AinAnj}) in the same way, to compute
\begin{equation}
\dfrac{D(\theta_{k}A_{kl}A_{lz})}{Dt} + 3 \theta_{m}A_{mk}A_{kl}A_{lz} -
\dfrac{2}{3}\theta_{k} A_{kz} A_{kl}A_{lk}  = -\dfrac{4}{3} N\theta_{z}
\theta_{n} A_{nz} + N A_{zk}A_{kl}A_{lz} .
\label{eq:Rtheta}
\end{equation}
To express this equation in terms of our scalar quantities, we use
again the Cayley-Hamilton theorem in Eq.~(\ref{eq:CHA}) for 
$A_{mk}A_{kl}A_{lz}=-QA_{mz}-R\delta_{mz}$ and
$A_{zk}A_{kl}A_{lz}=-QA_{zz}-R$. Then, the evolution equation for
$R_{\theta}$ results
$D_{t}{R_{\theta}}= 5QT/3 + 3RS- 4NST/3 - NQA -NR$. Finally, the
equation for $A$ is obtained by simply taking
$j=i=z$ in Eq.~(\ref{eq:AijH0}) , resulting in
$D_{t}{A}=-B-2Q/3- 2NS/3$, and the equation for $S$ is obtained by
taking $j=z$ in Eq.~(\ref{eq:tita_j}), resulting in $D_{t}{S}=NA-T$.

The resulting reduced model for the Lagrangian evolution of field
gradients of stratified turbulence can be summarized as
\begin{equation}
\begin{split}
\label{eq:ODEs}
&D_{t}{Q}=-3R+NT , \\
&D_{t}{R}=2Q^{2}/3+2NSQ/3+NR_{\theta} , \\
&D_{t}{R_{\theta}}= 5QT/3 + 3RS- 4NST/3 - NQA -NR , \\
&D_{t}{B}= 2QA/3+2R-NAS/3-NT , \\
&D_{t}{T}=-2R_{\theta} - 2SQ/3 + NB -2NS^{2}/3 , \\
&D_{t}{A}=-B-2Q/3- 2NS/3 , \\
&D_{t}{S}=NA-T .
\end{split}
\end{equation}
This system prescribes the evolution of field gradients along the
trajectories of the fluid particles. From the frame of reference of a
Lagrangian particle, these equations are a closed system of seven
ordinary differential equations. As in the reduced Euler model of
homogeneous and isotropic turbulence, the only approximations made to
derive the equations were to assume an ideal and unforced regime, and
to neglect the pressure Hessian $H_{ij}$. In the reduced Euler case,
dropping $H_{ij}$ makes field gradients to diverge and the system to
blow up in finite time. In our system stratification slows down the
blow up, and can prevent it in some cases if $N$ is large enough
\cite{sujovolsky2019invariant}. Nevertheless, the dynamics of this
system, even when it blows up at finite time, provides significant
information on the dynamics of the full set of partial differential
Eqs.~(\ref{eq:n-s_strat})-(\ref{eq:incomp}), as we will show in the
following sections.

%%%%%%%%%%%%%%%%%%%%%%%%%%%%%%%%%%%%%%%%%%
\begin{figure}
\includegraphics[width=5.9cm]{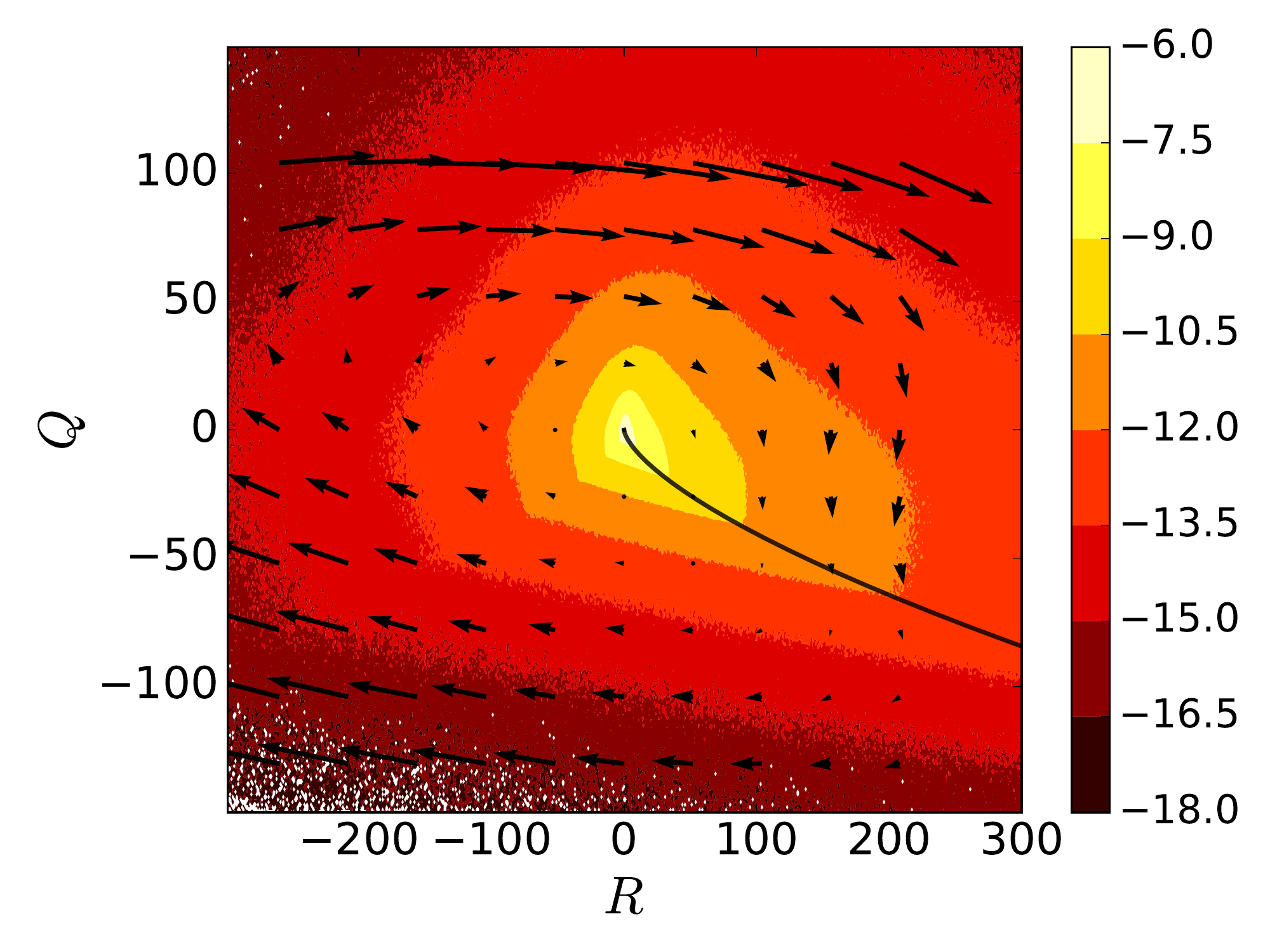}   
\includegraphics[width=5.9cm]{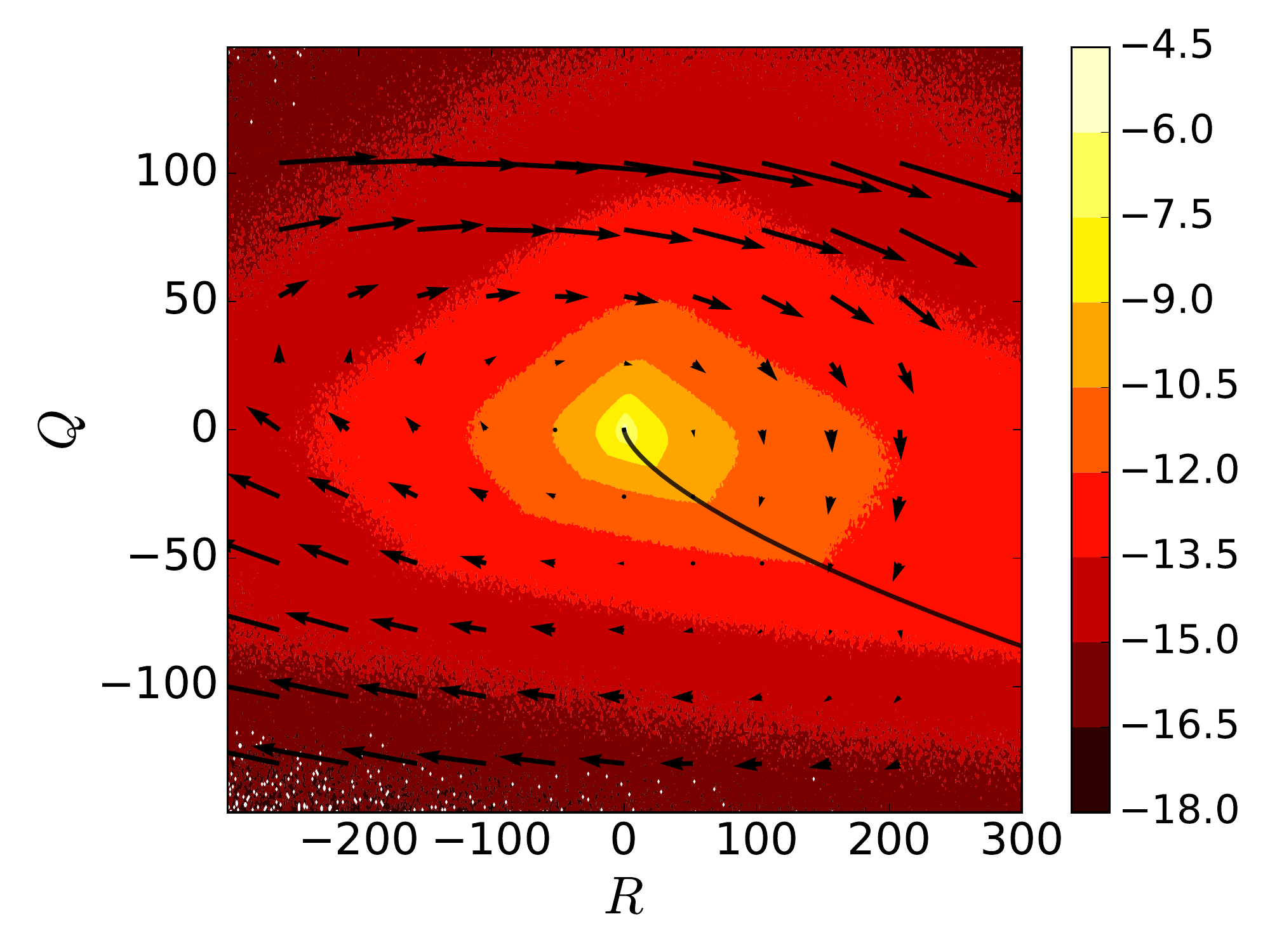}   
\includegraphics[width=5.9cm]{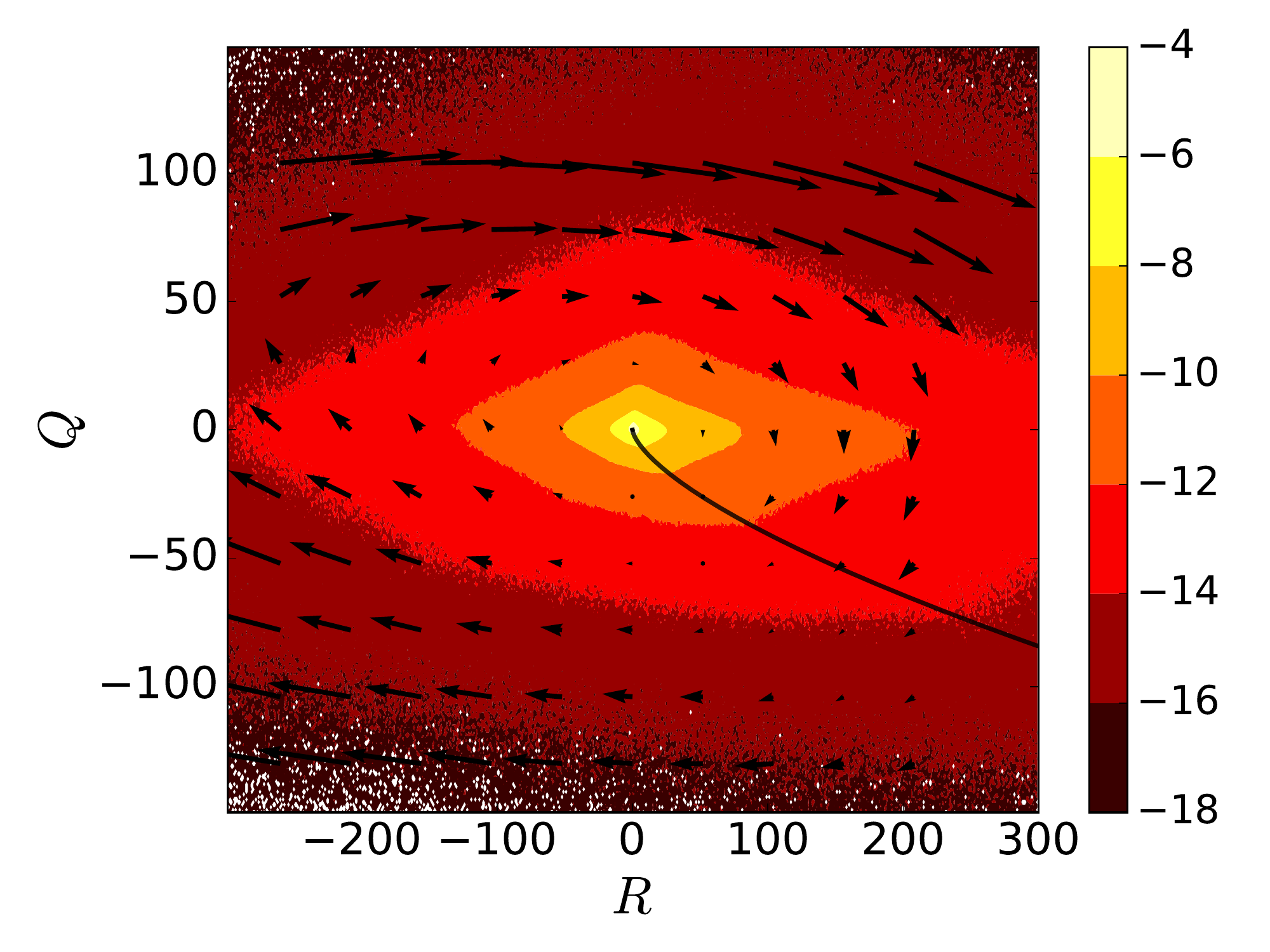}   
\caption{Joint probability density function of $Q$ and $R$ for $N=4$,
  $8$, and $12$ (from left to right). Colors represent the logarithm
  of the probability density of finding fluid elements with the
  corresponding values of $Q$ and $R$ in the numerical simulations of
  the Boussinesq equations, while the arrows indicate the averaged
  direction and speed in which fluid elements evolve. The Vieillefosse
  tail $Q=-(27/4 R^{2})^{1/3}$ is shown as reference with solid
  lines. For homogeneous and isotropic turbulence, this curve
  corresponds to an invariant manifold of the $Q$-$R$ reduced Euler
  model.}
\label{f:RQ}
\end{figure}
%%%%%%%%%%%%%%%%%%%%%%%%%%%%%%%%%%%%%%%%%%

\section{Evolution of fluid elements in phase space \label{sec:phase}}

In this section we characterize the topological properties of the
phase space of the reduced system in Eq.~(\ref{eq:ODEs}), and we
compare its predictions with those obtained from the time evolution of
field gradients in the direct numerical simulations of the full
Boussinesq equations. From the simulations in table \ref{tab:param} we
compute $Q$, $R$, $R_{\theta}$, $T$, $B$, $A$, and $S$ for all fluid
elements, and their time derivatives, and embed these quantities in
the phase space defined by the reduced model. Compared with previous
results presented in \cite{sujovolsky2019invariant}, here we will not
only study whether fluid elements in the full system accumulate in
certain regions of the phase space of the reduced system, but also how
fast or slow they evolve depending on the region in which they are.

This section is organized as follows: First we introduce the fixed
points of the system, for which the time derivatives of all the
scalars are zero. Associated to fixed points, dynamical systems can
have invariant manifolds, which are constrained regions of the phase
space where points (i.e., states of the system) can only evolve to
other points in the same manifold. In these manifolds, the
dimensionality of the system is further reduced
\cite{wiggins2003introduction}. We consider the invariant manifold
that is known to play a relevant role in the homogeneous and isotropic
case, and study how stratification modifies this manifold and
introduces new invariant manifolds. Finally, after discussing the role
of these manifolds and their physical relevance, we propose a
simplified picture of the phase space of stably stratified
turbulence.

%%%%%%%%%%%%%%%%%%%%%%%%%%%%%%%%%%%%%%%%%%
\begin{figure}
\includegraphics[width=5.9cm]{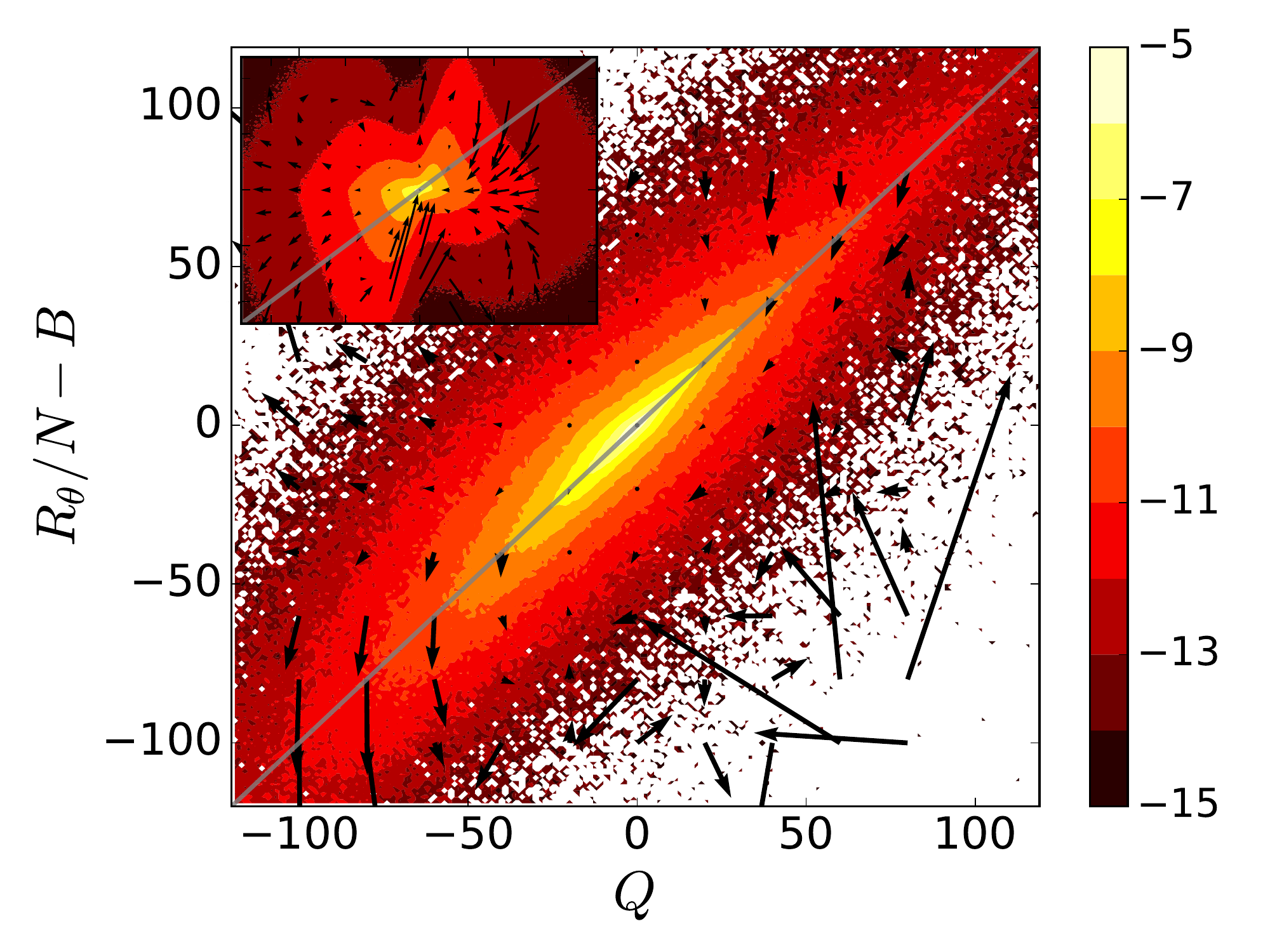}
\includegraphics[width=5.9cm]{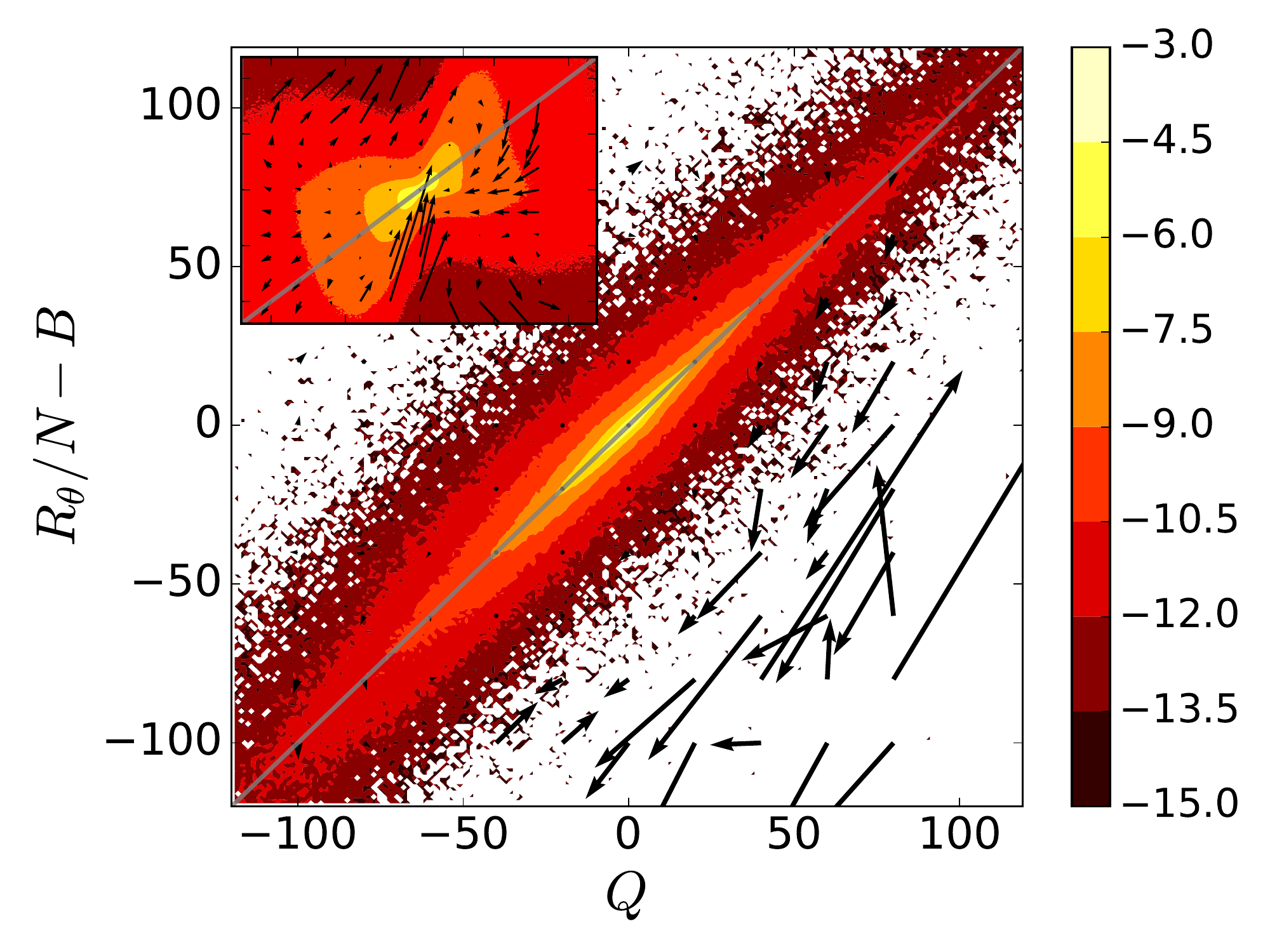}
\includegraphics[width=5.9cm]{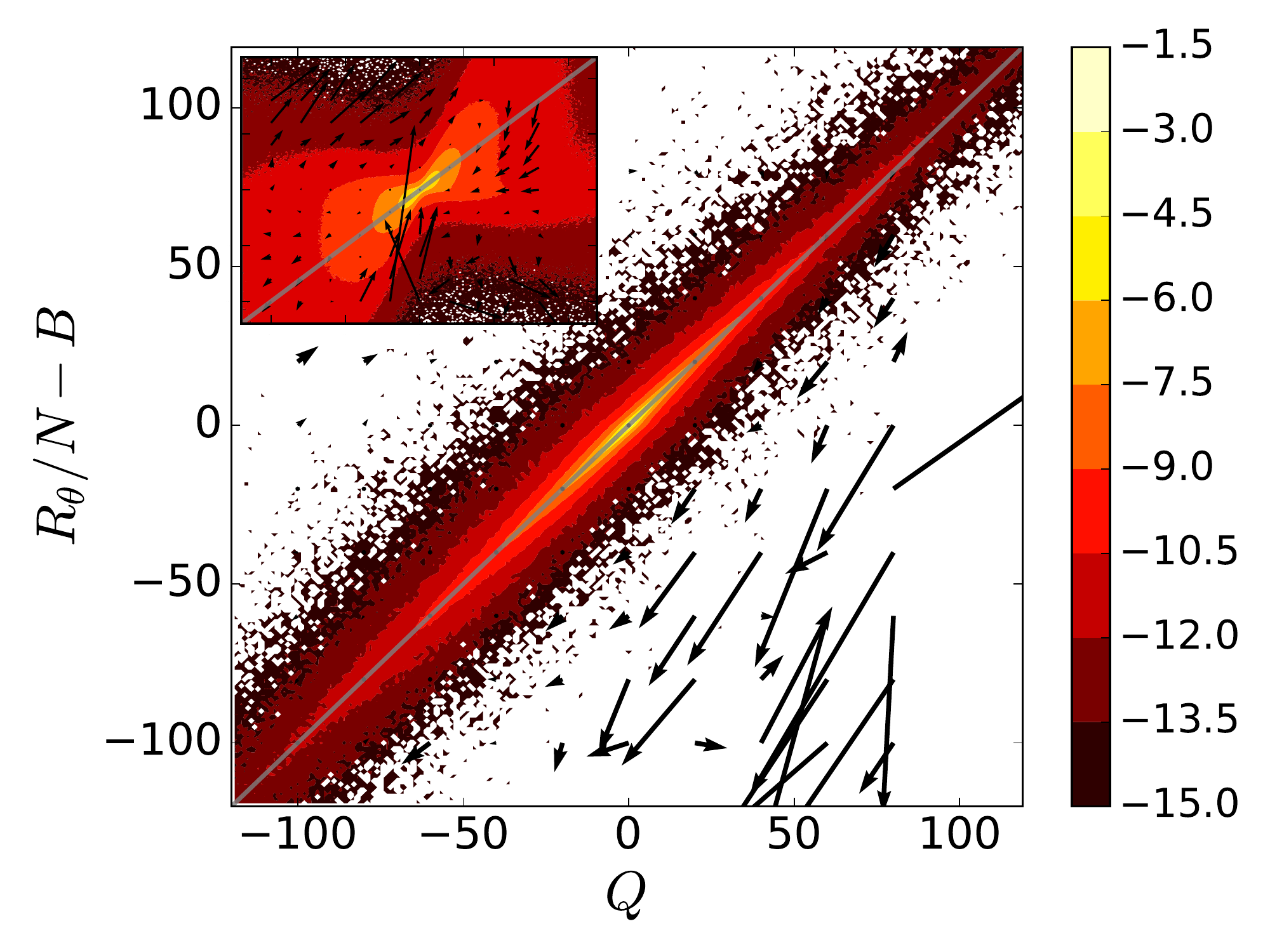}
\caption{Joint probability density function of
  $R_{\theta}/N-B$ and $Q$ for simulations of the full Boussinesq
  equations with $N=4$, 8, and 12 (from left to right), restricted to
  fluid elements with $S\approx 0$. Arrows indicate the mean rate of
  change of these quantities in the direct numerical simulations. The
  invariant manifold $\Sigma_\RN{1}$ with $Q=R_{\theta}/N-B $ is shown
  as a reference by solid lines. Insets show the same quantities but
  for all fluid elements (i.e., without any restriction on the
  possible value of $S$), and with the same ranges for the axes as the
  main figures. Insets in the following figures also follow this
  convention.}
\label{f:Q_RT-B}
\end{figure}
%%%%%%%%%%%%%%%%%%%%%%%%%%%%%%%%%%%%%%%%%%

\subsection{Fixed points}

We start by listing the fixed points of the reduced model for the
Lagrangian evolution of field gradients. The system in
Eq.~(\ref{eq:ODEs}) has two sets of fixed points,
\begin{equation}
\begin{split}
\RN{1}:\ &Q=R=R_{\theta}=T=B=A=S=0 ,  \\
\RN{2}:\ &R_{\theta}=\frac{2N^{2}Q-6Q^{2}}{3N},\
  B=\frac{2N^{2}}{3}-2Q, \ T=\frac{3R}{N}, \ A=\frac{3R}{N^{2}},\
  S=\frac{2Q}{N}-N , \ Q \textrm{ and } R \textrm{ free.} \\
\end{split}
\end{equation}

Fixed point $\RN{1}$ corresponds to null gradients of ${\bf u}$ and
$\theta$. Unlike the fixed point $Q=R=0$ in the reduced Euler
model for homogeneous and isotropic turbulence (i.e., for $N=0$, see
\cite{vieillefosse_local_1982, cantwell_exact_1992,
  chevillard2006lagrangian, meneveau2011lagrangian} and the
discussion below), this fixed point is not obviously unstable. When
the system in Eq.~(\ref{eq:ODEs}) is linearised around this fixed
point, the linear system of ordinary differential equations can be
represented by a matrix whose eigenvalues provide information of how
fluid elements evolve when perturbed in its vicinity. The array has
one eigenvalue equal to 0, four eigenvalues equal to $\pm iN$, and two
eigenvalues equal to $\pm i\sqrt{6}N/2$. As all eigenvalues are zero or
purely imaginary, small perturbations around fixed point $\RN{1}$
results in oscillations (between linear combinations of $Q$, $T$, $B$,
and $S$ in one case, and $A$ and $S$ in the other). Also because all
eigenvalues are zero or purely imaginary, center manifold theory can
be used to further reduce the dimensionality of the system
\cite{wiggins2003introduction} (see
Sec.~\ref{sec:local1}). Physically, this fixed point corresponds to
the stratified fluid at equilibrium, and linearly perturbing this
solution results in the excitation of internal gravity waves as
follows from Eq.~(\ref{eq:disp}). The behavior described here is the
counterpart for field gradients of the well known fixed point of the
full Boussinesq partial differential equations for ${\bf u}=\theta=0$ 
\cite{davidson_turbulence_2013}. Of course, if the fluid elements are 
perturbed far away from this equilibrium, nonlinearities will become
relevant and fluid elements may run away in phase space.

The other solution, fixed point $\RN{2}$, actually corresponds to a
manifold of fixed points, as $Q$ and $R$ are free (i.e., for each
value of these variables we have a fixed point in the system). To
confirm their presence in the full Boussinesq partial differential
equations, Fig.~\ref{f:RQtheta} shows the joint probability density
function of $R_{\theta}$ and $Q$ (i.e., the probability density of
finding fluid elements with different values of these two variables)
obtained from the direct numerical simulations, with arrows indicating
the mean direction and speed at which fluid elements evolve in this
plane (also obtained from the full Boussinesq system). Superimposed,
we show the relation $R_{\theta}=(2N^{2}Q-6Q^{2})/(3N)$ which
corresponds to fixed points $\RN{2}$ of the reduced model. There is a
correlation between points in the direct numerical simulations and
this relation, in the sense that arrows are small in the vicinity of
the fixed points (albeit there is no significant accumulation of fluid
elements in the manifold defined by these fixed points). Other lobes
showing significant accumulation of probability, as well as other
regions with slow evolution (i.e., with small arrows, or equivalently,
with small $DR_{\theta}/Dt$ and $DQ/Dt$) are associated to projections
into this $R_{\theta}$-$Q$ plane of slow manifolds in the reduced
system that will be discussed next (note these figures, and all
following figures, correspond to projections into planes of a 7
dimensional phase space).

%%%%%%%%%%%%%%%%%%%%%%%%%%%%%%%%%%%%%%%%%%
\begin{figure}
\includegraphics[width=5.9cm]{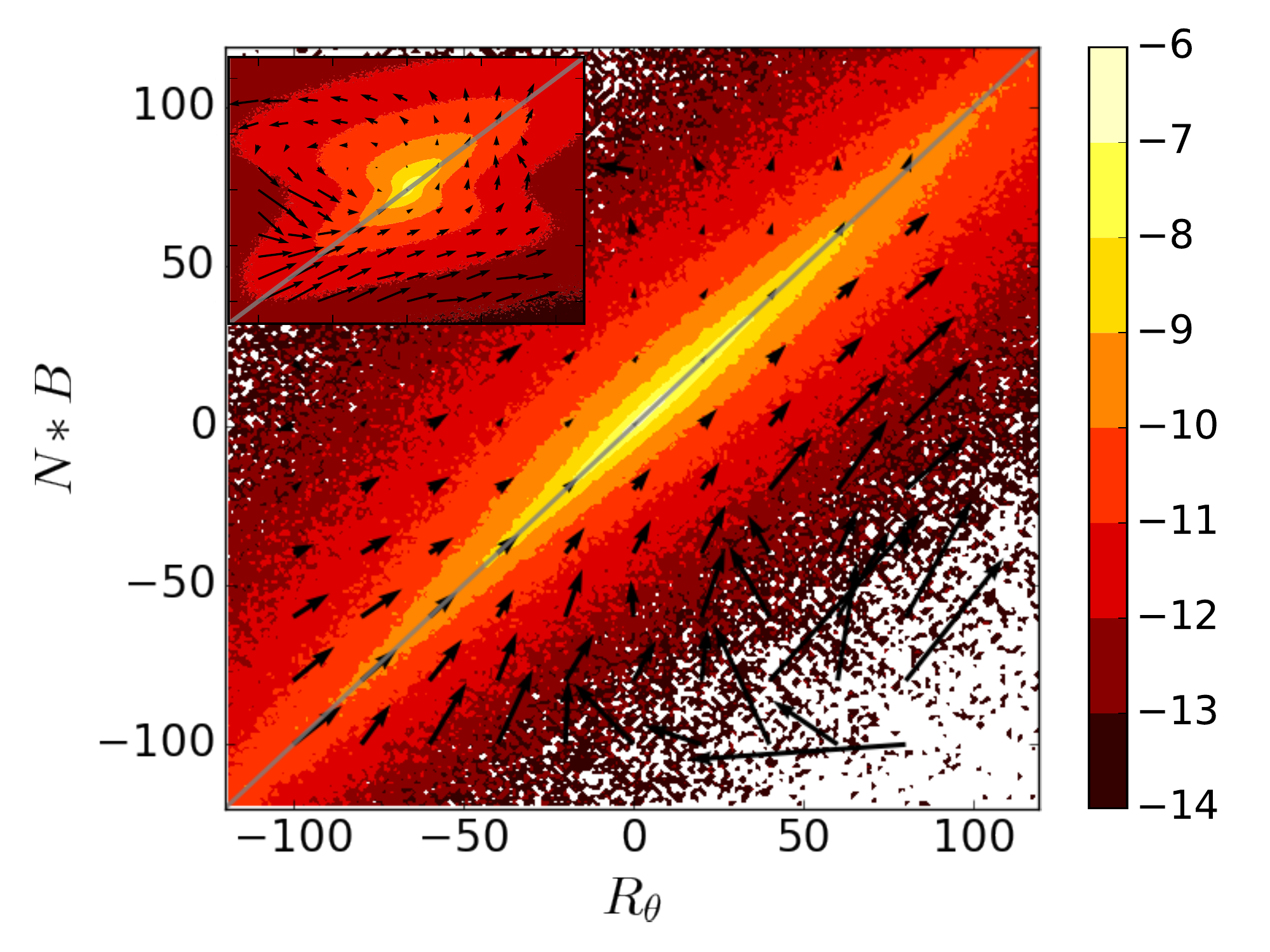}
\includegraphics[width=5.9cm]{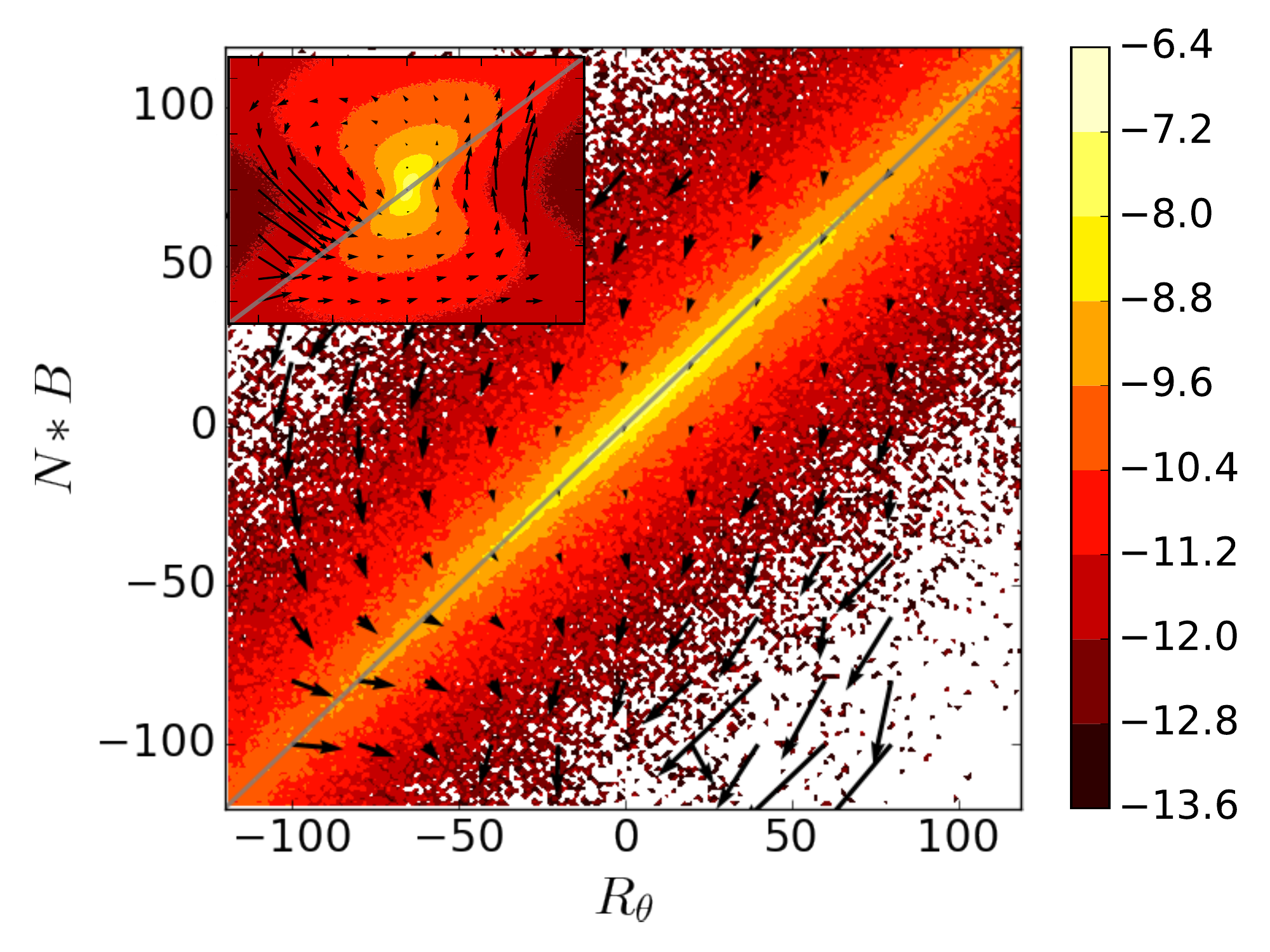}
\includegraphics[width=5.9cm]{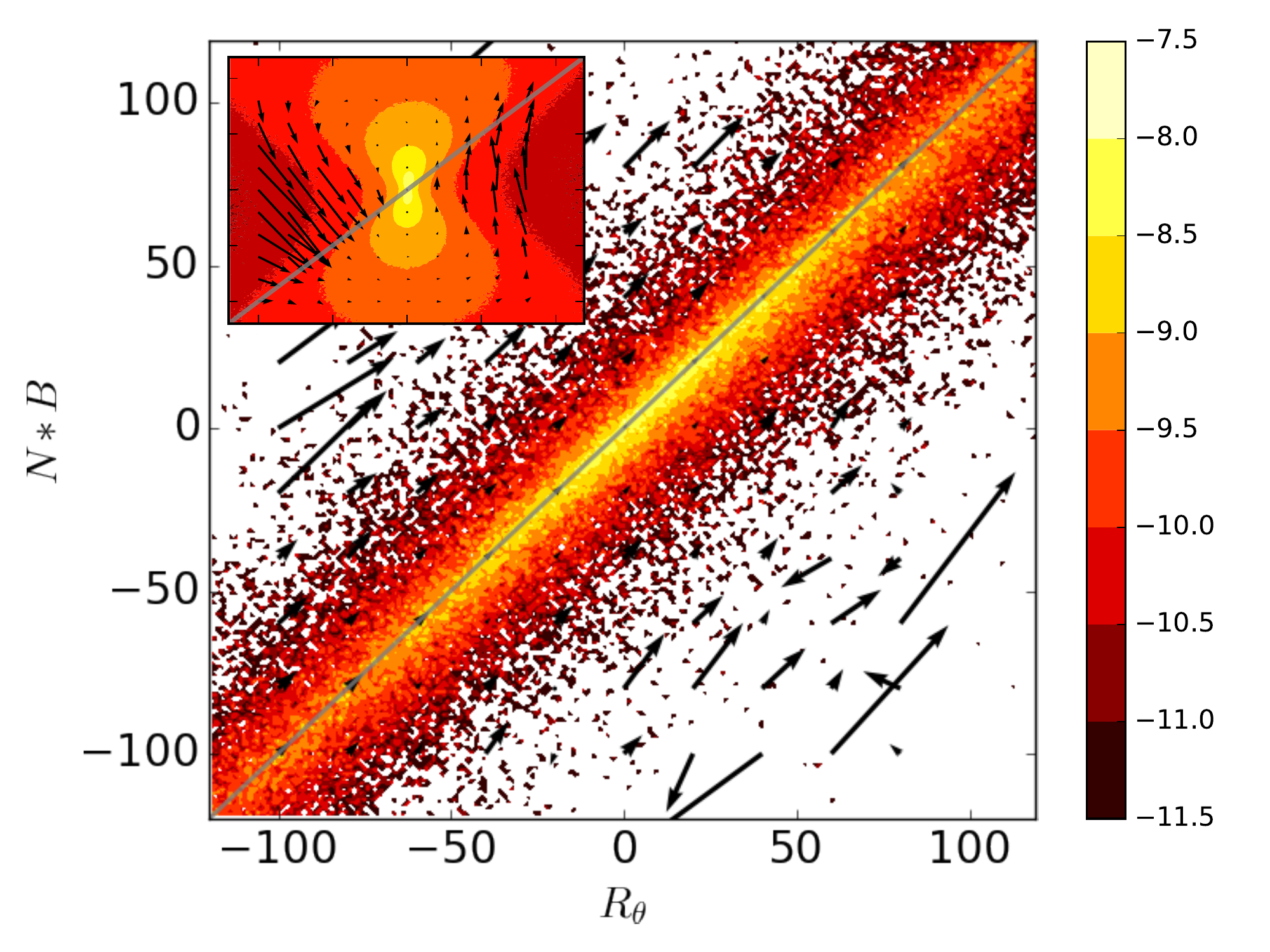}
\caption{Joint probability density function of $NB$ and $R_{\theta}$
  for direct numerical simulations with $N=4$, 8, and 12 (from left to
  right), together with arrows indicating mean rate of change of these
  quantities, restricted to fluid elements with $S\approx N$. The curve
  $NB=R_{\theta}$, corresponding to the invariant manifold
  $\Sigma_0$, is indicated as a reference by the solid lines.
  Insets show the same quantities but for all fluid elements (without
  any restriction on the possible value of $S$).}
\label{f:RT_NB}
\end{figure}
%%%%%%%%%%%%%%%%%%%%%%%%%%%%%%%%%%%%%%%%%%

%%%%%%%%%%%%%%%%%%%%%%%%%%%%%%%%%%%%%%%%%%
\begin{figure}
\includegraphics[width=5.9cm]{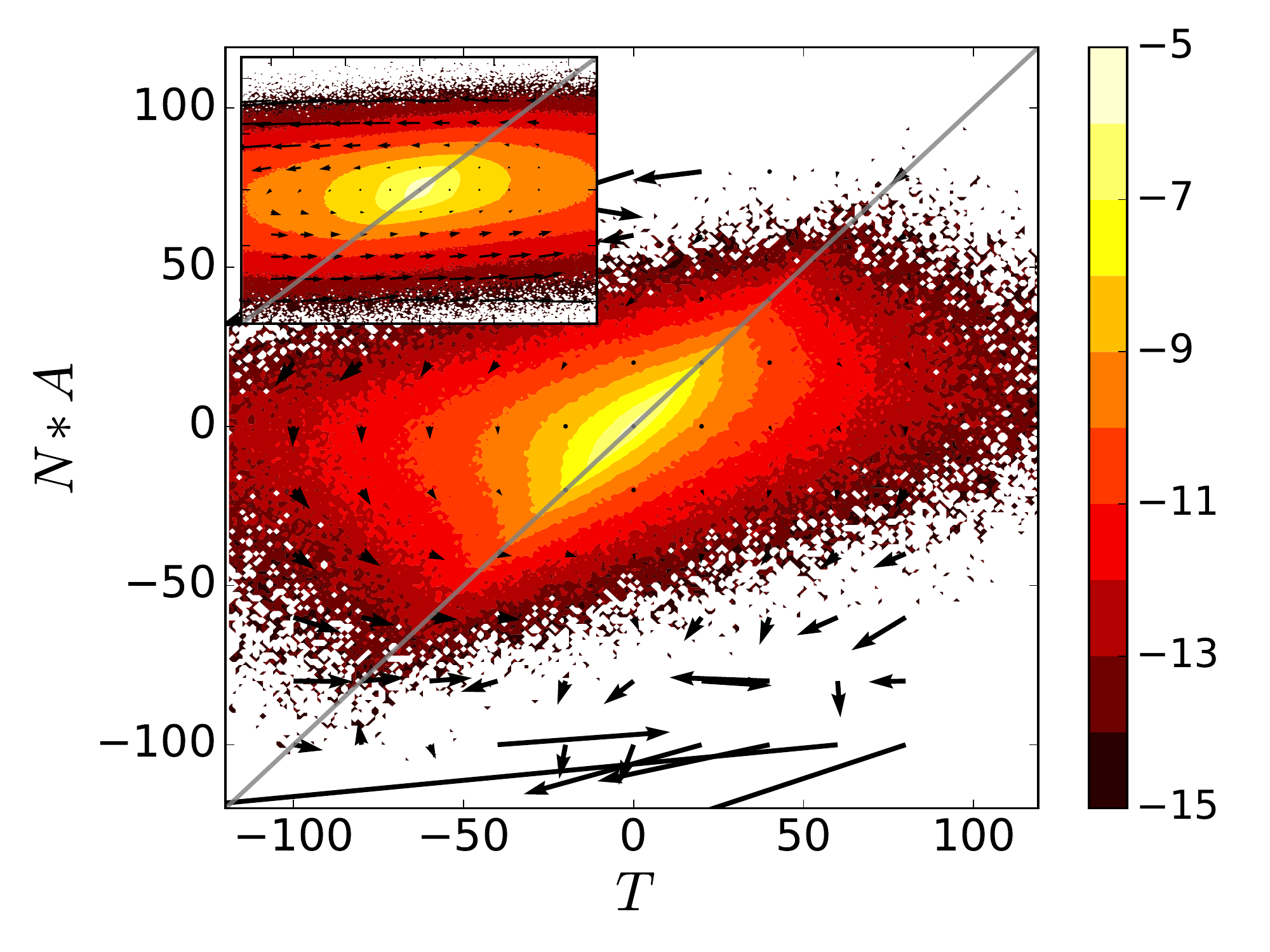}
\includegraphics[width=5.9cm]{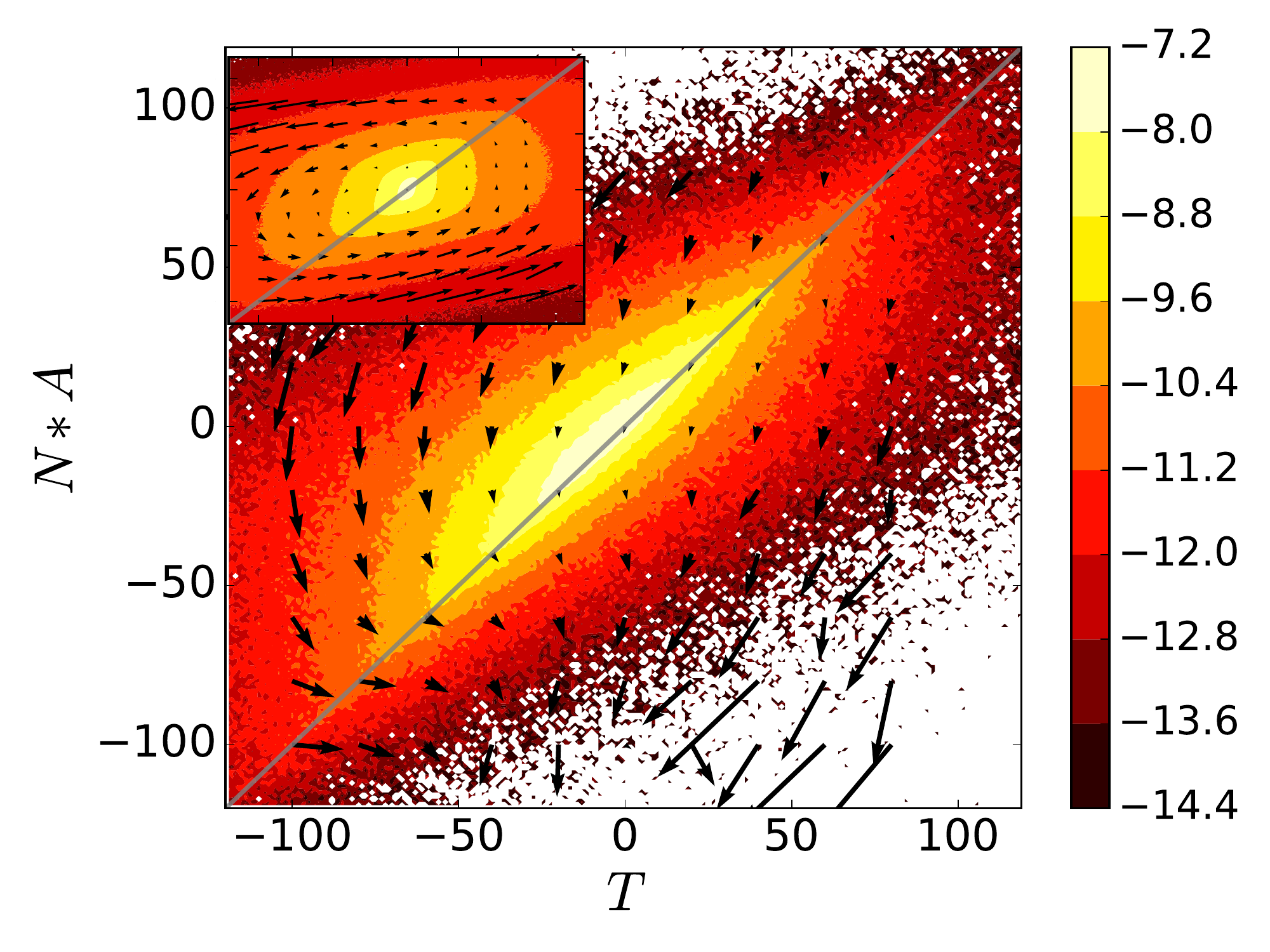}
\includegraphics[width=5.9cm]{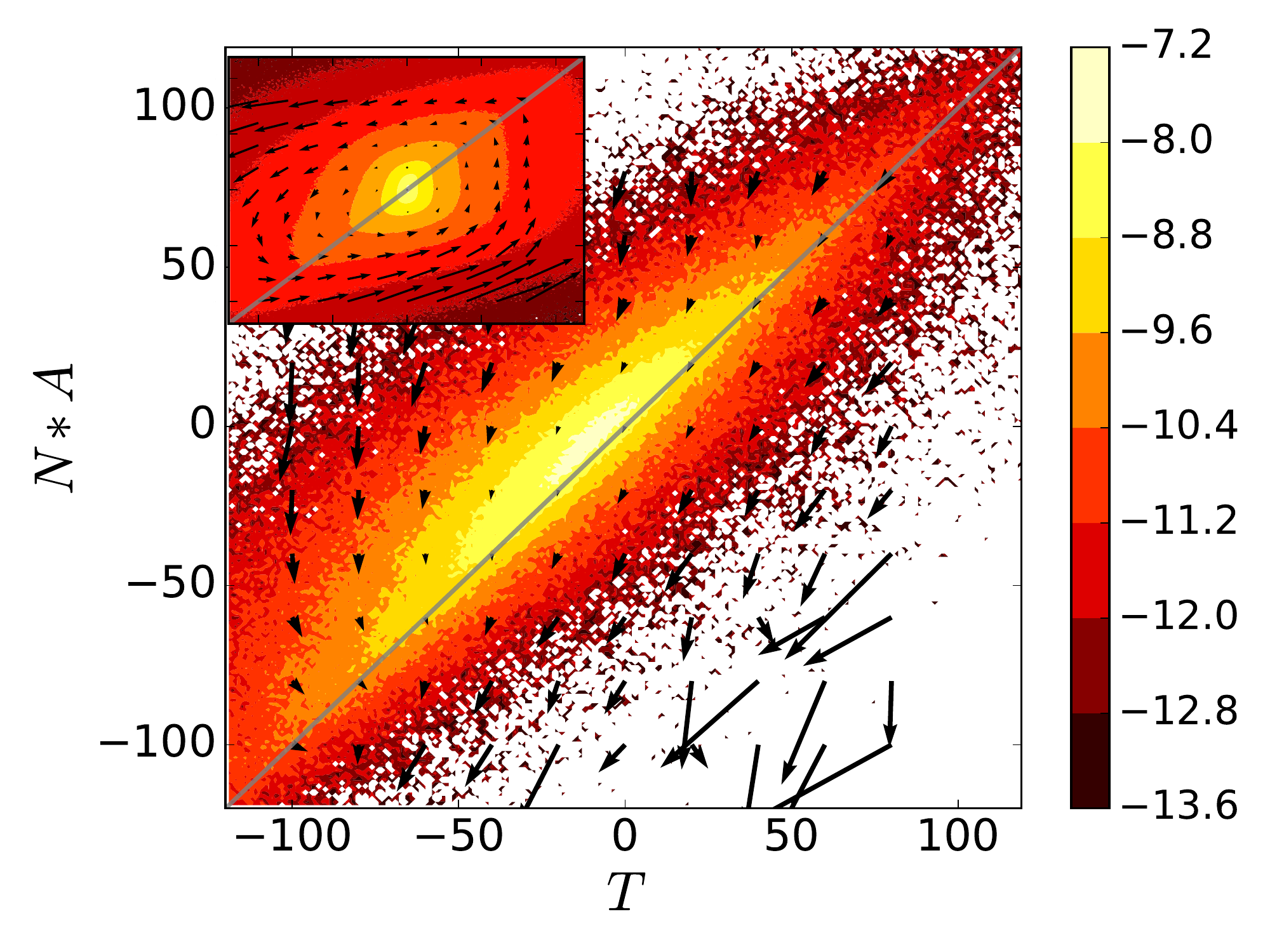}
\caption{Joint probability density function of $NA$ and $T$ for direct
  numerical simulations with $N=4$, 8, and 12 (from left to right),
  together with arrows indicating mean rate of change of these
  quantities, restricted to fluid elements with $S\approx N$. The
  curve $NA=T$, corresponding to the invariant manifold $\Sigma_0$, is
  indicated as a reference by the solid lines. Insets show the same
  quantities but for all fluid elements (without any restriction on
  the possible value of $S$).}
\label{f:T_NA}
\end{figure}
%%%%%%%%%%%%%%%%%%%%%%%%%%%%%%%%%%%%%%%%%%

\subsection{The Vieillefosse manifold \label{sec:manifolds}}

As mentioned before, an invariant manifold is a region of phase space
that is invariant under the action of the dynamical system (i.e., such
that initial conditions in this manifold remains in the same manifold
as time evolves). These manifolds are often constructed by perturbing
dynamical systems around fixed points, although global invariant
manifolds (i.e., valid up to any order in the nonlinearity) can also
exist.

In the case with $N=0$ (no stratification, homogeneous and isotropic
flows), the reduced system in Eq.~(\ref{eq:ODEs}) further reduces to
the so-called Vieillefosse or reduced Euler model for $Q$ and $R$
(which are the two rotationally invariant scalars obtained from the
traces of ${\bf A}^2$ and ${\bf A}^3$, and proportional to sums and
products of the eigenvalues of the $A_{ij}$ tensor)
\cite{vieillefosse_local_1982, cantwell_exact_1992,
  chevillard2006lagrangian, meneveau2011lagrangian},
\begin{equation}
\begin{split}
\label{eq:QR}
&D_{t}{Q}=-3R, \\
&D_{t}{R}=2Q^{2}/3 ,
\end{split}
\end{equation}
and also to a reduced system of ordinary differential equations for
the gradients of an isotropic passive scalar. The system in
Eqs.~(\ref{eq:QR}) has only one fixed point, $Q=R=0$, and one
invariant manifold, the so-called Vieillefosse tail given by
$Q=-(27/4 R^{2})^{1/3}$ \cite{vieillefosse_local_1982,
  meneveau_lagrangian_2011}, as it can be shown that
$D_t(4Q^3/27+R^2)= 0$. The reduced Euler system blows up
following this manifold, with gradients growing to arbitrarily large
(negative) values of $Q$. However, this manifold plays a crucial role
in the dynamics even in the viscous and forced Navier-Stokes case, as
it can be shown that its existence is associated with the observed
alignment of vorticity with an eigenvector of the strain-rate tensor,
with the phenomenon of vortex stretching, and with the development of
extreme events in the flow \cite{Chong_1990, gulitski_velocity_2007,
  meneveau_lagrangian_2011, Dallas_2013}. 

In the stratified case ($N \neq 0$), from the reduced system in
Eq.~(\ref{eq:ODEs}) this relation is replaced by
\begin{equation}
\frac{D}{Dt} \left( \frac{4Q^3}{27}+R^2 \right)= \frac{3N}{2}
  \left[ 3NR (3B + 2Q) + 2Q^2T \right] ,
\end{equation}
and thus, as $N$ increases from zero, this manifold stops being
invariant and it becomes less relevant for the dynamics. Indeed, in the
direct numerical simulations we observe that for small values of $N$
some fluid elements still accumulate near this manifold, but that this
accumulation decreases as $N$ increases. To illustrate this,
Fig.~\ref{f:RQ} shows the isocontours of the joint probability density
functions of $Q$ 
and $R$ for all simulations. Besides the change in the PDFs, there is
also a change in the rate of change of $Q$ and $R$: For $N=4$
vectors are small near the Vieillefosse tail, and tend to align with
this manifold in its vicinity. Instead, for $N=8$ and 12, although
fluid elements still slow down near the manifold, they cross it and
they orbit around the point $Q=R=0$. This corresponds, as mentioned
above, to the wave-like motions near fixed point $\RN{1}$, and also
indicates that the mechanism of vortex stretching (see, e.g.,
\cite{Chong_1990, Dallas_2013}) is significantly affected by the
stratification, in agreement with the well known change in the
geometry of vortical structures in stably stratified turbulence from
tubes to pancakes. Indeed, note that $Q$ can be written as
\cite{Dallas_2013}
\begin{equation}
  Q = \frac{1}{4}\left[ {\boldsymbol \omega}^2 -
    2 \, \textrm{Tr} ({\bf s}^2) \right] ,
\end{equation}
where ${\boldsymbol \omega} = {\boldsymbol \nabla} \times {\bf u}$ is
the vorticity, Tr denotes the trace, and
${\bf s}=({\bf A} + {\bf A}^T)/2$ is the strain-rate tensor. As a
result, fluid elements with $Q>0$ are dominated by vorticity, and
fluid elements with $Q<0$ are dominated by strain. Also, for $R>0$
flow topologies are stable, while for $R<0$ flow topologies are
unstable. Finally, it can be also shown (see, e.g.,
\cite{Dallas_2013}) that regions in Fig.~\ref{f:RQ} with $R>0$ and
$Q>0$ correspond to unstable compression of vortices, regions with
$R>0$ and $Q$ below the Vieillefosse tail correspond to unstable
vortex sheet structures, and regions with $R<0$ correspond to stable
tube-like structures (which can be also stretched depending on the 
value of $Q$). Thus, the apparent accumulation (for sufficiently large
$N$) of fluid elements near $Q\approx 0$ and with any sign of $R$
seems to indicate structures alternate between all these possible
configurations in the stably stratified case (the local structure of
the flow will be discussed in more detail in
Sec.~\ref{sec:alignment}).

%%%%%%%%%%%%%%%%%%%%%%%%%%%%%%%%%%%%%%%%%%
\begin{figure}
\includegraphics[width=8.5cm]{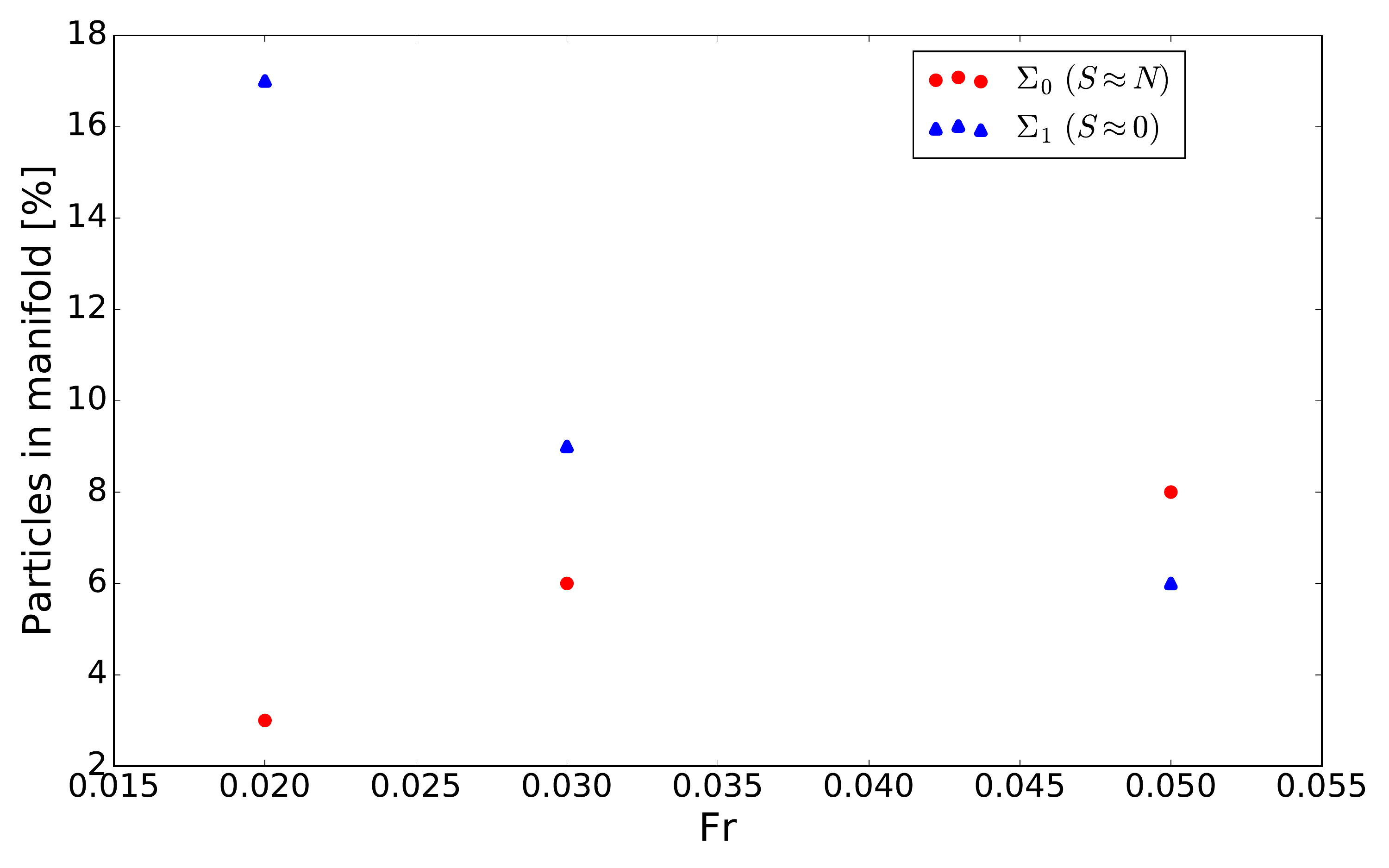}   
\caption{Fraction of fluid elements (in percentage) in manifolds
  $\Sigma_0$ (with $S\approx N$) and $\Sigma_\RN{1}$ (with
  $S\approx 0$) as a function of Fr, for all simulations in table
  \ref{tab:param} (i.e., for $N=12$, 8, and 4, in that order with
  increasing Fr). Notice how the percentage of fluid elements in the
  wave-like invariant manifold ($\Sigma_\RN{1}$) decrease with
  increasing Fr, while the percentage of fluid elements at the onset
  of convection (i.e., in $\Sigma_0$) increase with increasing Fr, at
  least for the values of Re considered here.}
\label{f:S0_SN}
\end{figure}
%%%%%%%%%%%%%%%%%%%%%%%%%%%%%%%%%%%%%%%%%%

\subsection{The invariant manifold in the vicinity of fixed point
  $\RN{1}$  \label{sec:local1}}

An expansion of the ordinary differential equations in
Eq.~(\ref{eq:ODEs}) in the vicinity of fixed point $\RN{1}$
shows that 
\begin{equation}
\Sigma_{\RN{1}}: \frac{D}{Dt} \left( \frac{R_{\theta}}{N}-B-Q \right)
  = 0 ,
\end{equation}
i.e., $R_{\theta}/N-B-Q=0$ is a local invariant manifold of the
system.

Figure \ref{f:Q_RT-B} shows the joint probability density functions of
$R_{\theta}/N-B$ and $Q$ for the direct numerical simulations,
together with arrows indicating the mean rate of change of these
quantities in the phase space, only for fluid elements with $S\approx 0$
(as expected near fixed point $\RN{1}$), using the same criteria for
the selection of fluid elements as in \cite{sujovolsky2019invariant}
(briefly, here and in the following conditions $S\approx 0$ or
$S\approx N$ mean all particles satisfying any of these conditions
within 10\% of the value of $N$ were chosen). The insets in
Fig.~\ref{f:Q_RT-B} show the same probability density functions but
without any restriction on the values of $S$. For the restricted
cases, and for all simulations with different values of $N$, a strong
correlation between $R_{\theta}/N-B$ and $Q$ is seen, as evidenced by
the larger probability densities near the  manifold
$R_{\theta}/N-B=Q$. Moreover, the correlation increases with
increasing $N$. And this correlation is seen even for large values of
$R_{\theta}/N-B$ and $Q$, in spite of the fact that the invariant
manifold is local (i.e., only expected to be valid in the vicinity of
the fixed point $\RN{1}$).

Rates of change are very small near the invariant manifold
$R_{\theta}/N-B=Q$, and large far from it, more clearly in the
simulation with $N=12$ (note the disparate size of the arrows in
different regions of phase space in Fig.~\ref{f:Q_RT-B}). This
suggests that fluid elements evolve slowly near this manifold, or in
other words, that the local manifold $\Sigma_{\RN{1}}$ is not only
preserved in the full Boussinesq system, but is also a slow manifold
of the system.

\subsection{The invariant manifold in the vicinity of fixed point
  $\RN{2}$  \label{sec:local2}}

Only valid in the vicinity of fixed point $\RN{2}$, the system in
Eq.~(\ref{eq:ODEs}) has two invariant manifolds,
\begin{equation}
\begin{split}
\Sigma_{\RN{2},a}: \ &Q=\frac{N^{2}}{3}, \ R \textrm{ free}, \
  \frac{D}{Dt} \left( -4R+NT+N^{2}A \right)=0, \\
\Sigma_{\RN{2},b}: \ &Q=-\frac{N^{2}}{3}, \ R \textrm{ free}, \
  \frac{D}{Dt} \left( R_{\theta}-NB + \frac{8NQ}{3} \right)=0.
\end{split}
\end{equation}

But in order to reach fixed points $\RN{2}$, the system needs a value
of $Q$ or $B$ of $O \left(N^{2}\right)$, so this set of fixed points
are hard to access. Indeed, analysis of the direct numerical
simulations indicate that the invariant manifolds associated with
fixed points $\RN{2}$ do not seem to have a relevant role in the
dynamics of the fluid elements.

%%%%%%%%%%%%%%%%%%%%%%%%%%%%%%%%%%%%%%%%%%
\begin{figure}
\includegraphics[width=8.5cm]{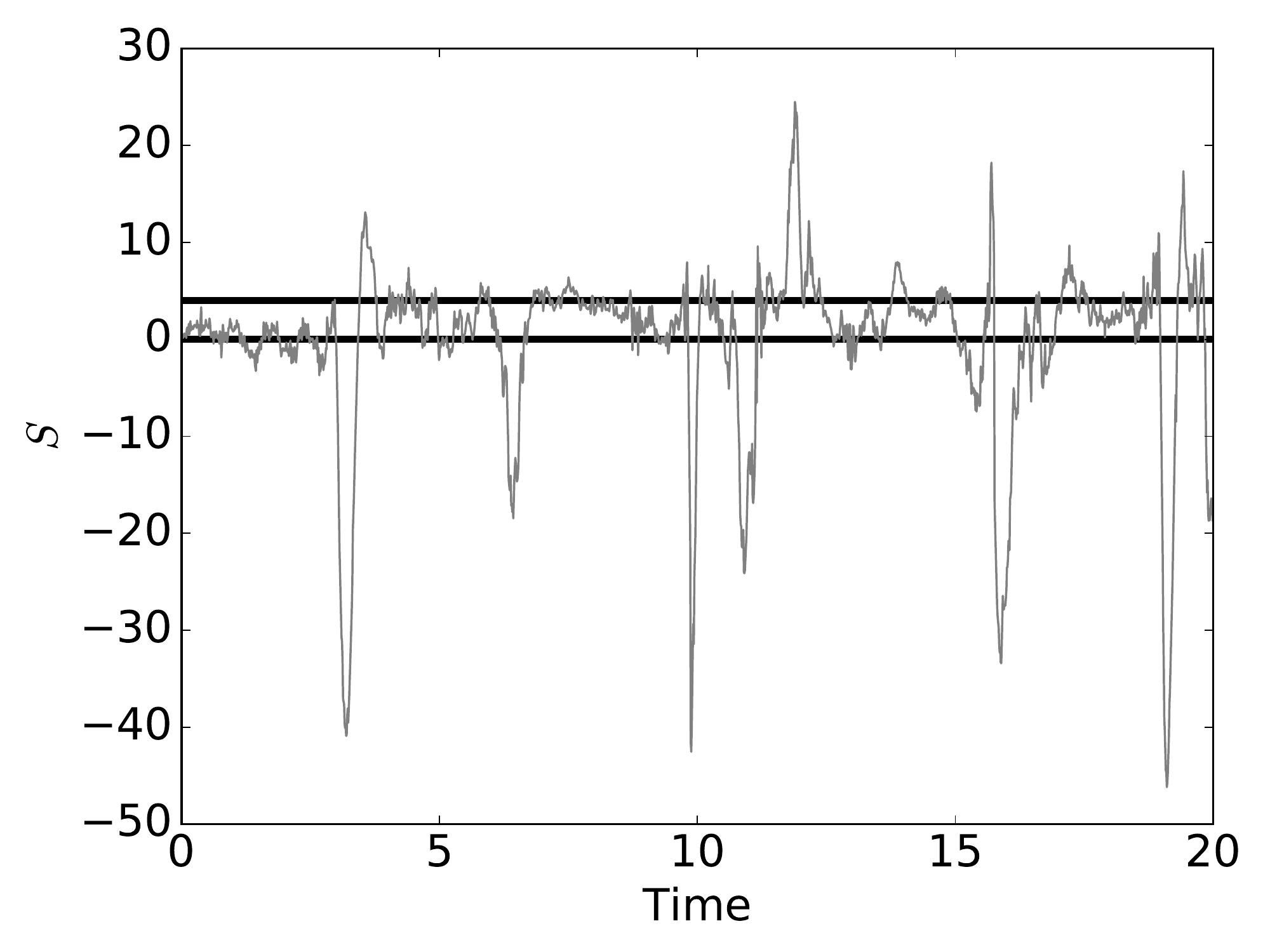}   
\includegraphics[width=8.5cm]{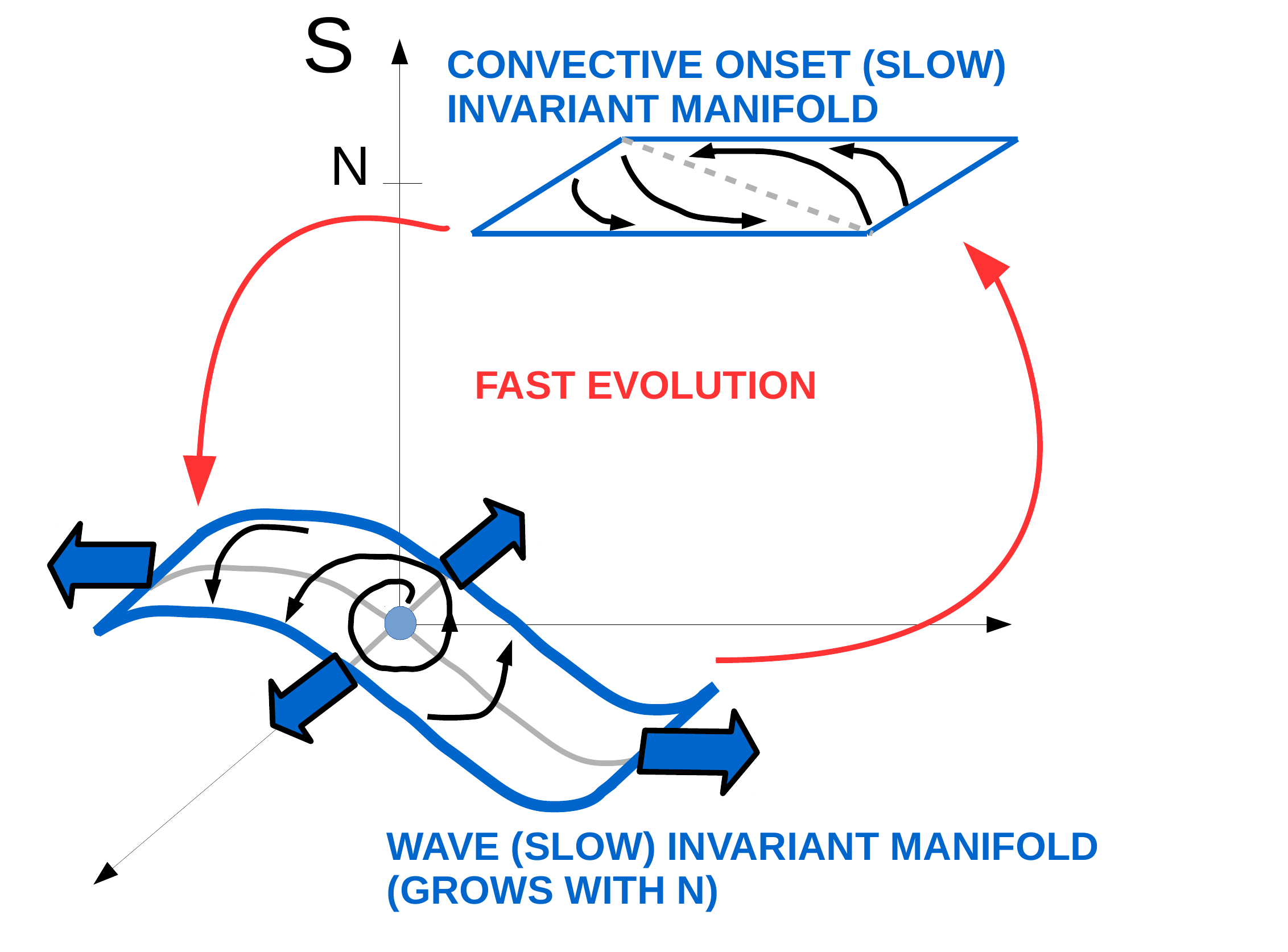}   
\caption{{\it Left:} Value of $S$ as a function of time for one
  particle in the simulation with $N=4$. Horizontal lines with $S=0$
  and $S=N=4$ are shown as references. The particle spends more time
  near these two values, with fast bursts and excursions in
  between. {\it Right:} Schematic of the proposed dynamics of fluid
  elements of the full Bousinnesq system embedded in the phase space of
  the reduced system. For simplicity, the 7 dimensional phase space is
  represented with only 3 axes. Fluid elements spend a long time in a
  ``wavy'' state in the slow invariant manifold $\Sigma_\RN{1}$ with
  $S\approx 0$ and $Q\approx R_\theta/N-B$ (whose size grows with
  $N$). As their energy increases, fluid elements eventually escape
  from $\Sigma_\RN{1}$ and evolve fast towards the global invariant
  manifold $\Sigma_0$ with $S\approx N$, $T\approx NA$, and
  $R_\theta \approx NB$. Fluid elements in this manifold are at the
  brink of the convective instability, and any perturbation results in
  efficient dissipation and in the fast return of fluid elements to
  the first invariant manifold.}
\label{f:scheme}
\end{figure}
%%%%%%%%%%%%%%%%%%%%%%%%%%%%%%%%%%%%%%%%%%

\subsection{The global invariant manifold}

From the system in Eq.~(\ref{eq:ODEs}), it can be verified that
$T=NA$, $S=N$, and $R_{\theta}=NB$ is a global invariant manifold as
\begin{equation}
\Sigma_0: D_{t}(T-NA)=0, \ D_{t}S=0, \ D_{t}(R_{\theta}-NB)=0.
\end{equation}
This particular manifold is valid for all orders of the nonlinearity,
and inside it, the system in Eq.~(\ref{eq:ODEs}) reduces to a
dynamical system with just 4 degrees of freedom,
\begin{equation}
\begin{split}
&D_{t}{Q}=-3R+NT, \\
&D_{t}{R}=N^{2} B + 2N^{2} Q/3 +  2Q^{2}/3, \\
&D_{t}{B}= -4N^{2}T/3 + 2NR + 2QT/3, \\
&D_{t}{T}=-NB-2N^{3}3-2NQ/3.
\end{split}
\end{equation}

This manifold has a rich physical interpretation, since as
$S=\partial_{z} \theta$, then for $\partial_{z} \theta=N$ the gradient
Richardson number in Eq.~(\ref{eq:rig}) becomes zero, and fluid
elements in this manifold are at the onset of the convective
instability becoming vertically unstable at the slightest
perturbation. Thus, and although stably stratified turbulence is
expected to display low vertical mixing \cite{lindborg_vertical_2008,
  aartrijk_single-particle_2008} and wave-like solutions, fluid
elements in $\Sigma_0$ can display a marginal instability as reported
for stratified turbulence in \cite{Smyth_2013, Smyth_2019}, go through
bursts and trigger sudden and intermittent local convective processes
as observed in \cite{rorai_turbulence_2014, Pearson_2018,
  feraco_vertical_2018, pouquet2019linking}, and significantly enhance 
vertical dispersion as reported in \cite{sujovolsky2019vertical}.

Figure \ref{f:RT_NB} shows the joint PDFs of $R_\theta$ and $NB$,
while Fig.~\ref{f:T_NA} shows the joint PDFs of $NA$ and $T$, obtained
from the direct numerical simulations, and superimposed with mean
rates of change of all quantities. As references, we also indicate in
these figures the relations $T=NA$ and $R_{\theta}=NB$ of the global
invariant manifold $\Sigma_0$. The correlation of the dynamics of
fluid elements in the numerical simulations with this manifold
improves when fluid elements are restricted to cases with
$S \approx N$ (as expected for $\Sigma_0$). Moreover, fluid elements 
again evolve slowly in the vicinity of this manifold. A slow
evolution, and the accumulation of fluid elements near this manifold, 
can be  explained as the evolution of the convective instability
(which takes place in the order of the turnover time) is slower than
the fast internal gravity waves and other physical processes in the
flow.

\subsection{Overall dynamics}

The data from the direct numerical simulations presented so far
indicates that fluid elements spend a significant time exploring two
invariant manifolds. Moreover, the data also shows that as $N$
increases, the local invariant manifold $\Sigma_{\RN{1}}$ holds a
larger fraction of particles, as opposed to the global manifold which
holds a smaller fraction of fluid elements. This is shown in
Fig.~\ref{f:S0_SN}, which shows the percentage of fluid elements at
any moment with $S\approx 0$ or with $S\approx N$ in all simulations
in table \ref{tab:param}, as a function of Fr. However, note that
previous studies \cite{feraco_vertical_2018, sujovolsky2019invariant}
suggest that this behavior may be monotonous only for sufficiently
large $N$ (or sufficiently small $\mathrm{Fr}$), while for
intermediate values of $N$ fluid elements could escape more rapidly
(and non-monotonously with $N$ of Fr) from the local invariant
manifold.

The present analysis also indicates that once fluid elements escape
from any of these two manifolds, the evolution of fluid elements is
fast as they move in phase space from one manifold to the other. This
is not only suggested by the reduced model, but also seen in the data
from the simulations of the full Boussinesq equations. Indeed, as
shown above, the invariant manifolds are not only partially preserved
in the full system, but are also slow, as rates of change of field
gradients in phase space decrease dramatically in the vicinity of
these manifolds. Moreover, fluid elements display fast and extreme
values of the gradients when $S \neq 0$ or $N$, and in these cases the
fluid elements seem to rapidly recover one of these two values. This
is also illustrated in Fig.~\ref{f:scheme}, where the value of $S$ is
shown as a function of time for an individual fluid element in the
simulation with $N=4$. Note $S$ fluctuates around $0$ or $4$ for long
times, and changes between these two values are often mediated by a
long and fast excursion in the value of $S$ (we will see similar
alternations between slow and fast evolution in the dynamics of the
potential vorticity in Sec.~\ref{sec:PV}).

Following these results, we propose the following scenario: The phase 
space of stably stratified turbulence is composed of two main slow
invariant manifolds. The local manifold $\Sigma_{\RN{1}}$, valid in
the proximity of the flow at equilibrium (fixed point $\RN{1}$)
corresponds to the case in which waves dominate the system
dynamics. When energy in the waves grows too much (either from
nonlinear amplification, from the wave turbulence cascade, or from
effects neglected in the reduced model such as forcing or pressure
gradients), fluid elements can escape this local manifold and explore
(fast) the phase space until finding the global slow manifold
$\Sigma_0$. When this happens, the wave turbulence solutions
break, and the finding of $\Sigma_0$ can be interpreted as the
result of fluid elements looking for a different surface of solutions
in phase space to efficiently dissipate their energy. Manifold
$\Sigma_0$ is at the brink of the local convective instability. Any
perturbation from this manifold excites local convection, and the full
Boussinesq equations can dissipate energy through strong turbulence
mechanisms. Then, particles can rapidly go back to the first local
manifold $\Sigma_{\RN{1}}$ and be dominated again by the waves. These
two manifolds, which the particles can inhabit for long periods of
time, are embedded in a phase space in which particles evolve fast,
and which fills the gap between the stable wave-like regime and the
efficient dissipation of accumulated energy (see the diagram in
Fig.~\ref{f:scheme}). In this scenario, as $N$ increases, the size and 
stability of $\Sigma_{\RN{1}}$ can be also expected to increase (see
Fig.~\ref{f:Q_RT-B} and \cite{sujovolsky2019invariant} for a
confirmation of this behavior), resulting in less and less excursions
to the strongly nonlinear region of phase space associated with the
$\Sigma_0$ manifold.

%%%%%%%%%%%%%%%%%%%%%%%%%%%%%%%%%%%%%%%%%%
\begin{figure}
\includegraphics[width=8.9cm]{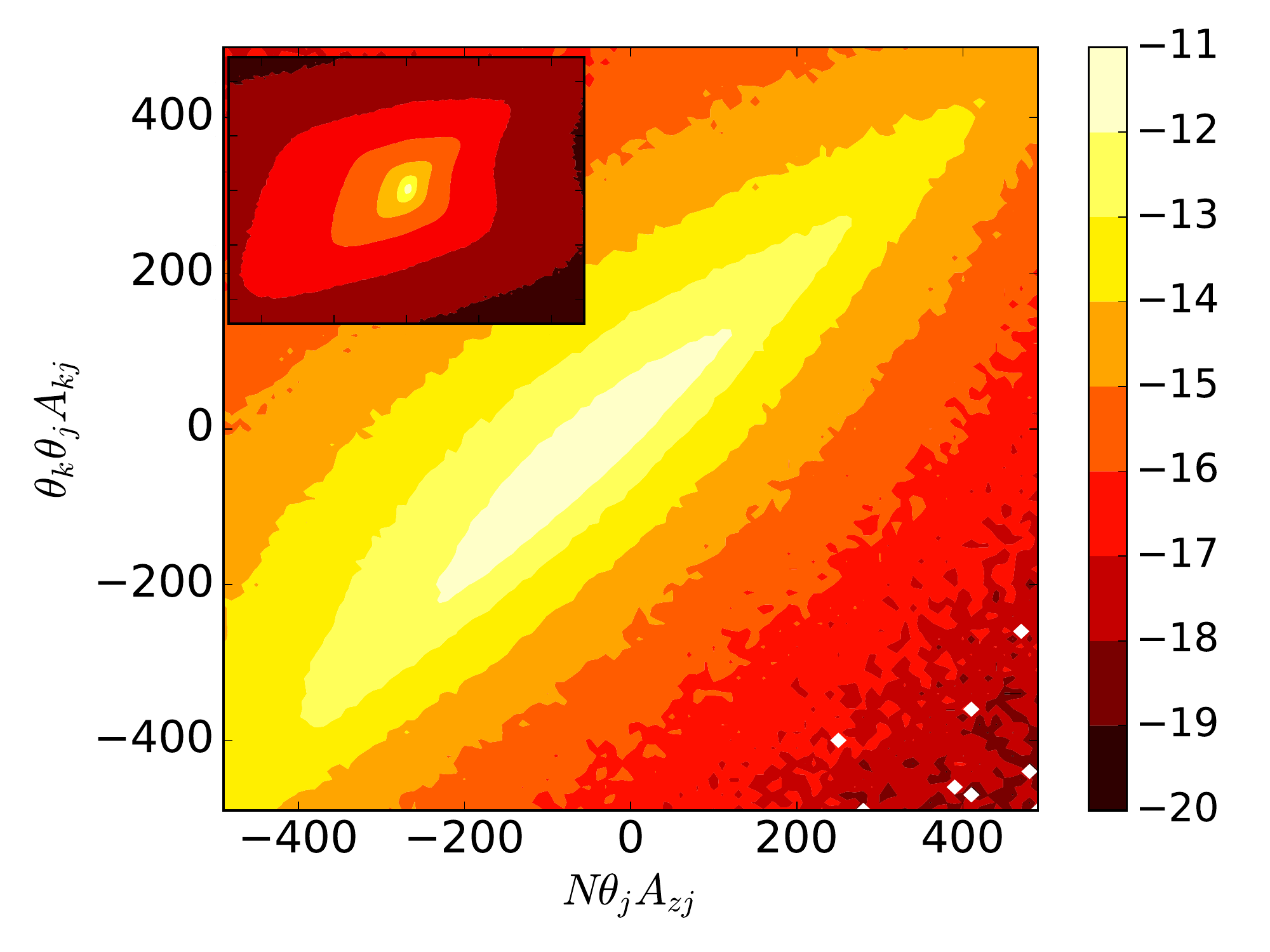}   
\includegraphics[width=8.9cm]{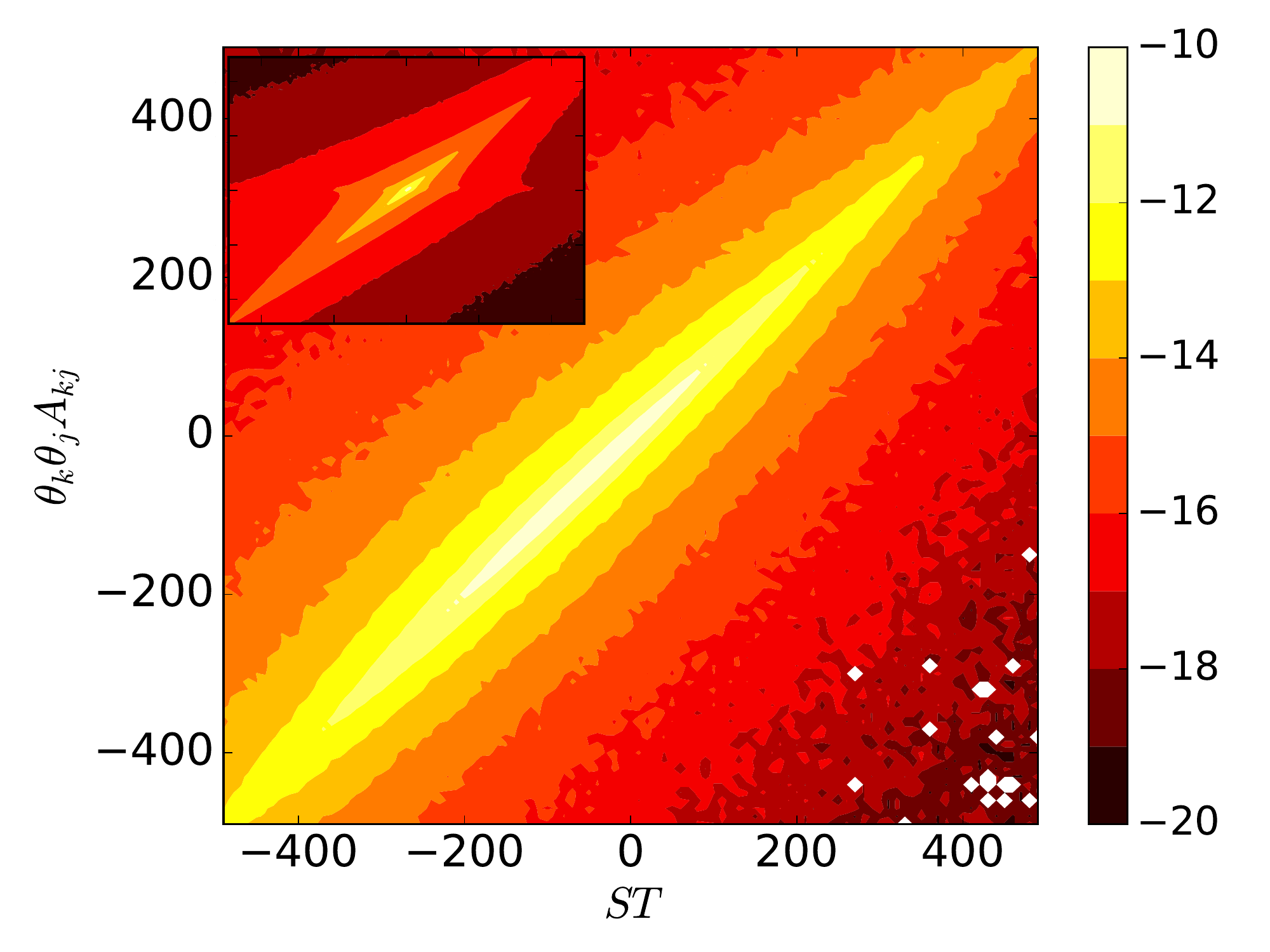}
\caption{Joint probability density function of 
  $\theta_{k}\theta_{j}A_{kj}$ and $N \theta_{j} A_{zj}$ ({\it left})
  and of $\theta_{k}\theta_{j}A_{kj}$ and $ST$ ({\it right}), for
  fluid elements with $S \approx N$ in the simulations with $N=8$. The
  insets show the same quantities but for fluid elements without any
  restriction on the values $S$ can take.}
\label{f:TGS}
\end{figure}
%%%%%%%%%%%%%%%%%%%%%%%%%%%%%%%%%%%%%%%%%%

\section{Turbulent production of vertical buoyancy
  fluctuations \label{sec:balance}}

We can now discuss some of the physical implications of the behavior
reported in the previous section for the modeling of stably stratified
turbulence, for the formation of flow structures, and for the
evolution of important quantities in geophysical flows such as the
potential vorticity. Let's start with turbulent production of buoyancy
fluctuations, which affects dissipation and is important for mixing
and for subgrid scale models. The $z$ component of
Eq.~(\ref{eq:tita_j}) can be written as
\begin{equation}
\dfrac{D\theta_{z}}{Dt} = - A_{kz} \theta_{k} + N A_{zz} ,
\label{eq:tita_z}
\end{equation}
and gives the evolution of the vertical buoyancy gradients as seen
when following fluid element trajectories in the ideal Boussinesq
equations. In this equation, the production and destruction of 
vertical buoyancy gradients are controlled by strain and rotation
(associated with the vorticity) as described by the term
$-A_{kz}\theta_{z}=-T$, and by the buoyancy term $N A_{zz}=N A$. For
fluid elements in the global invariant manifold $\Sigma_{0}$ (i.e.,
for fluid elements at the brink of convection), the two terms are 
balanced (as $T=NA$), and Eq.~(\ref{eq:tita_z}) reduces to
$D_{t}\theta_{z}=0$. In physical terms, production of vertical
buoyancy gradients by turbulence is perfectly balanced by linear
(buoyancy) effects.

Moreover, in the global invariant manifold $\Sigma_0$, as
$\theta_z=S=N$ does not vary, we can derive another balance relation
for buoyancy gradients, which has important consequences for subgrid
modeling as it provides a condition over the turbulent dissipation of
spatial buoyancy variations. Multiplying Eq.~(\ref{eq:tita_j}) by
$\theta_{j}$ we obtain the following equation for the second-order
one-point correlation of buoyancy gradients,
\begin{equation}
\dfrac{1}{2}\dfrac{D (\theta_{j}\theta_{j})}{Dt} = -
  \theta_{k}\theta_{j}A_{kj} + N \theta_{j} A_{zj}.
\label{eq:energy_tita}
\end{equation}
For $i=j$ this equation gives the evolution of
$|{\boldsymbol \nabla} \theta|^2$ which (after integrating over the
whole volume) controls the total dissipation of potential energy in
the flow $E_P = \int \theta^2/2 \, dV$ (see, e.g.,
\cite{Marino_2013}). As a result, this equation often appears in
subgrid models of stratified turbulence
\cite{gulitski_velocity_2007}.

The first term on the right hand side of Eq.~(\ref{eq:energy_tita}), 
$-\theta_{k}\theta_{j}A_{kj}$, corresponds to the turbulence
production term \cite{gulitski_velocity_2007}, while the second term,
$N \theta_{j} A_{zj}$, accounts again for buoyancy effects. We can
expect the time derivative on the left hand side of
Eq.~(\ref{eq:energy_tita}) to become small when a fluid element is
close to any of the two invariant manifolds, as time evolution becomes
slow in these manifolds. For the particular case of manifold
$\Sigma_{0}$, in which turbulence and field gradients can be expected
to be more relevant, neglecting the time derivative in
Eq. (\ref{eq:energy_tita}) implies that as long as particles remain in
this manifold, then
\begin{equation}
 \theta_{k}\theta_{j}A_{kj} \sim N \theta_{j} A_{zj} \sim S T .
\label{eq:energy_tita_2}
\end{equation}
This relation implies that also for the second-order one-point
correlation of buoyancy gradients, turbulent production is
approximately balanced by buoyancy effects. And if this relation holds
in the numerical data, it also confirms that
$|{\boldsymbol \nabla} \theta|^{2}$ varies slowly in the $\Sigma_{0}$
manifold.

Figure \ref{f:TGS} shows the joint probability density functions of
$\theta_{k}\theta_{j}A_{kj}$ and $N \theta_{j} A_{zj}$, and of
$\theta_{k}\theta_{j}A_{kj}$ and $TS$, for fluid elements close to the
$\Sigma_0$ invariant manifold in the simulation of the full Boussinesq
equations with $N=8$ (the insets show the same probability density
functions for all fluid elements in the flow, irrespectively of their
value of $S$). The observed correlations are in good agreement with
Eq.~(\ref{eq:energy_tita}), and indicate a possible way to estimate
relevant subgrid production terms using the reduced model.

%%%%%%%%%%%%%%%%%%%%%%%%%%%%%%%%%%%%%%%%%%
\begin{figure}
\includegraphics[width=5.9cm]{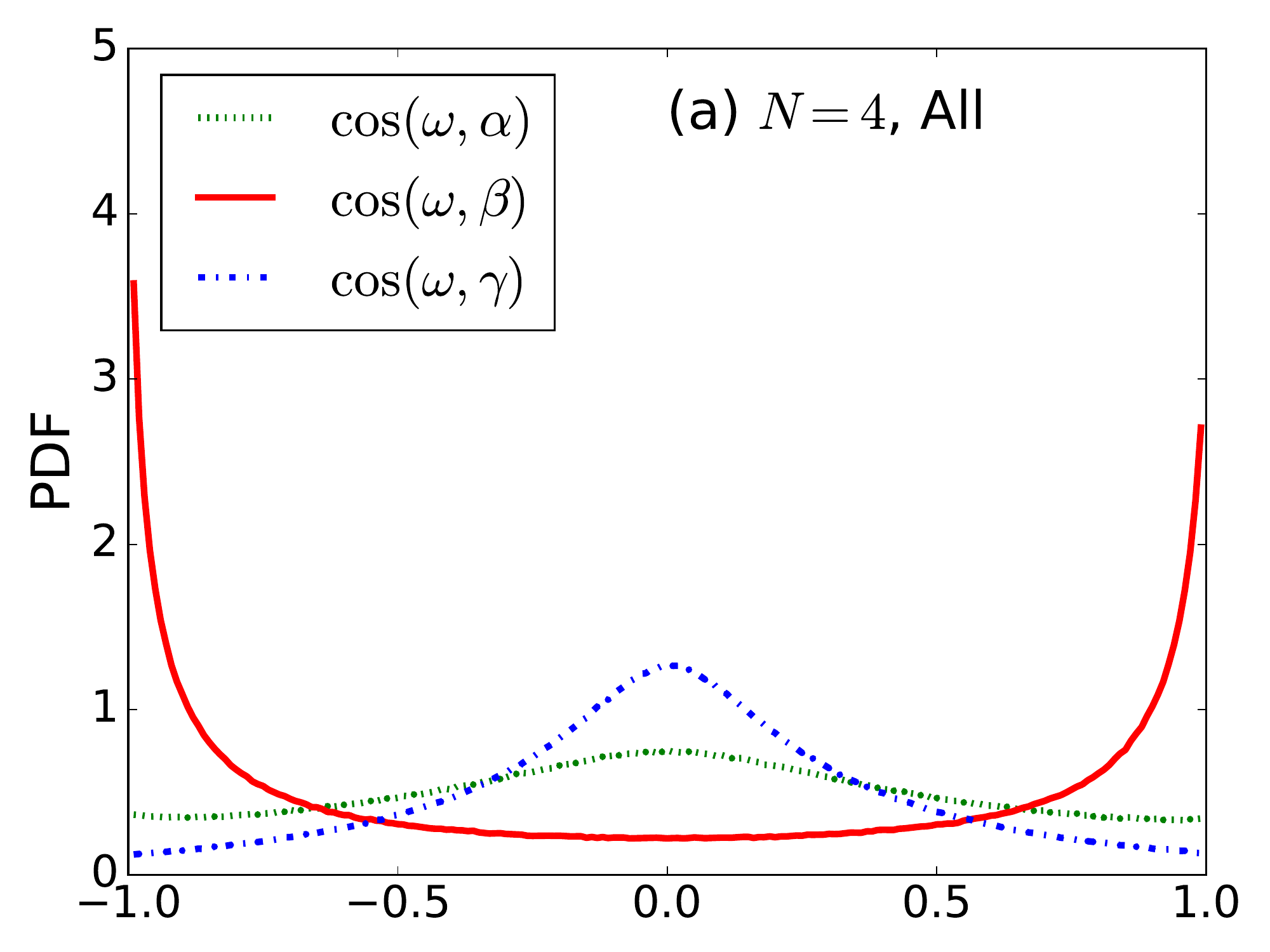}   
\includegraphics[width=5.9cm]{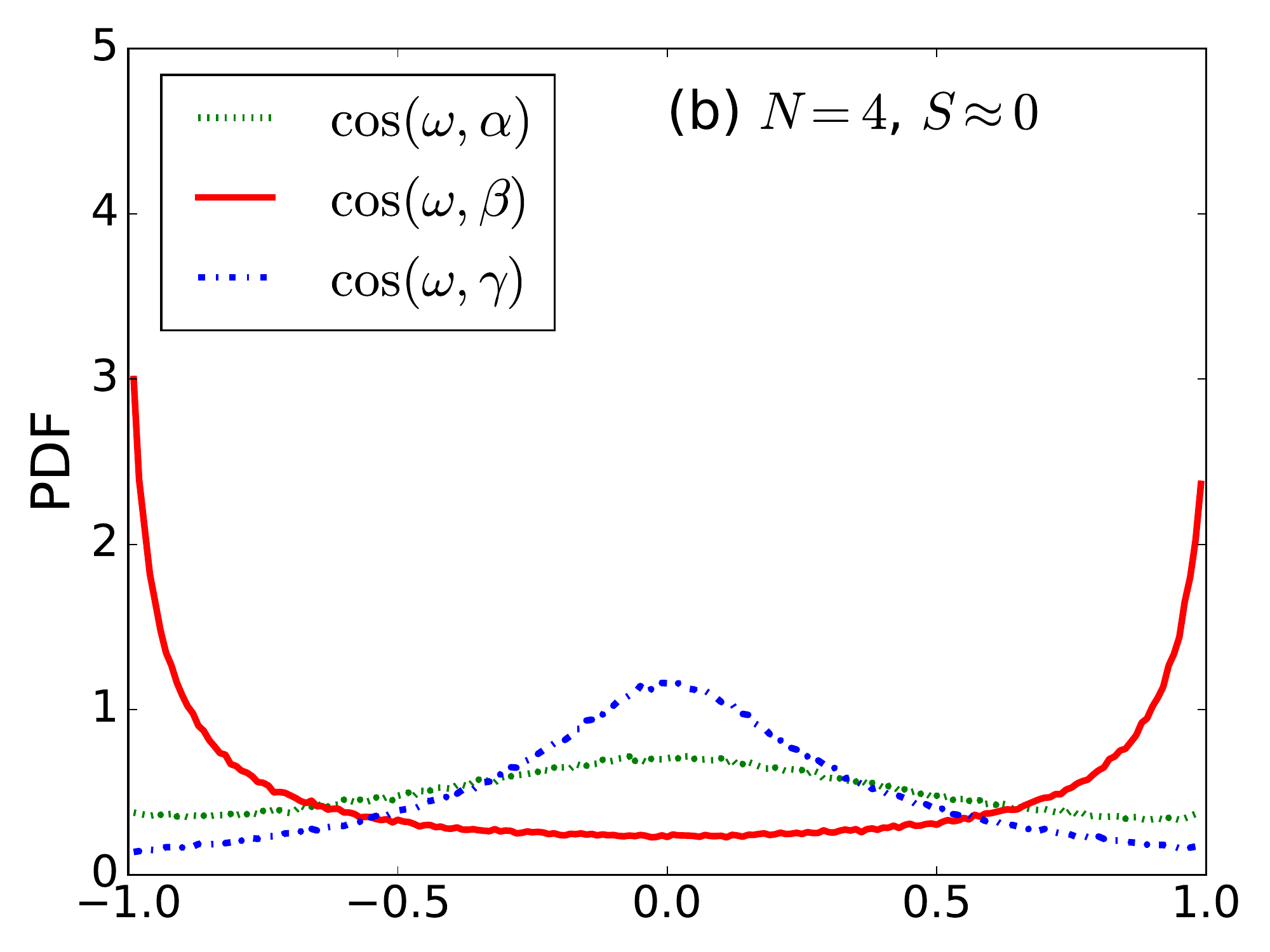}    
\includegraphics[width=5.9cm]{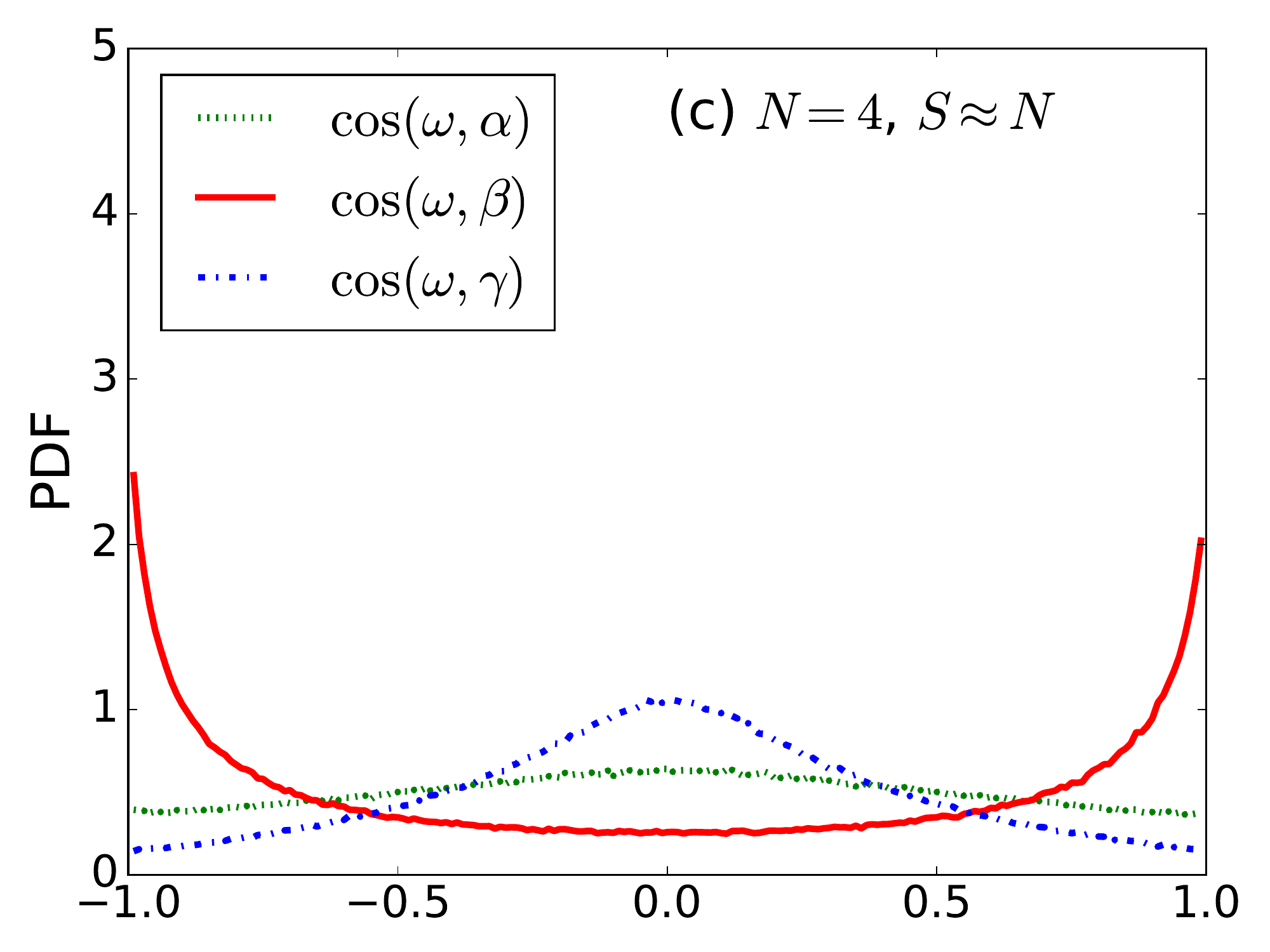} \\   
\includegraphics[width=5.9cm]{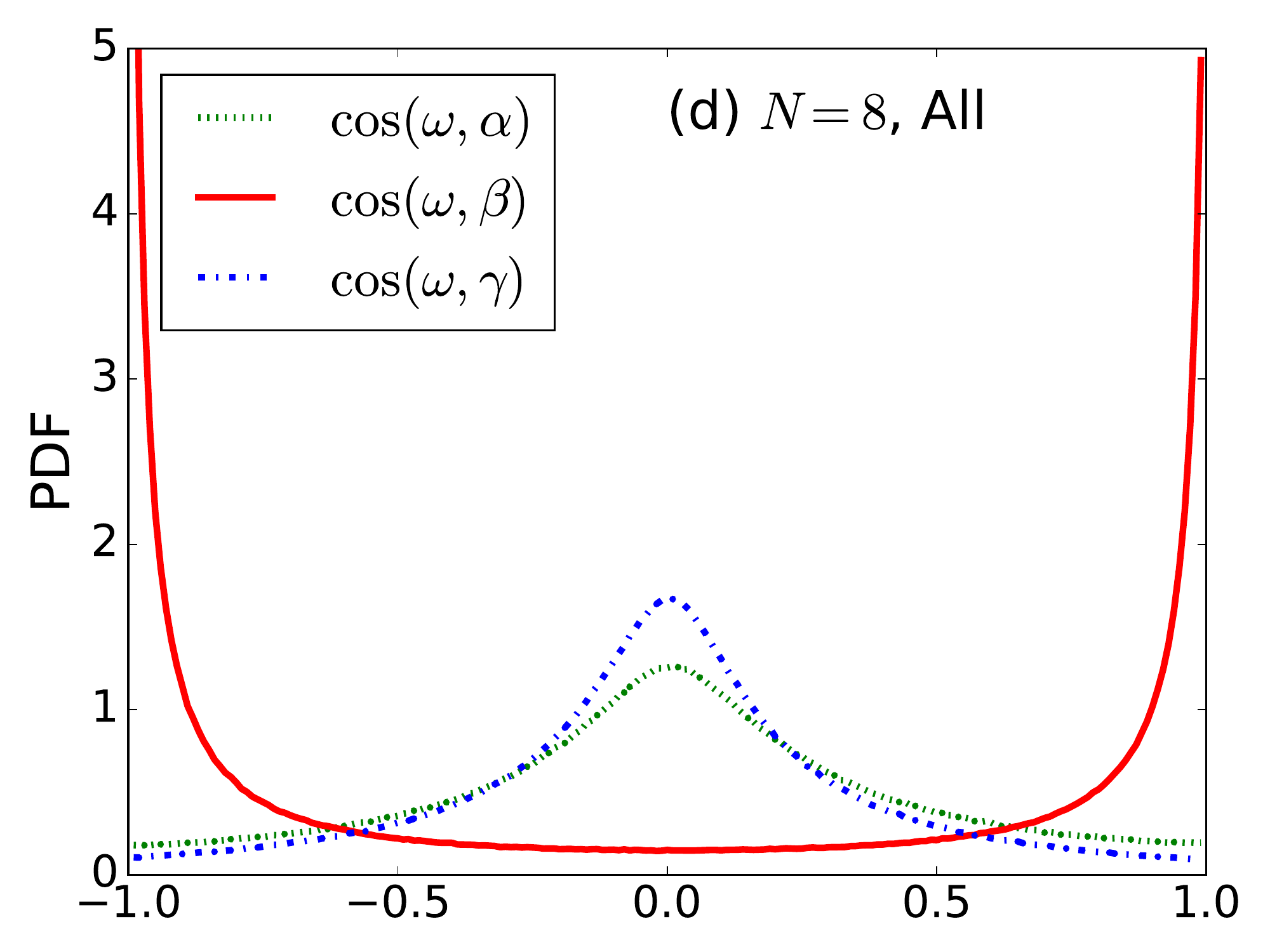}   
\includegraphics[width=5.9cm]{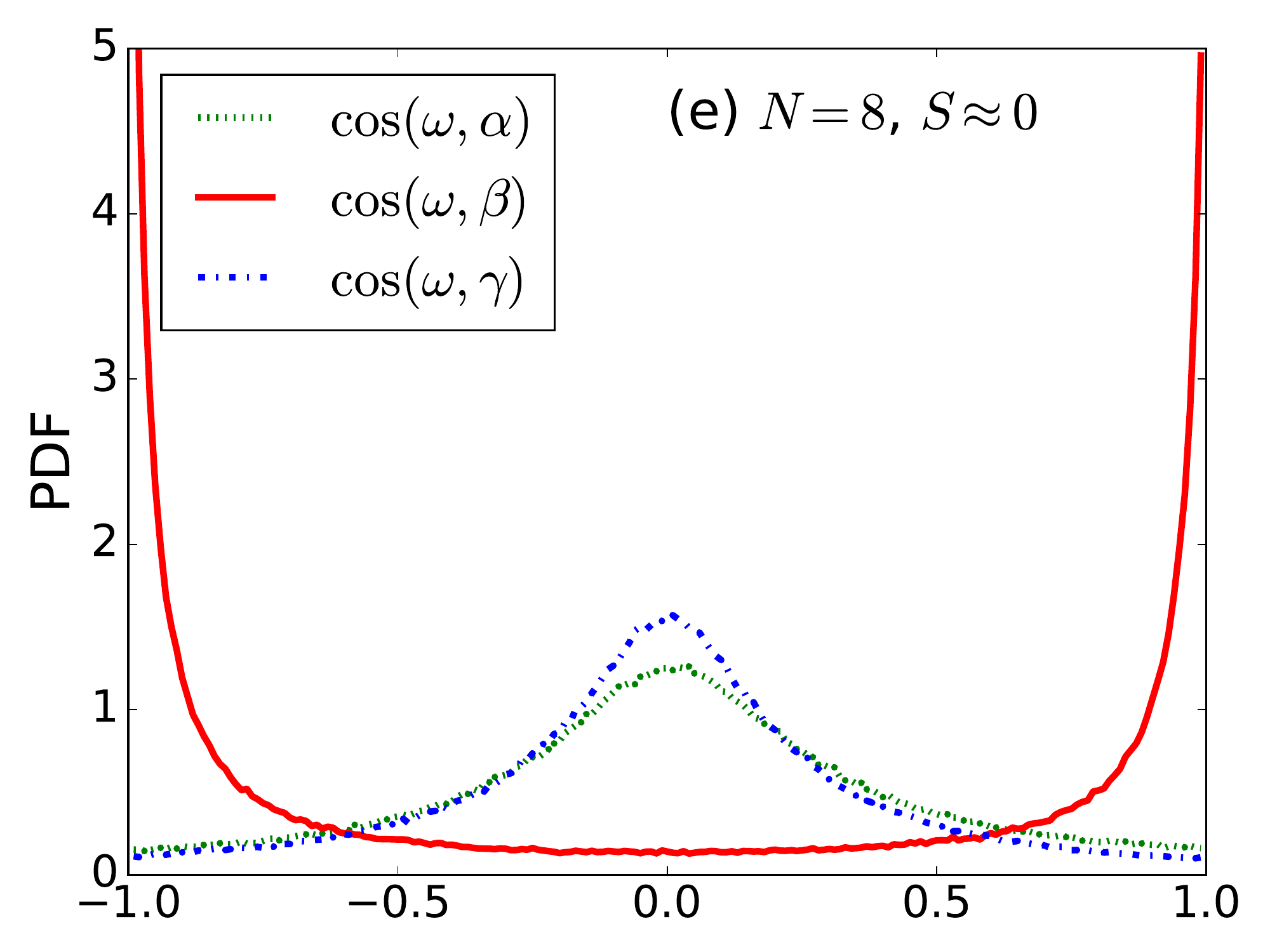}  
\includegraphics[width=5.9cm]{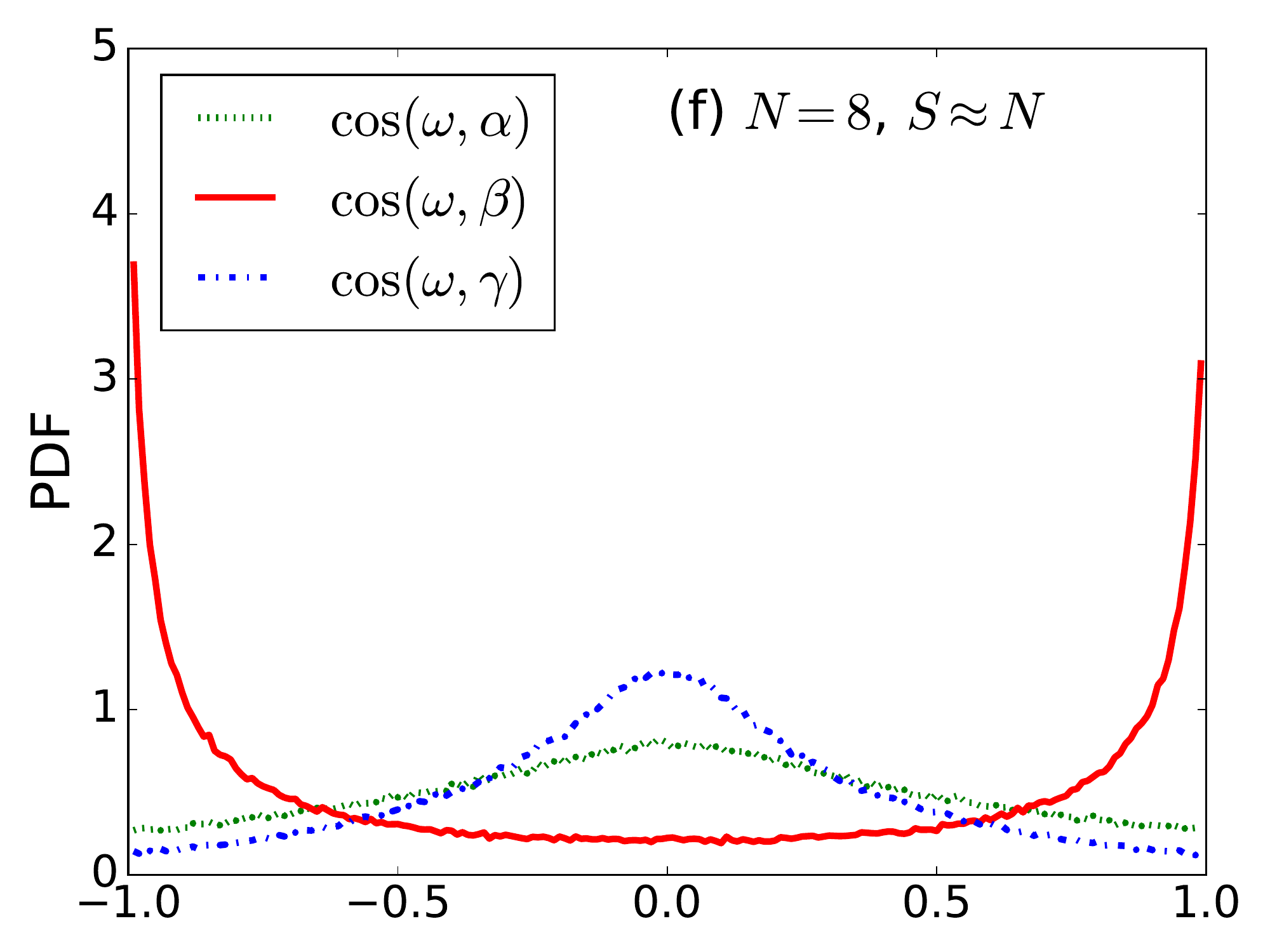} \\   
\includegraphics[width=5.9cm]{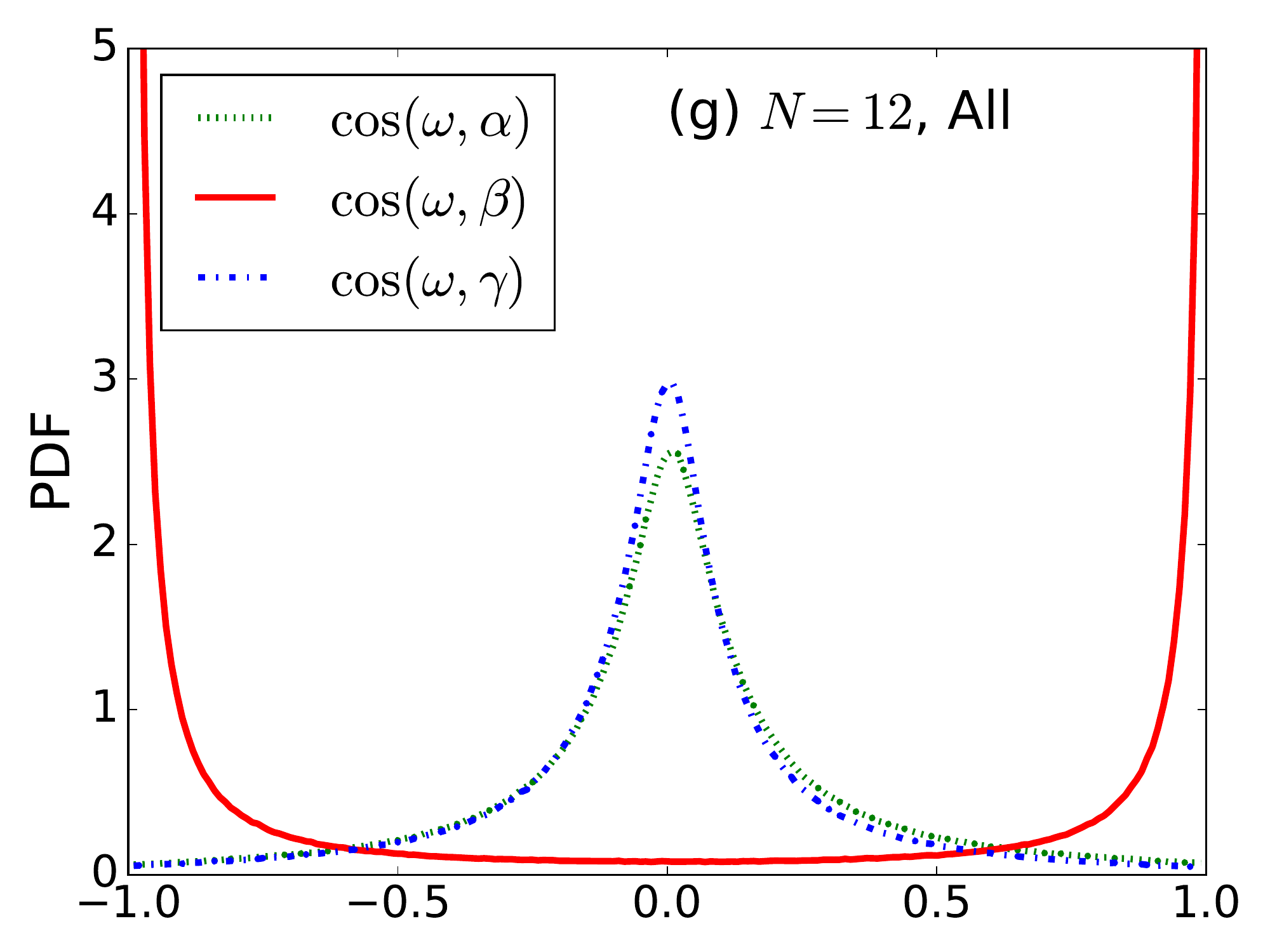}   
\includegraphics[width=5.9cm]{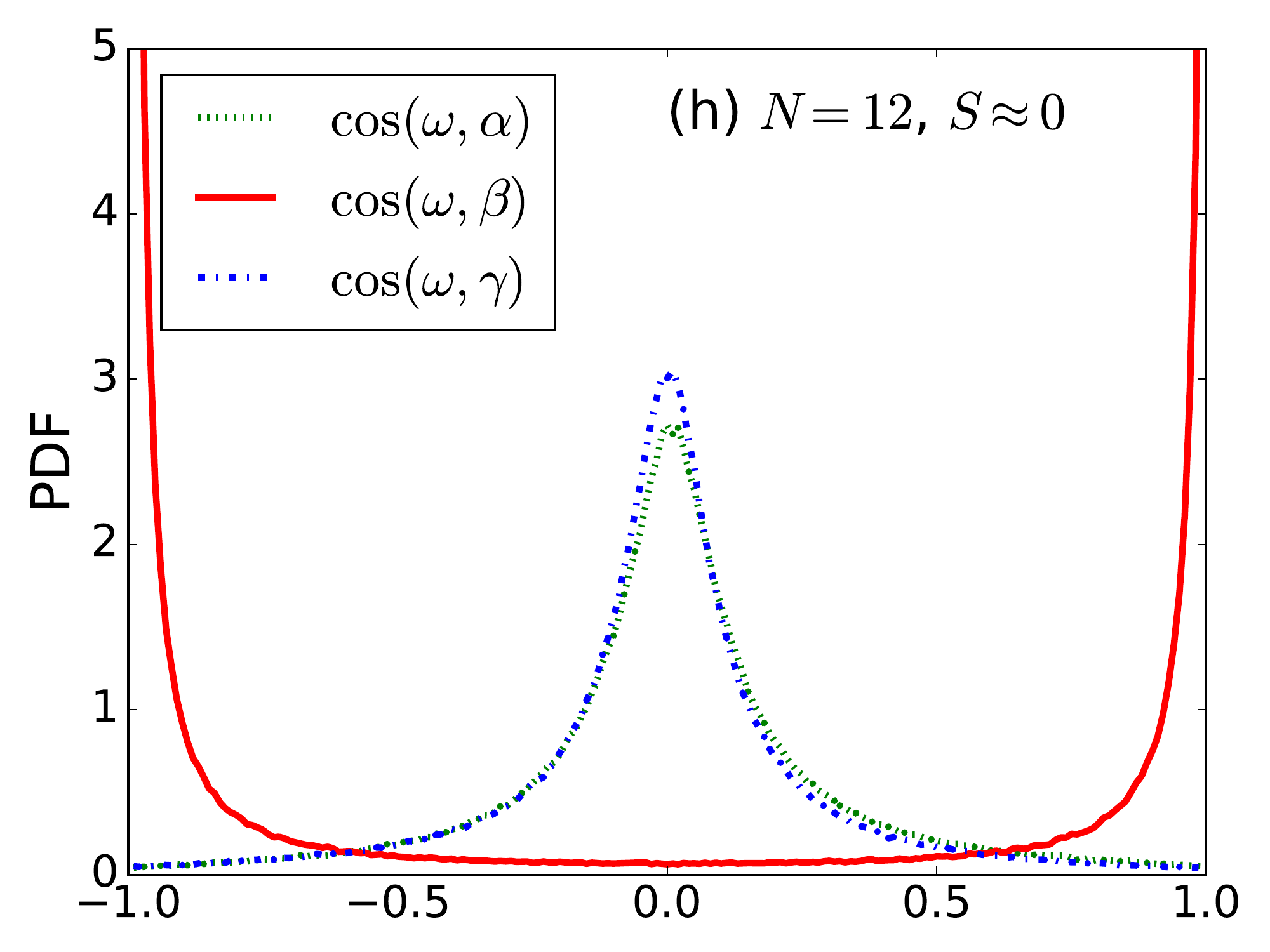}
\includegraphics[width=5.9cm]{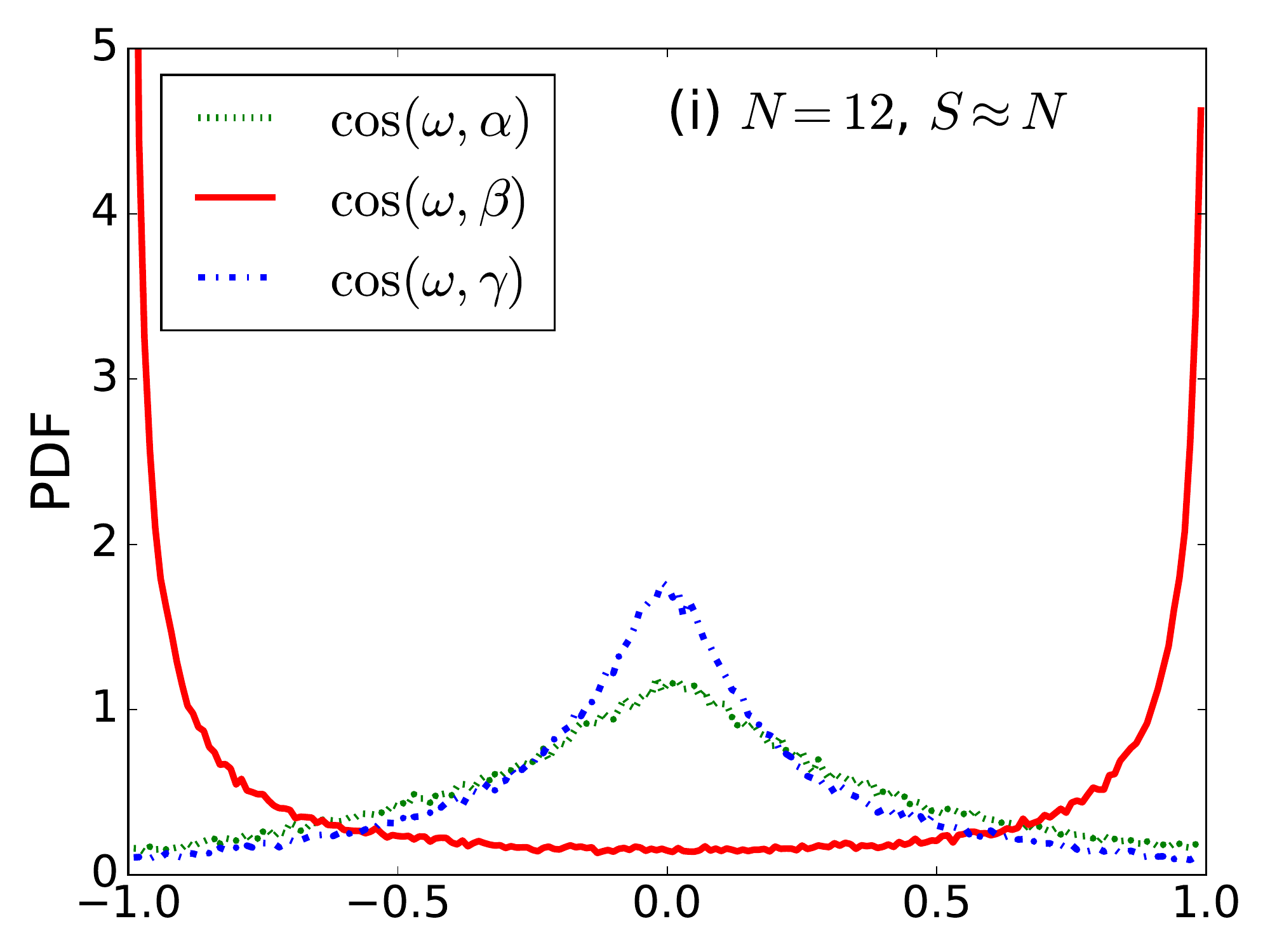}   
\caption{Probability density functions (PDFs) of the cosine of the
  angles between the vorticity ${\boldsymbol \omega}$ and the
  eigenvectors ${\boldsymbol \alpha}$, ${\boldsymbol \beta}$ and
  ${\boldsymbol \gamma}$ of the strain rate tensor $s_{ij}$. In the
  first row, panels (a), (b), and (c) correspond to simulations with
  $N=4$, with panel (a) showing the PDFs of all fluid elements (i.e.,
  without any restriction on the value of $S$), panel (b) showing
  fluid elements restricted to instants with $S\approx 0$, and panel
  (c) restricted to $S\approx N$. The second row, with panels (d),
  (e), and (f), shows the same PDFs for the simulation with $N=8$,
  while the third row, with panels (g), (h), and (i), shows the case
  with $N=12$.}
\label{f:w_abg}
\end{figure}
%%%%%%%%%%%%%%%%%%%%%%%%%%%%%%%%%%%%%%%%%%

\section{Alignment of field gradients and local flow
  geometry \label{sec:alignment}}

In homogeneous and isotropic turbulence, the existence of the
Vieillefosse invariant manifold in the restricted Euler model
(dependent only in the variables $Q$ and $R$) has implications for
the alignment of the vorticity with one of the principal axes of the
strain-rate tensor, and as a result, with the formation of vortical
structures through vortex stretching \cite{Chong_1990,
  meneveau2011lagrangian, Dallas_2013}. As was discussed in 
Sec.~\ref{sec:reduced}, the replacement of the Vieillefosse tail by
two other invariant manifolds in the system in Eq.~(\ref{eq:ODEs})
should have an important effect in vortex stretching and in the
development of structures in stably stratified flows. These effects
are discussed in this section, by studying the alignment of the
vorticity and of buoyancy gradients with the principal axes of the
strain-rate tensor, as well as with the Cartesian axes, paying special
attention to directions parallel and perpendicular to gravity.

%%%%%%%%%%%%%%%%%%%%%%%%%%%%%%%%%%%%%%%%%% 
\begin{figure}
\includegraphics[width=5.9cm]{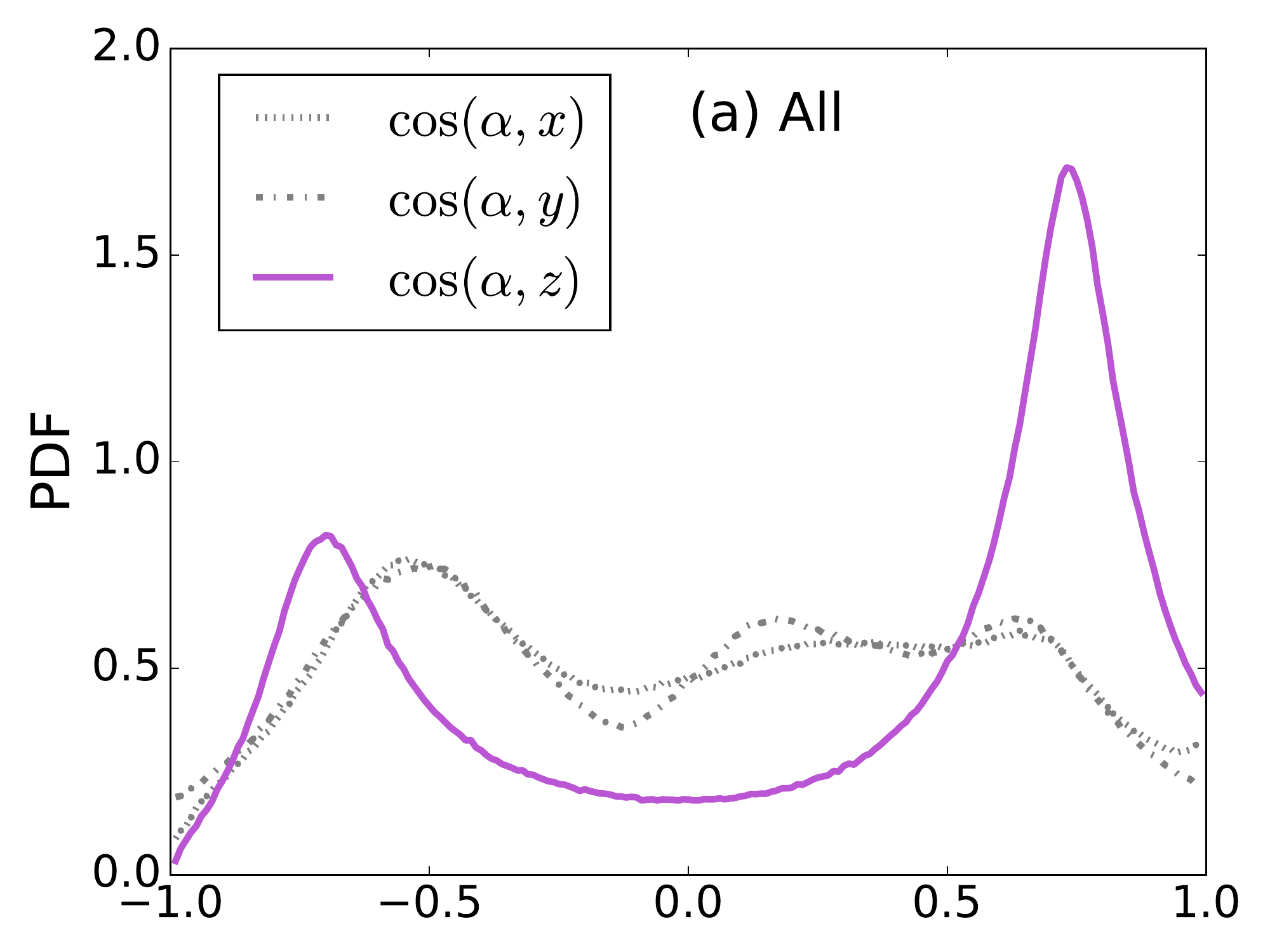}   
\includegraphics[width=5.9cm]{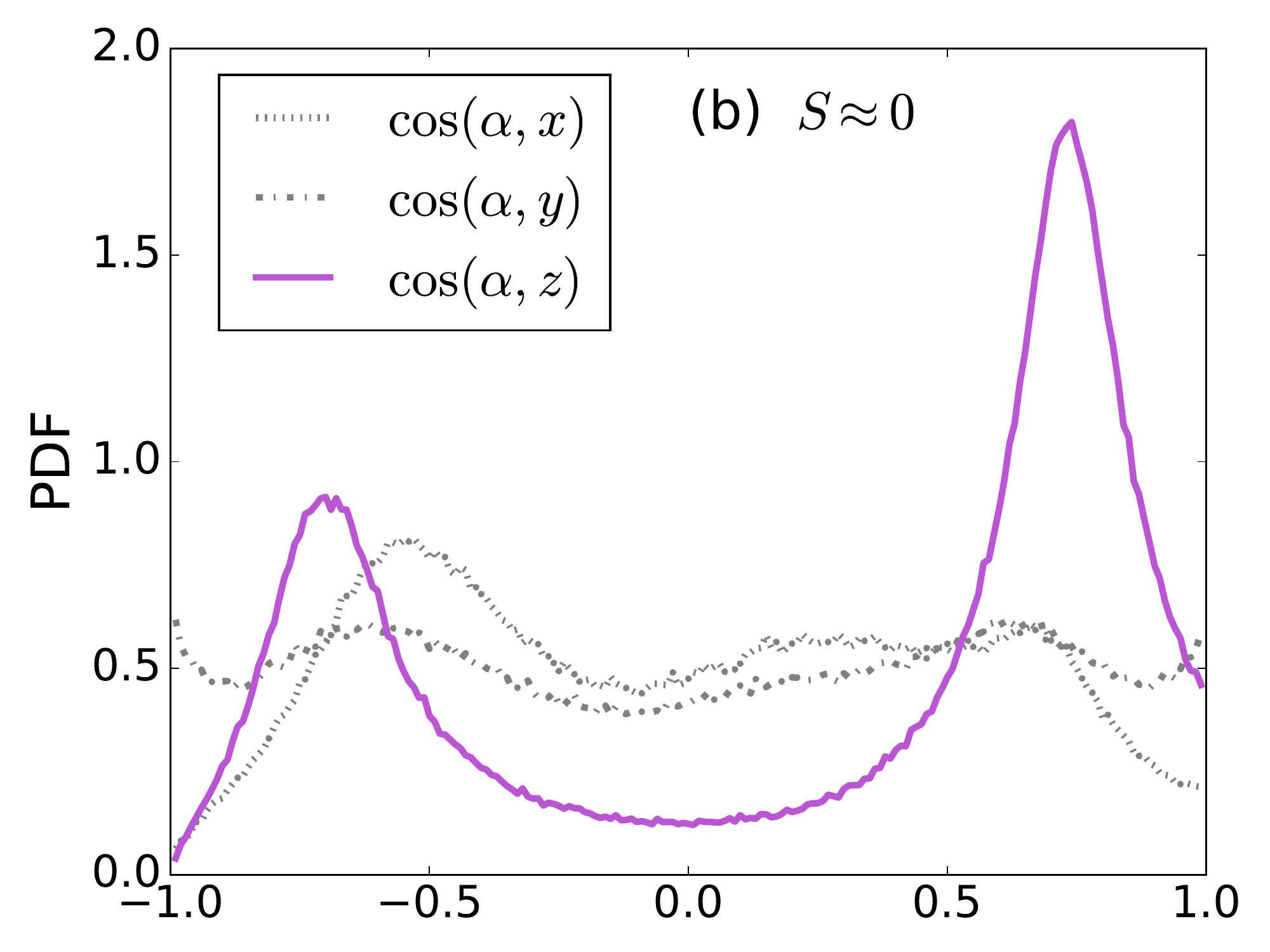}   
\includegraphics[width=5.9cm]{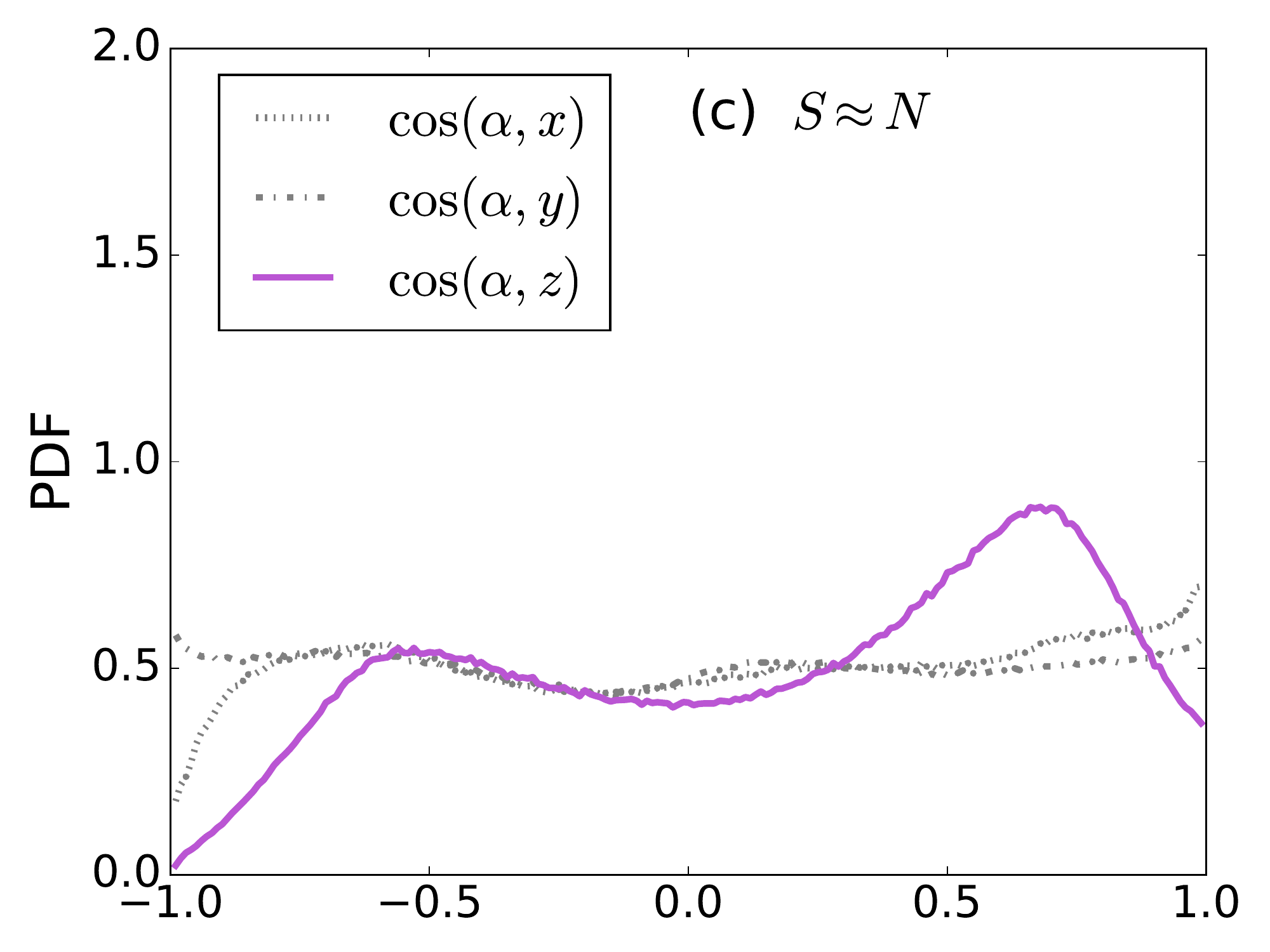} \\   
\includegraphics[width=5.9cm]{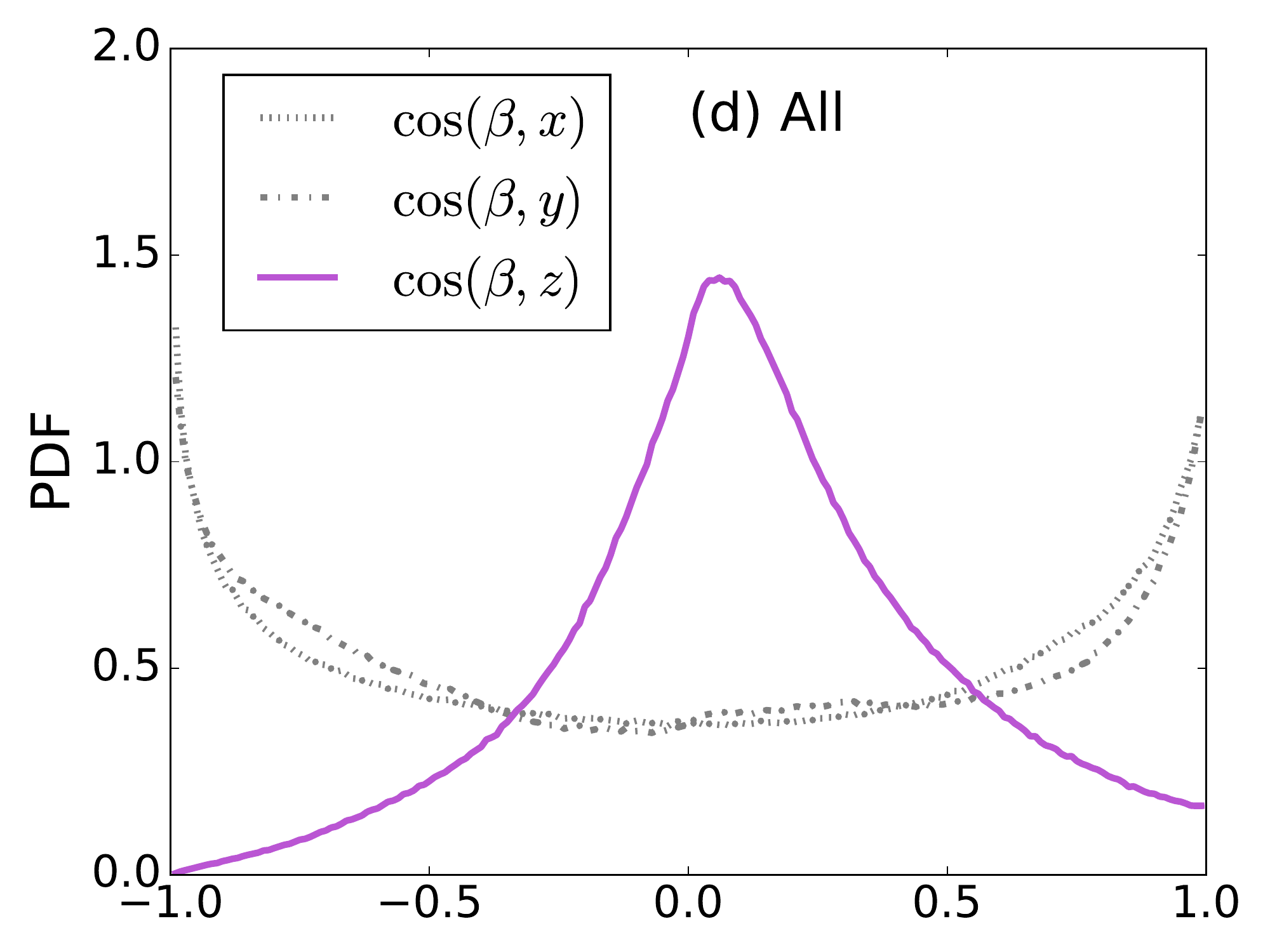}   
\includegraphics[width=5.9cm]{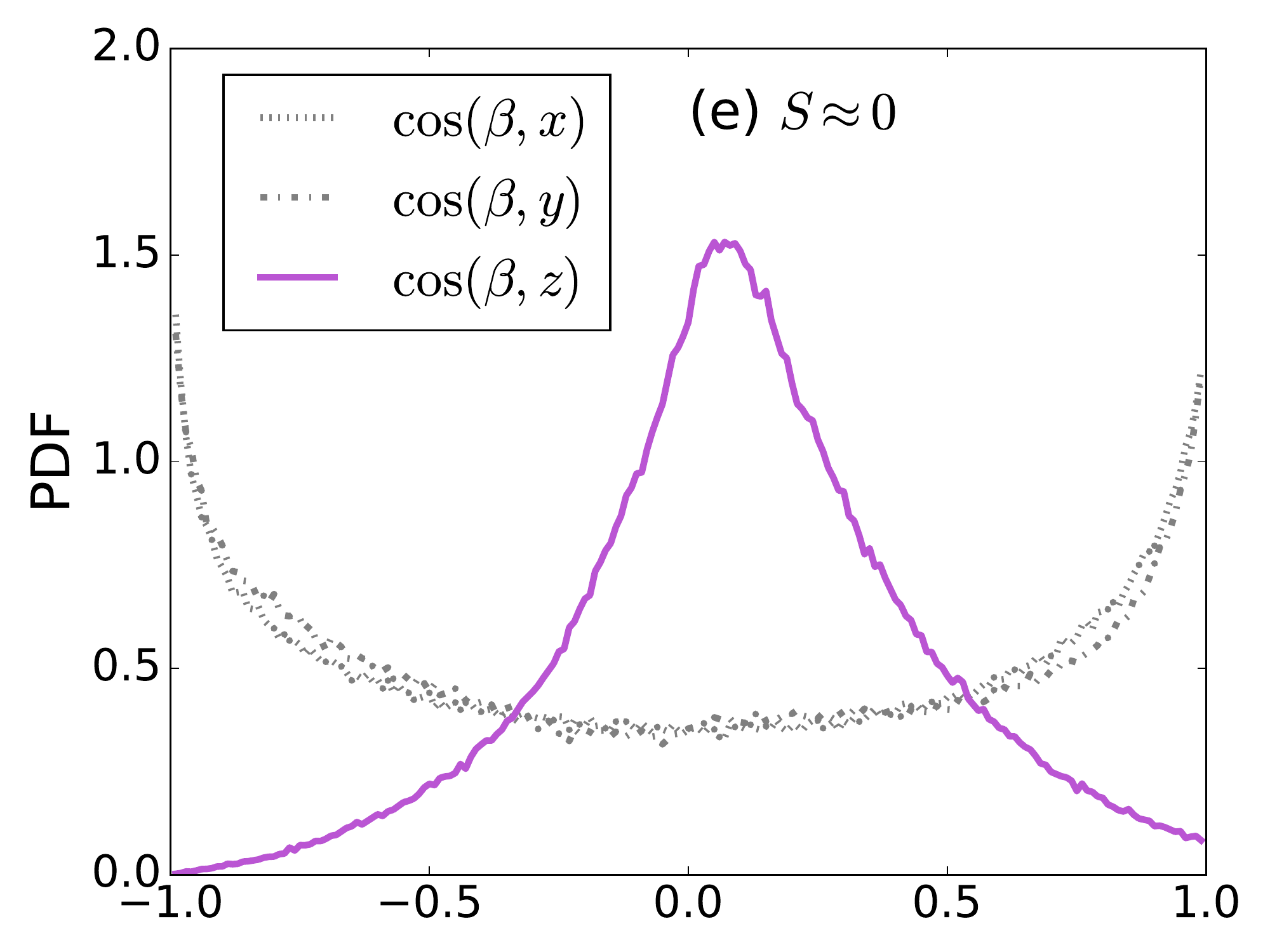}   
\includegraphics[width=5.9cm]{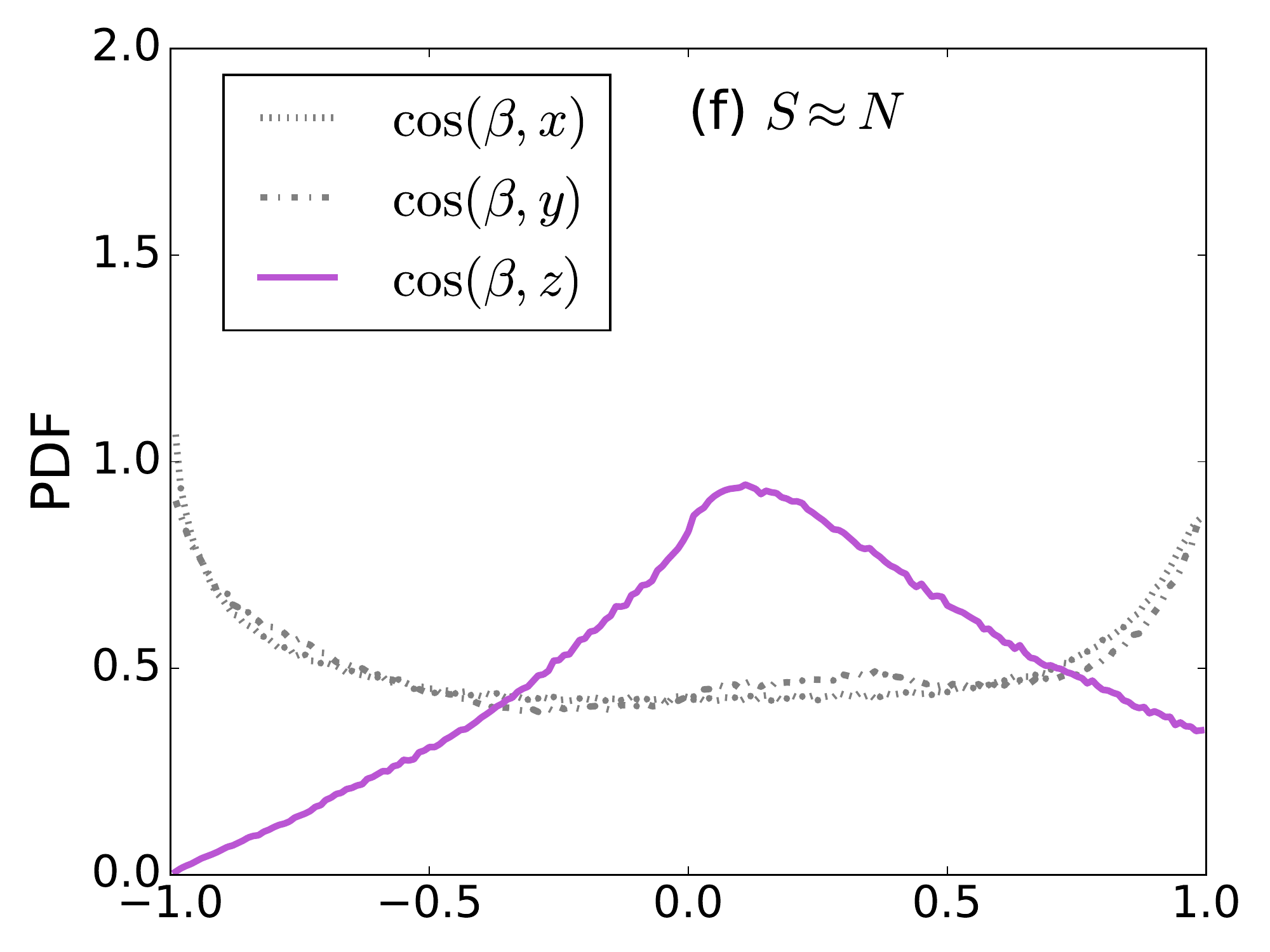} \\   
\includegraphics[width=5.9cm]{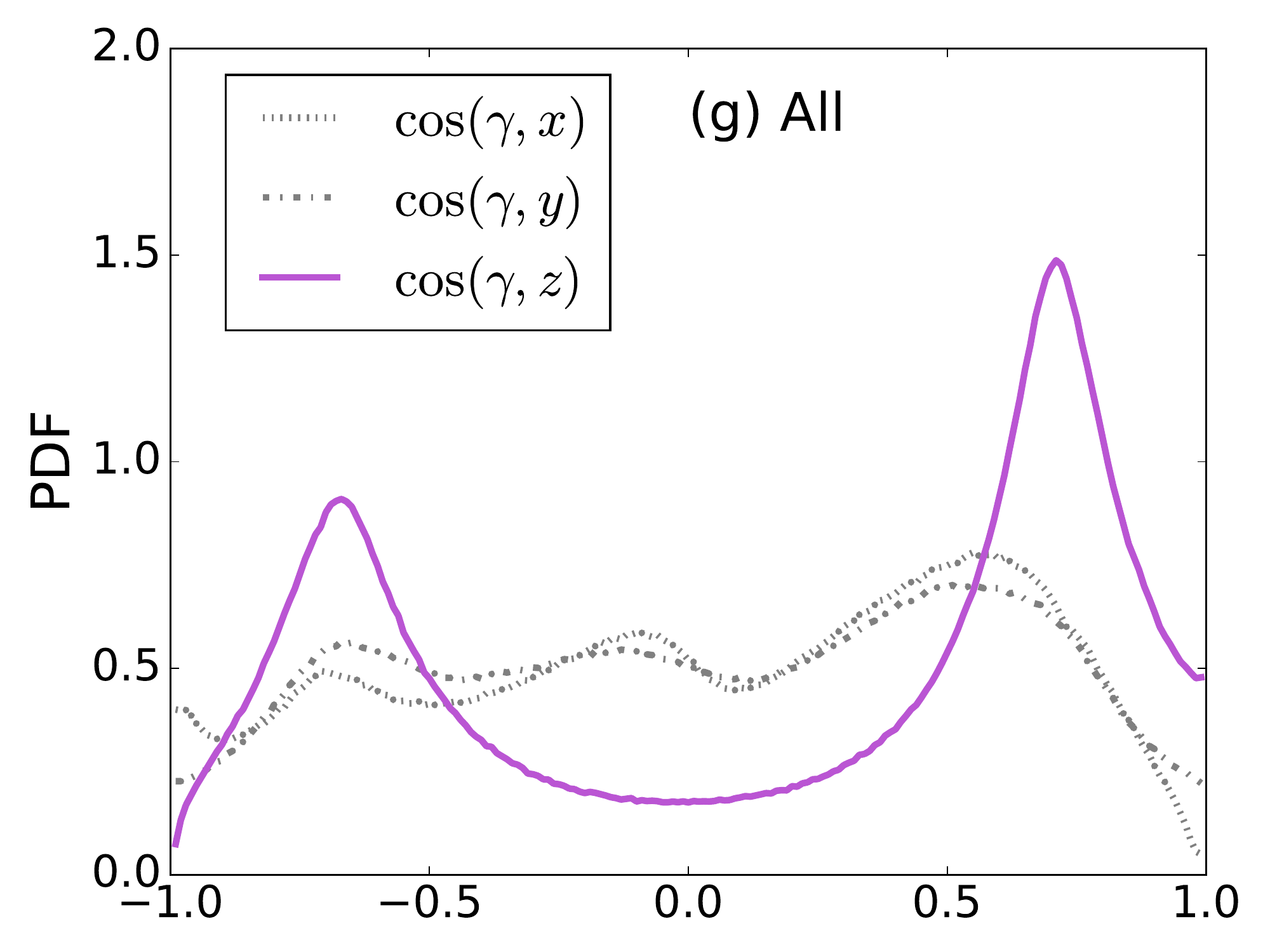}   
\includegraphics[width=5.9cm]{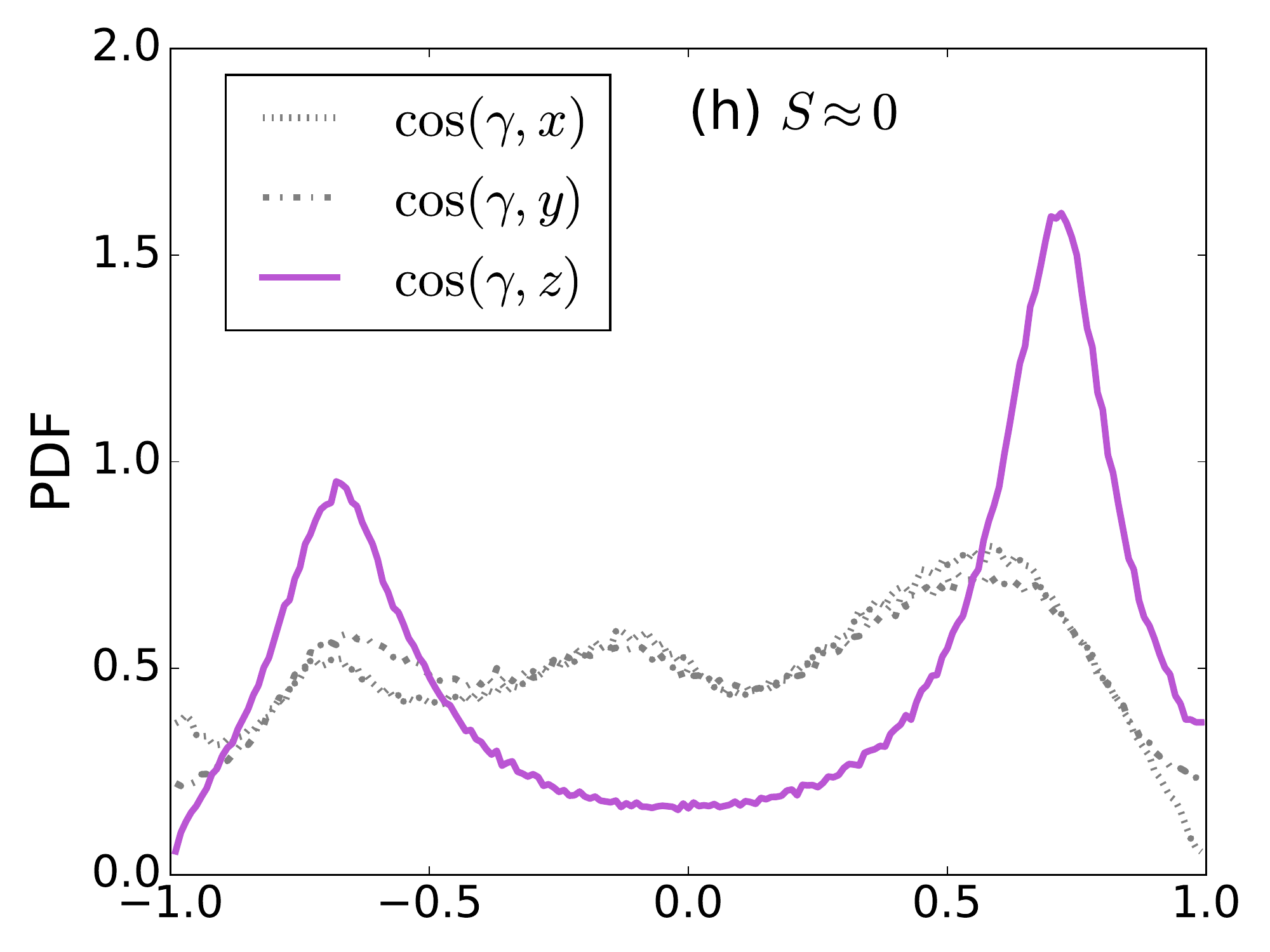}   
\includegraphics[width=5.9cm]{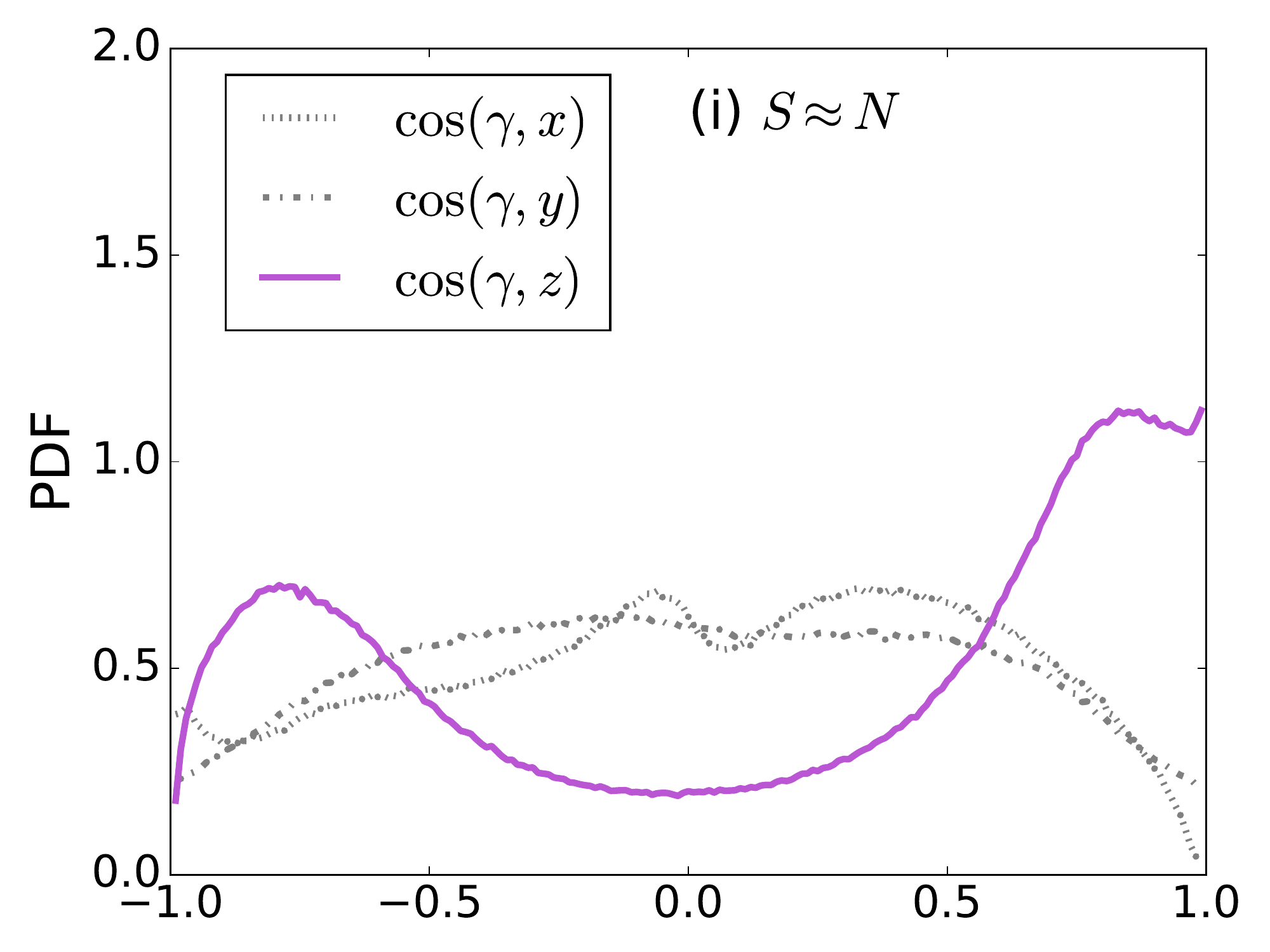}   
\caption{Probability density functions (PDFs) of the cosine of the
  angles between the eigenvectors ${\boldsymbol \alpha}$,
  ${\boldsymbol \beta}$, and ${\boldsymbol \gamma}$ of the strain rate
  tensor $s_{ij}$, and the Cartesian axes ${\bf x}$, ${\bf y}$, and
  ${\bf z}$, for the simulation with $N=8$. In the first row, panels
  (a), (b), and (c) show the PDFs of the alignment between
  ${\boldsymbol \alpha}$ and the Cartesian axes, for (a) all fluid
  elements, (b) fluid elements with $S\approx 0$, and (c) fluid
  elements with $S\approx N$. The second row, with panels (d), (e),
  and (f), shows the same for the ${\boldsymbol \beta}$ eigenvector,
  while the third row, with panels (g), (h), and (i), shows the same
  for ${\boldsymbol \gamma}$.}
\label{f:abg_xyz}
\end{figure}
%%%%%%%%%%%%%%%%%%%%%%%%%%%%%%%%%%%%%%%%%%

In the context of this analysis, we can study separately the two
relevant regions of the phase space of stably stratified turbulence by
discriminating fluid elements which have (instantaneously)
$S\approx 0$ or $S\approx N$. We can also study alignment globally
(i.e., for all fluid elements) by considering fluid elements in the
entire flow, without any restriction on the value of $S$. As mentioned
above, we will consider the vorticity ${\boldsymbol \omega}$ at the
position of each of these fluid elements, the density gradient
${\boldsymbol \nabla}\theta$, and the strain-rate tensor defined as in
Sec.~\ref{sec:manifolds} as
\begin{equation}
s_{ij} = \frac{1}{2} \left(A_{ij}+A_{ji} \right).
\end{equation}
This symmetric tensor describes the rate at which fluid elements are
stretched and sheared. It has three eigenvalues $s_{\alpha}$,
$s_{\beta}$, and ${s_{\gamma}}$ (with decreasing value, and with the
eigenvalue ${s_{\gamma}}$ being negative). Each eigenvalue is
associated to an eigenvector, respectively ${\boldsymbol \alpha}$,
${\boldsymbol \beta}$, and ${\boldsymbol \gamma}$, which are
orthogonal and define the principal axes of stretching and
shearing. In fact, the three eigenvalues correspond to the strain
along these principal axes, so that fluid elements along these
directions are locally strained but not rotated. Studying the
alignment between these quantities then reduces to studying the angles 
between the vectors ${\boldsymbol \omega}$,
${\boldsymbol \nabla}\theta$, ${\boldsymbol \alpha}$ ,
${\boldsymbol \beta}$, and ${\boldsymbol \gamma}$.

%%%%%%%%%%%%%%%%%%%%%%%%%%%%%%%%%%%%%%%%%% 
\begin{figure}
\includegraphics[width=5.9cm]{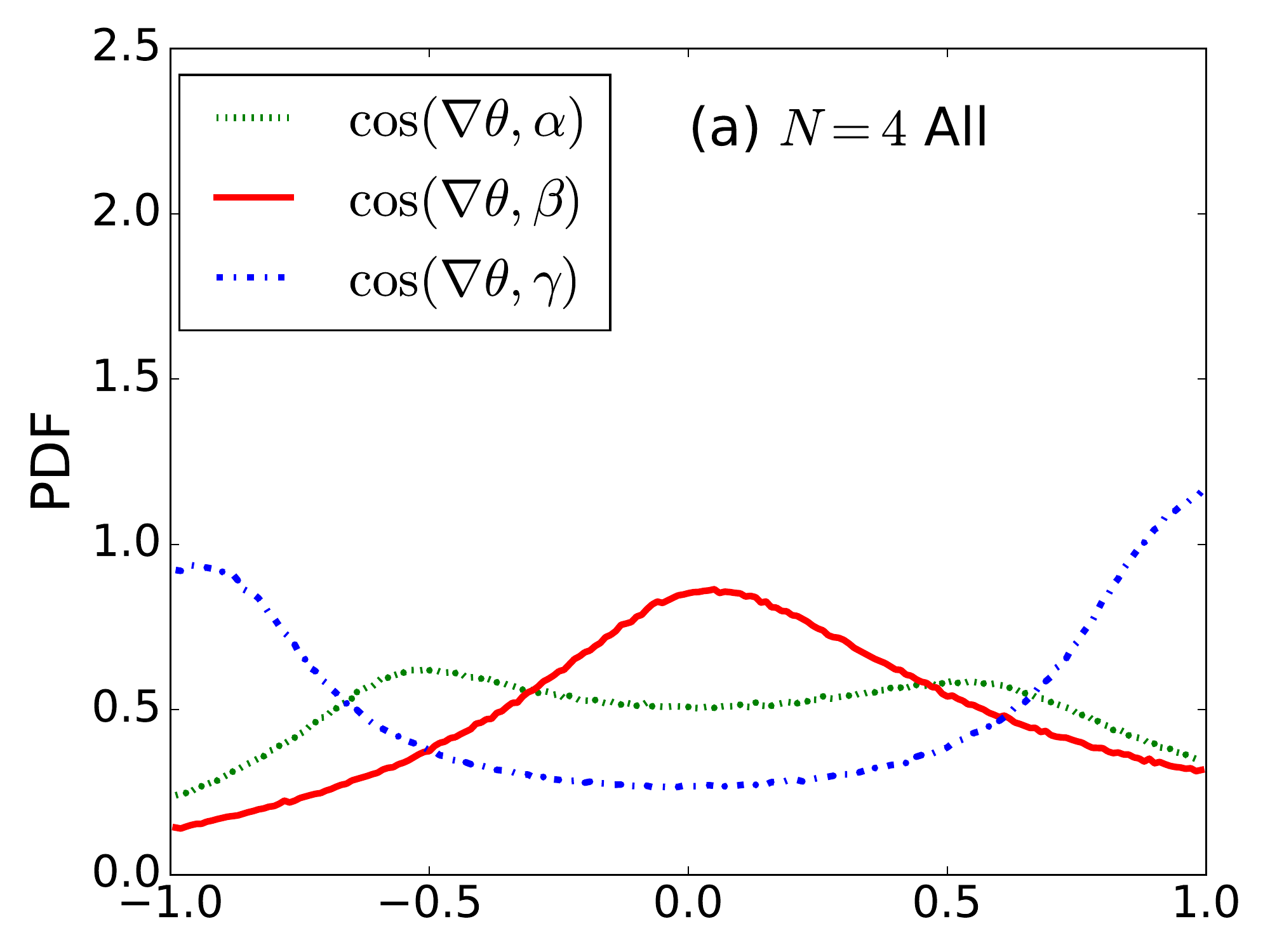}   
\includegraphics[width=5.9cm]{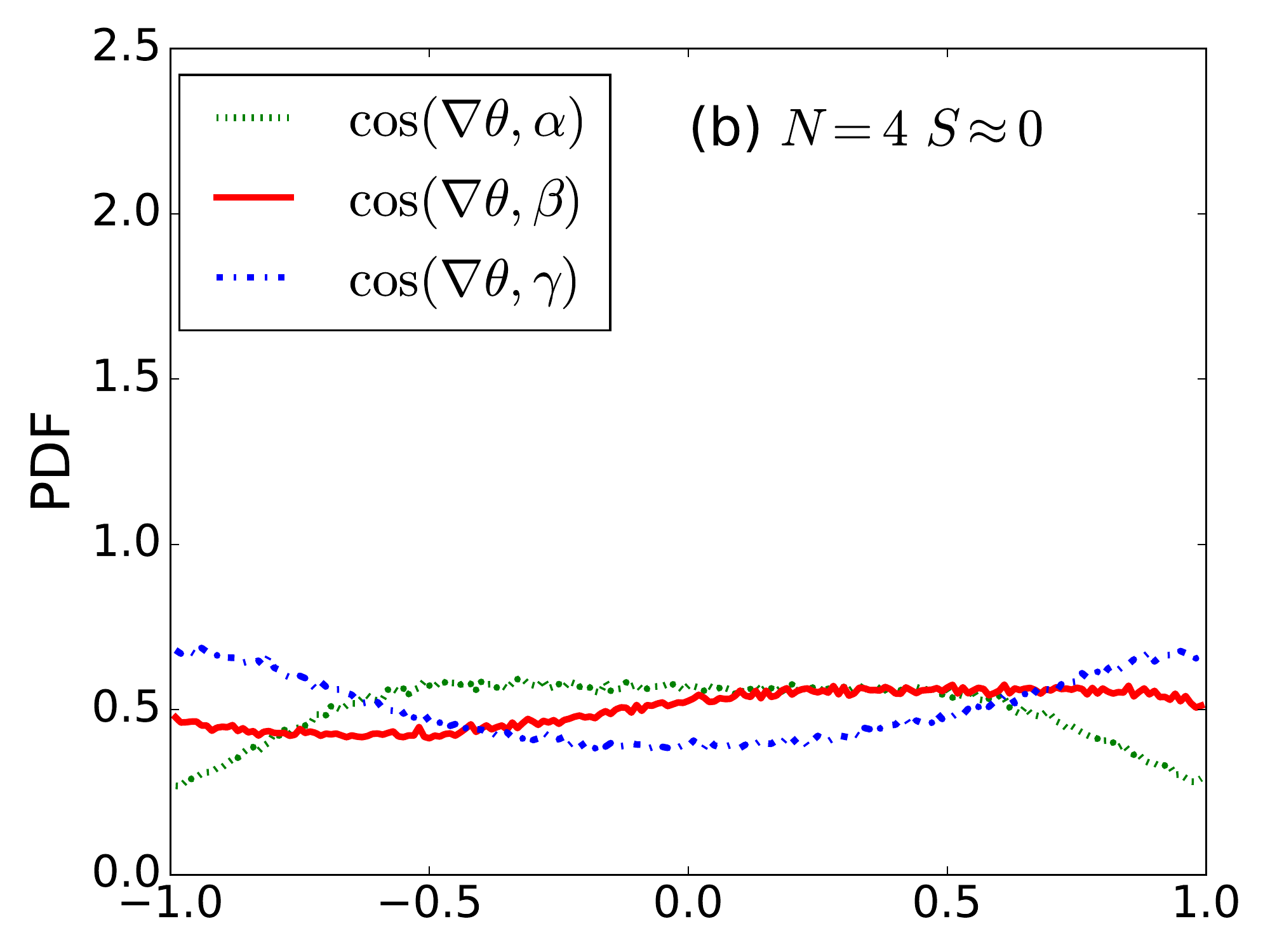}   
\includegraphics[width=5.9cm]{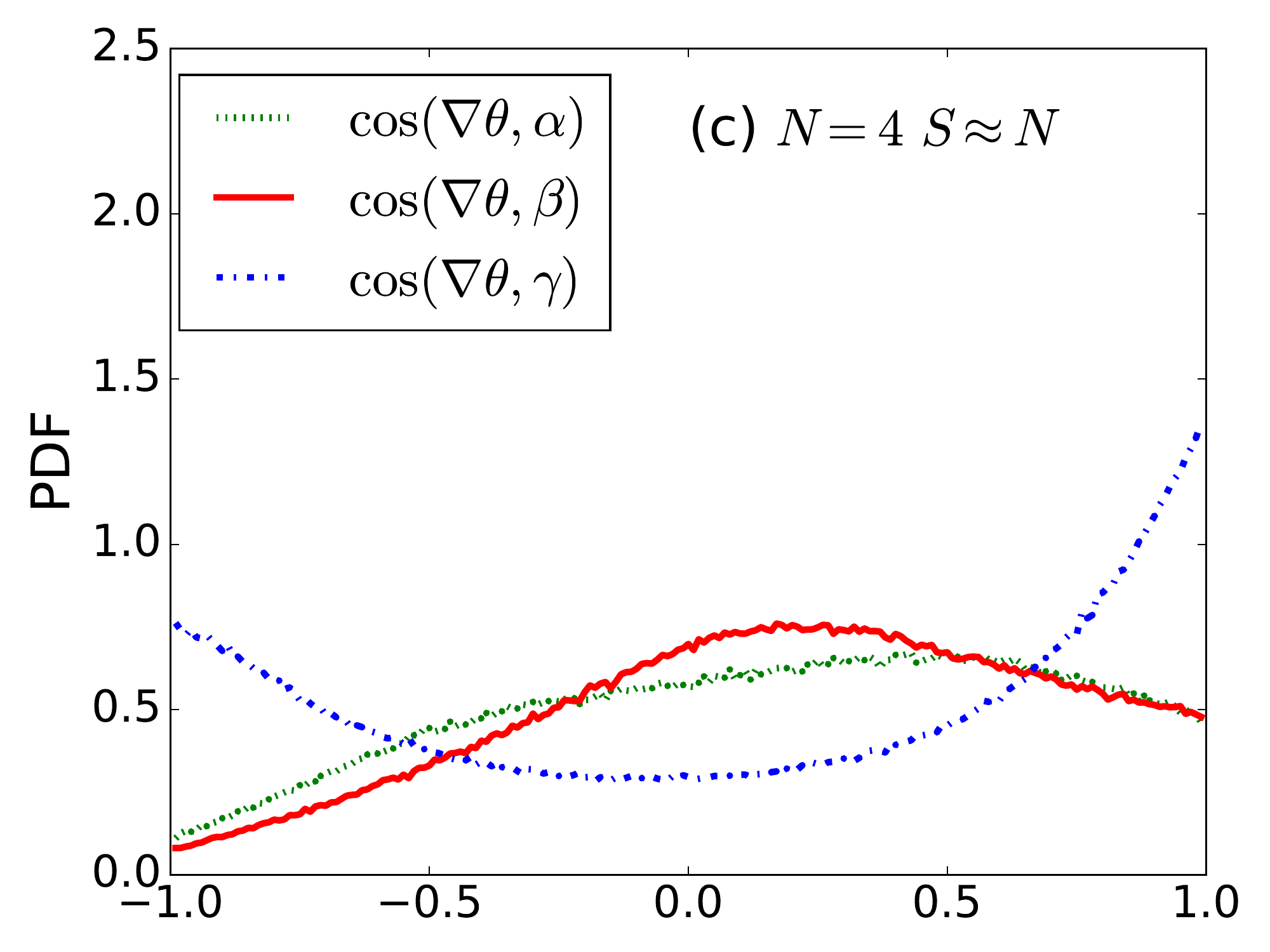} \\  
\includegraphics[width=5.9cm]{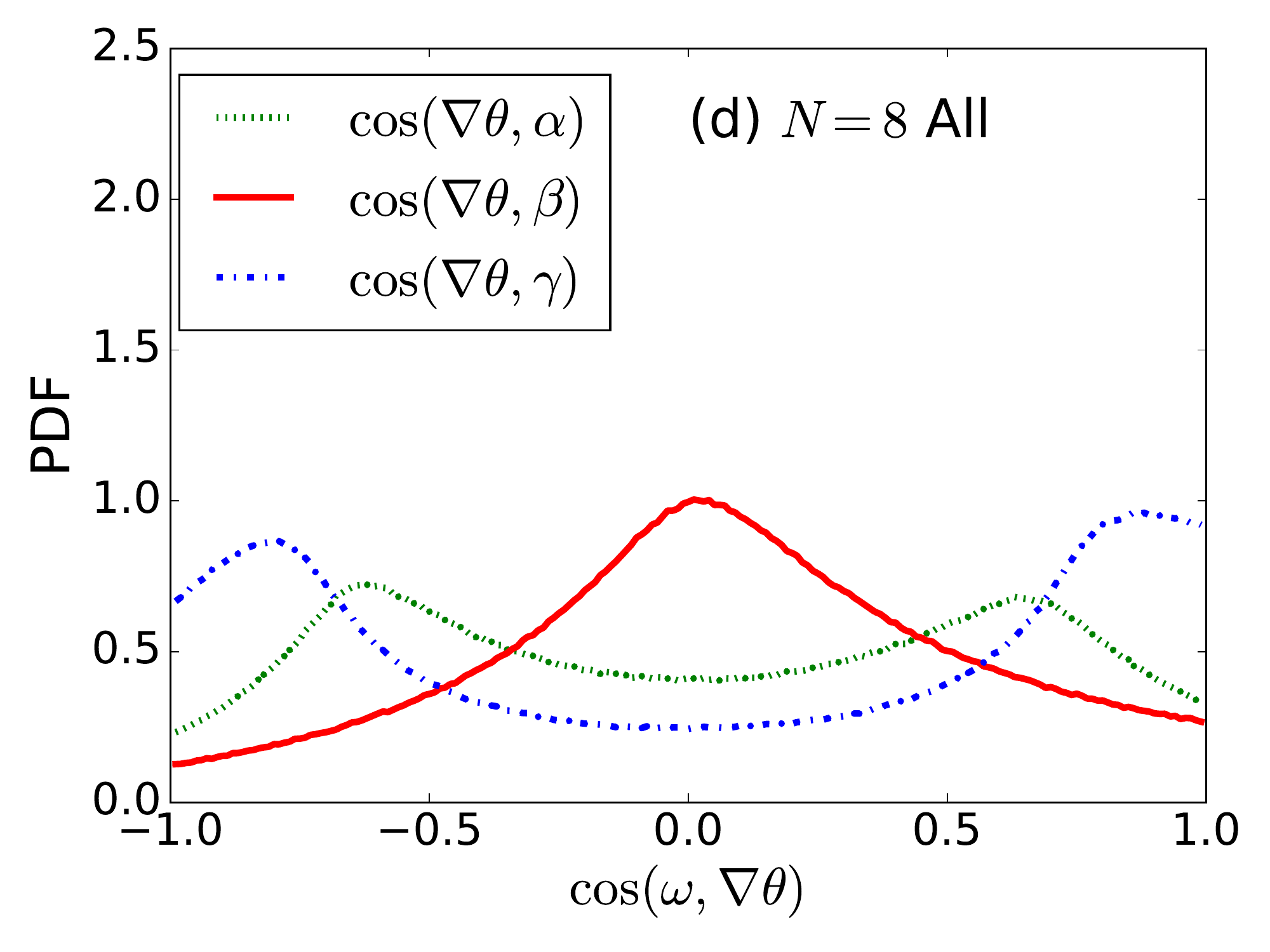}   
\includegraphics[width=5.9cm]{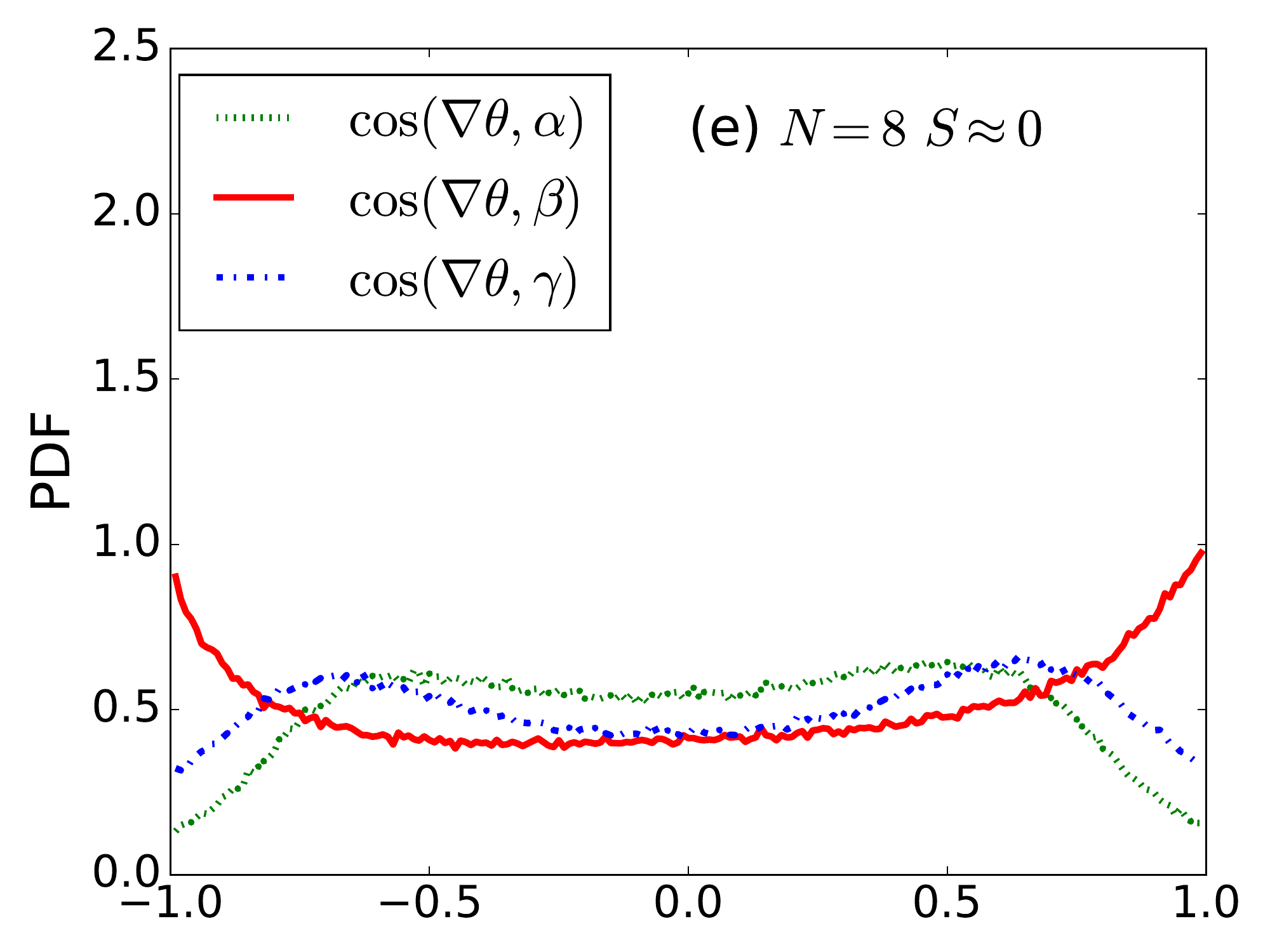}  
\includegraphics[width=5.9cm]{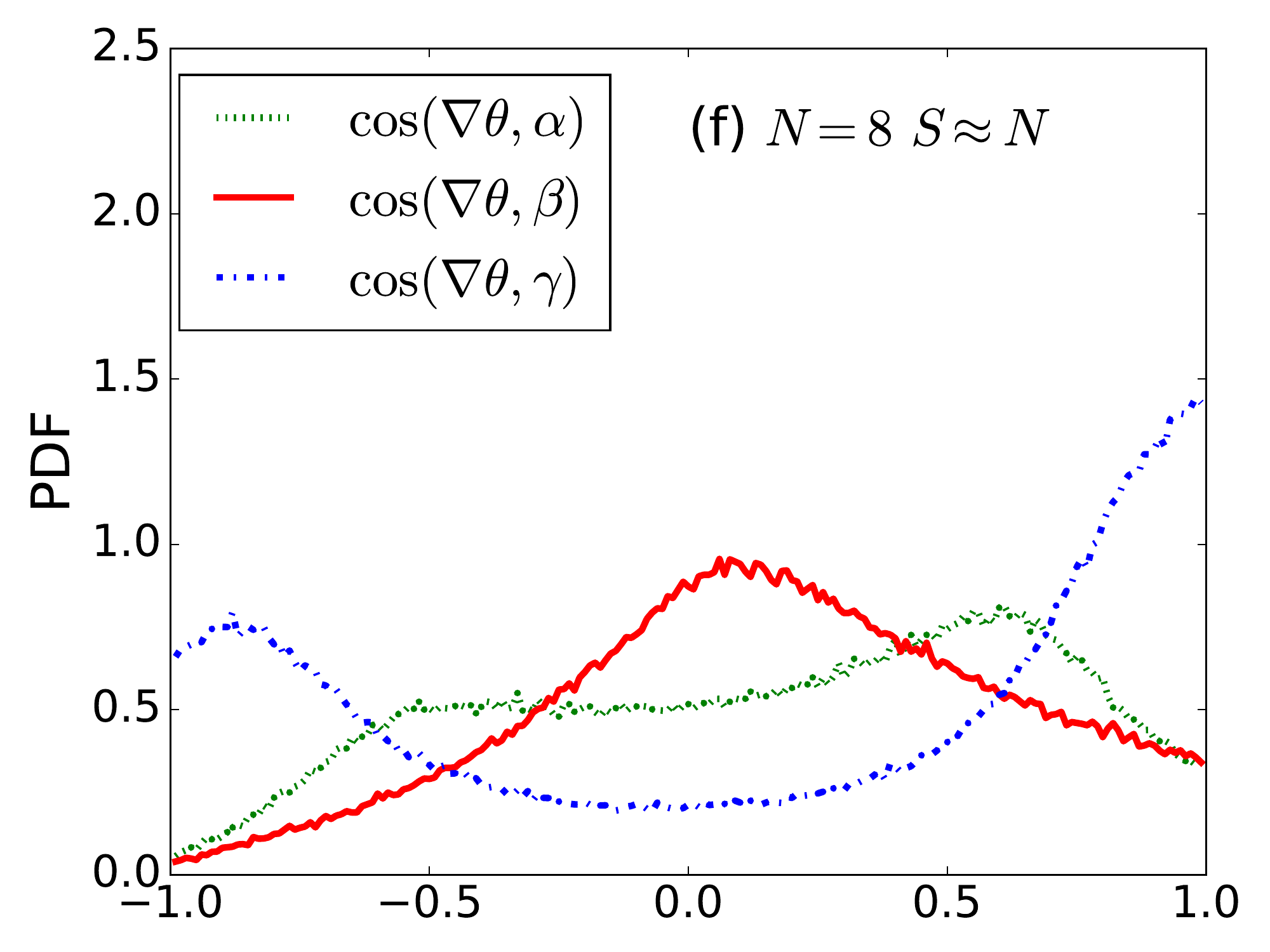}  \\   
\includegraphics[width=5.9cm]{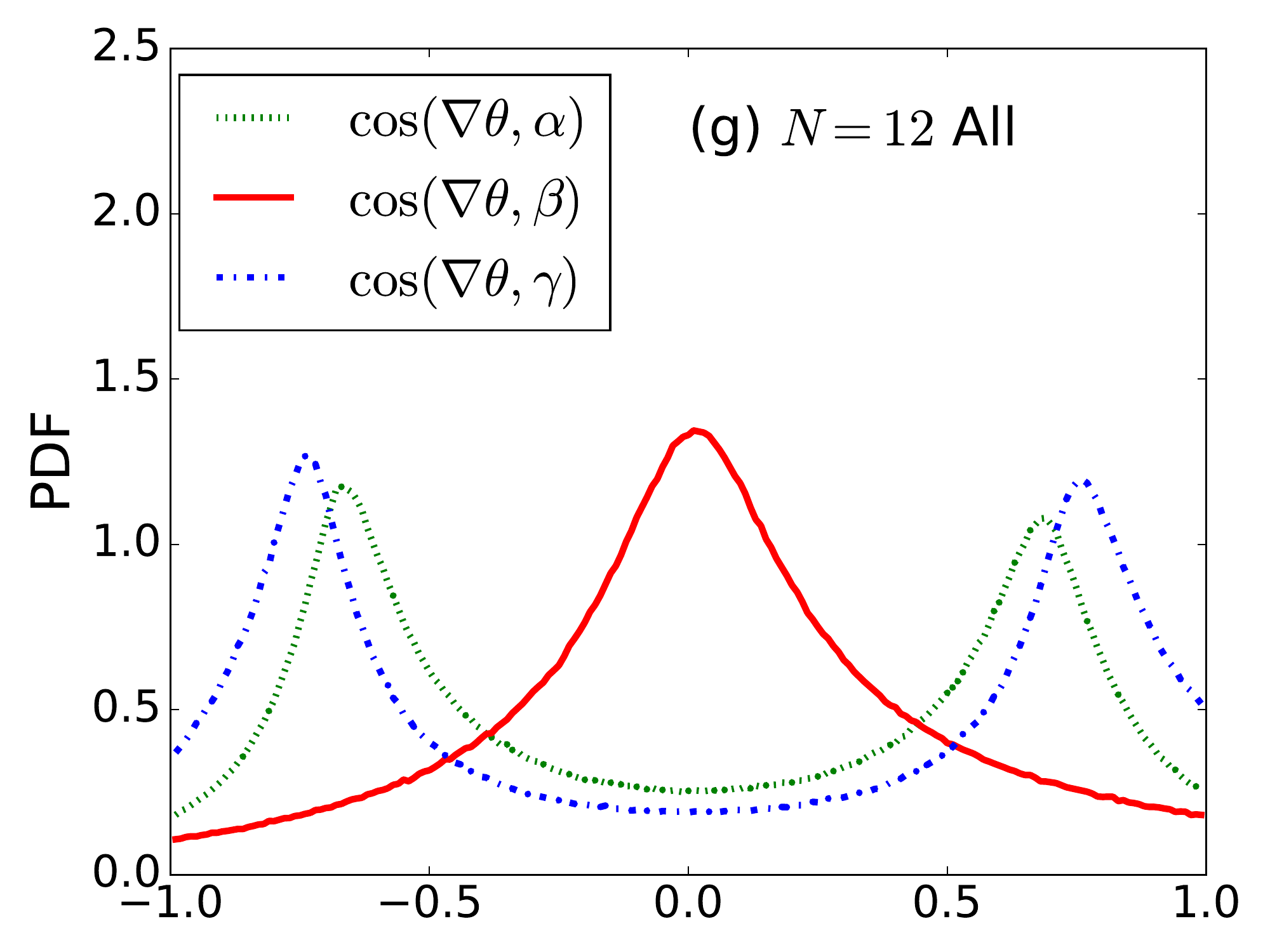}   
\includegraphics[width=5.9cm]{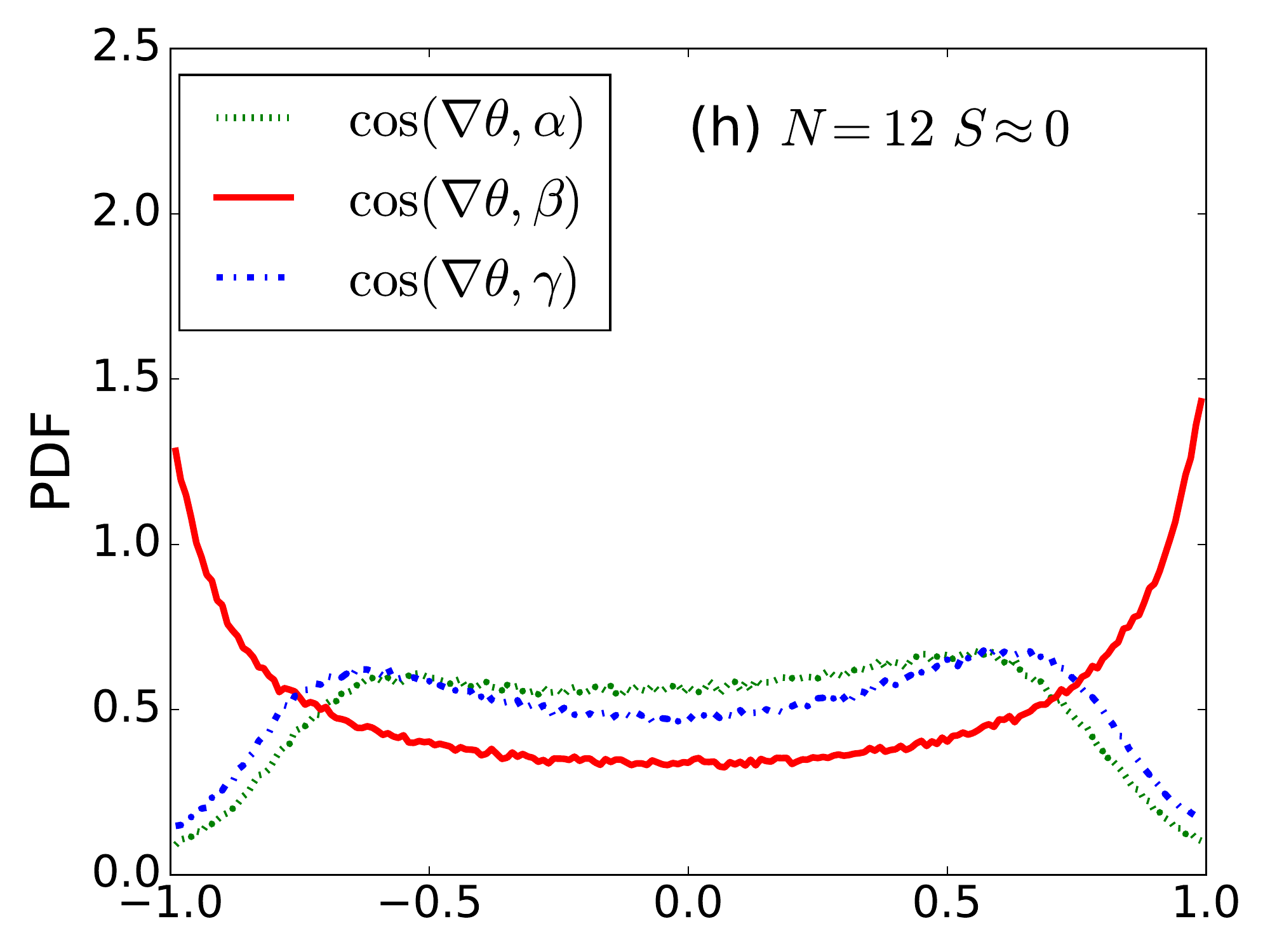}   
\includegraphics[width=5.9cm]{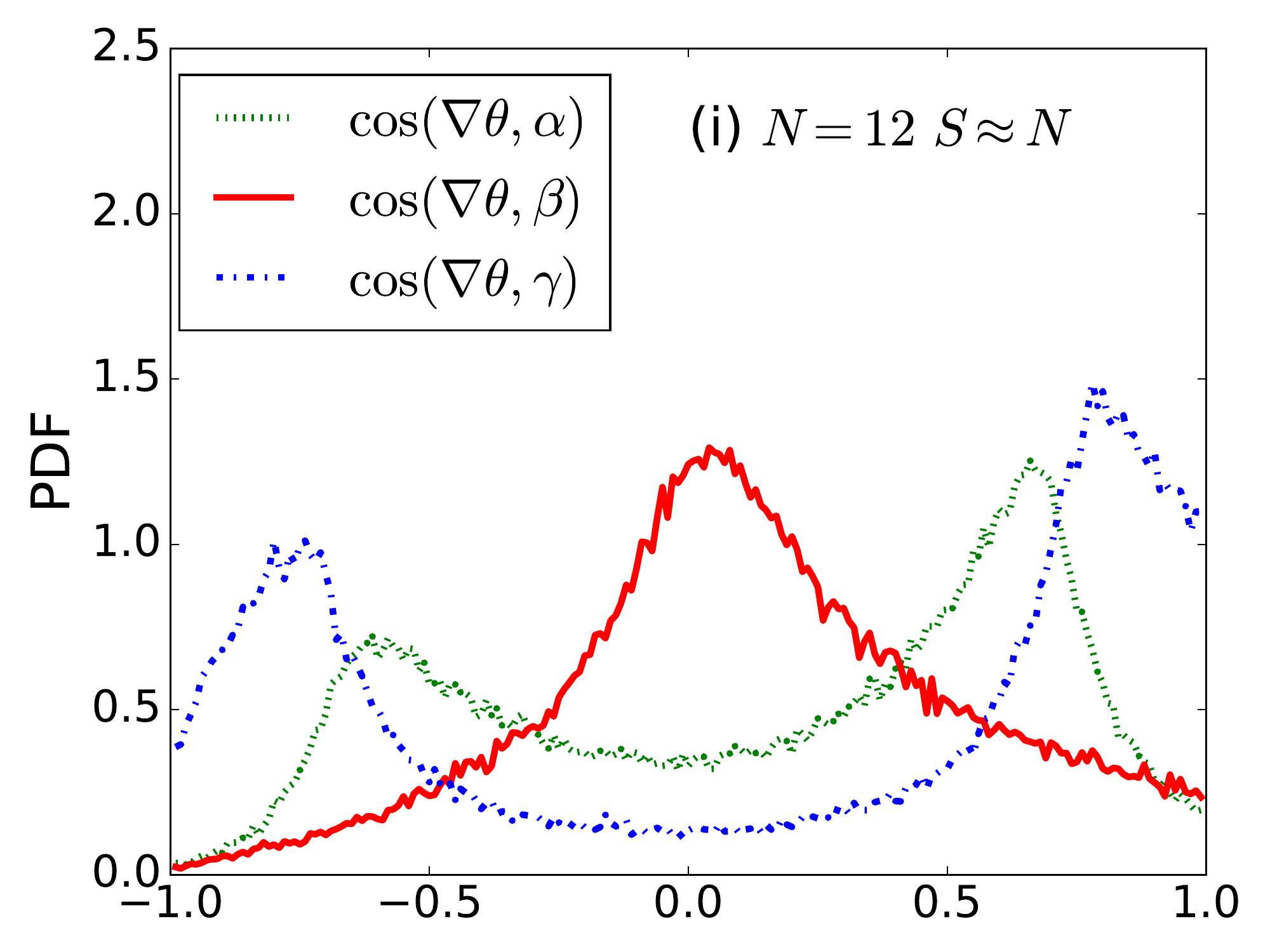}   
\caption{Probability density functions (PDFs) of the cosine of the
  angles between the density gradient ${\boldsymbol \nabla} \theta$
  and the eigenvectors ${\boldsymbol \alpha}$,
  ${\boldsymbol \beta}$, and ${\boldsymbol \gamma}$ of the strain rate
  tensor $s_{ij}$. In the first row, panels (a), (b), and (c) show the
  PDFs for the simulation with $N=4$, for (a) all fluid elements, (b)
  fluid elements with $S\approx 0$, and (c) fluid elements with
  $S\approx N$. The second row, with panels (d), (e), and (f), shows
  the same PDFs for $N=8$, while the third row, with panels (g), (h),
  and (i), shows the same for $N=12$.}
\label{f:th_abg}
\end{figure}
%%%%%%%%%%%%%%%%%%%%%%%%%%%%%%%%%%%%%%%%%%

In homogeneous and isotropic turbulence vorticity
${\boldsymbol \omega}$ aligns with the intermediate strain eigenvector
${\boldsymbol \beta}$. In other words, vorticity is driven away from
the direction of compression (associated to ${\boldsymbol \gamma}$),
resulting in vortex stretching at an intermediate rate (as in practice
$s_\beta>0$, but smaller than $s_\alpha$). A restricted Euler model of
the  vorticity-shear interaction suggest that, as this process takes
place in a time shorter than the eddy turnover time, the alignment of
${\boldsymbol \omega}$ with ${\boldsymbol \beta}$ is the result of
angular momentum conservation \cite{ashurst1987alignment,
  gulitski_velocity_2007, meneveau2011lagrangian}. This is also the
case for stably stratified turbulence (albeit with a dependence on the
level of stratification), as is shown in Fig.~\ref{f:w_abg}. The
figure shows the PDFs of the cosine of the angle between
${\boldsymbol \omega}$ and the eigenvectors
${\boldsymbol \alpha}$, ${\boldsymbol \beta}$, and
${\boldsymbol \gamma}$ of the strain-rate tensor, for the simulations
with $N=4$, $8$, and $12$, and for all fluid elements, as well as for
fluid elements restricted to cases with $S\approx 0$ or with
$S \approx N$. The most probable values of the cosine of the angle
between ${\boldsymbol \omega}$ and ${\boldsymbol \beta}$ peak always
at $\pm 1$, while the probability density of the cosine between
${\boldsymbol \omega}$ and ${\boldsymbol \alpha}$ or
${\boldsymbol \gamma}$ peaks at zero (and slightly more for
${\boldsymbol \gamma}$). The alignment between ${\boldsymbol \omega}$
and the intermediate eigenvector ${\boldsymbol \beta}$ increases with
the stratification, and takes place in all fluid elements
irrespectively of the region of the phase space they are visiting
(although the alignment is better for fluid elements with $S\approx 0$
than for fluid elements at the brink of convection with
$S\approx N$).

%%%%%%%%%%%%%%%%%%%%%%%%%%%%%%%%%%%%%%%%%%
\begin{figure}
\includegraphics[width=5.9cm]{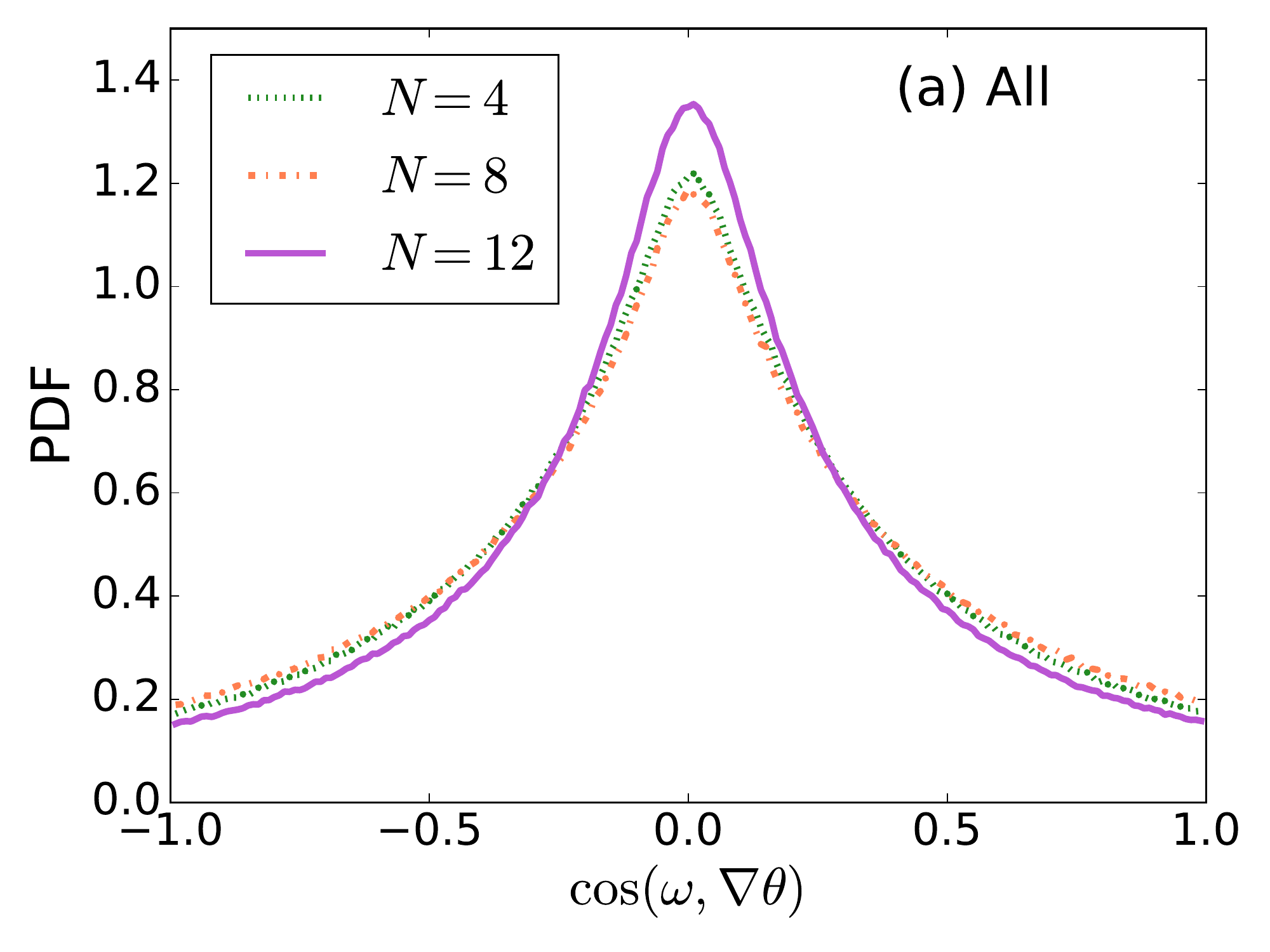}   
\includegraphics[width=5.9cm]{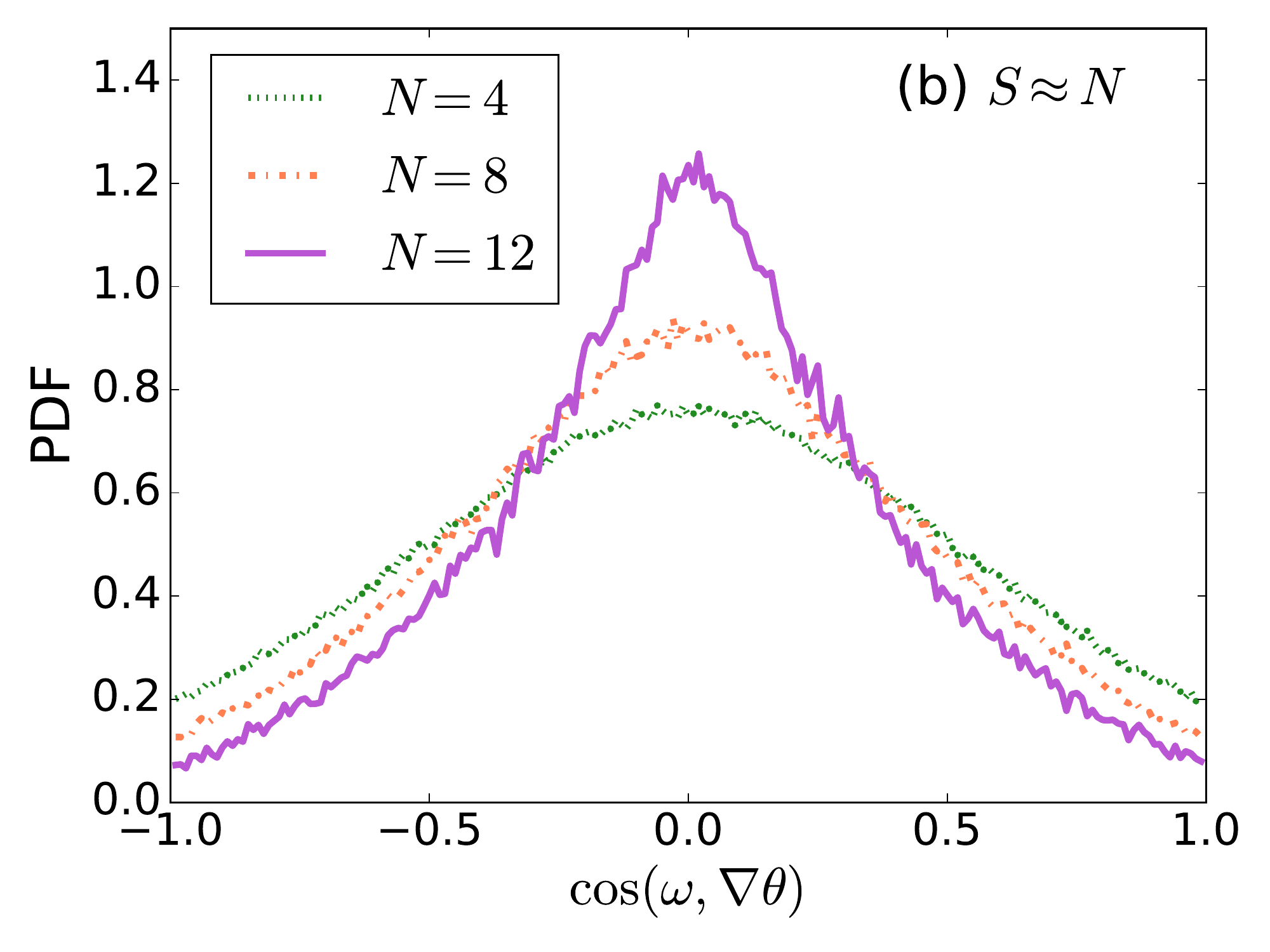}   
\includegraphics[width=5.9cm]{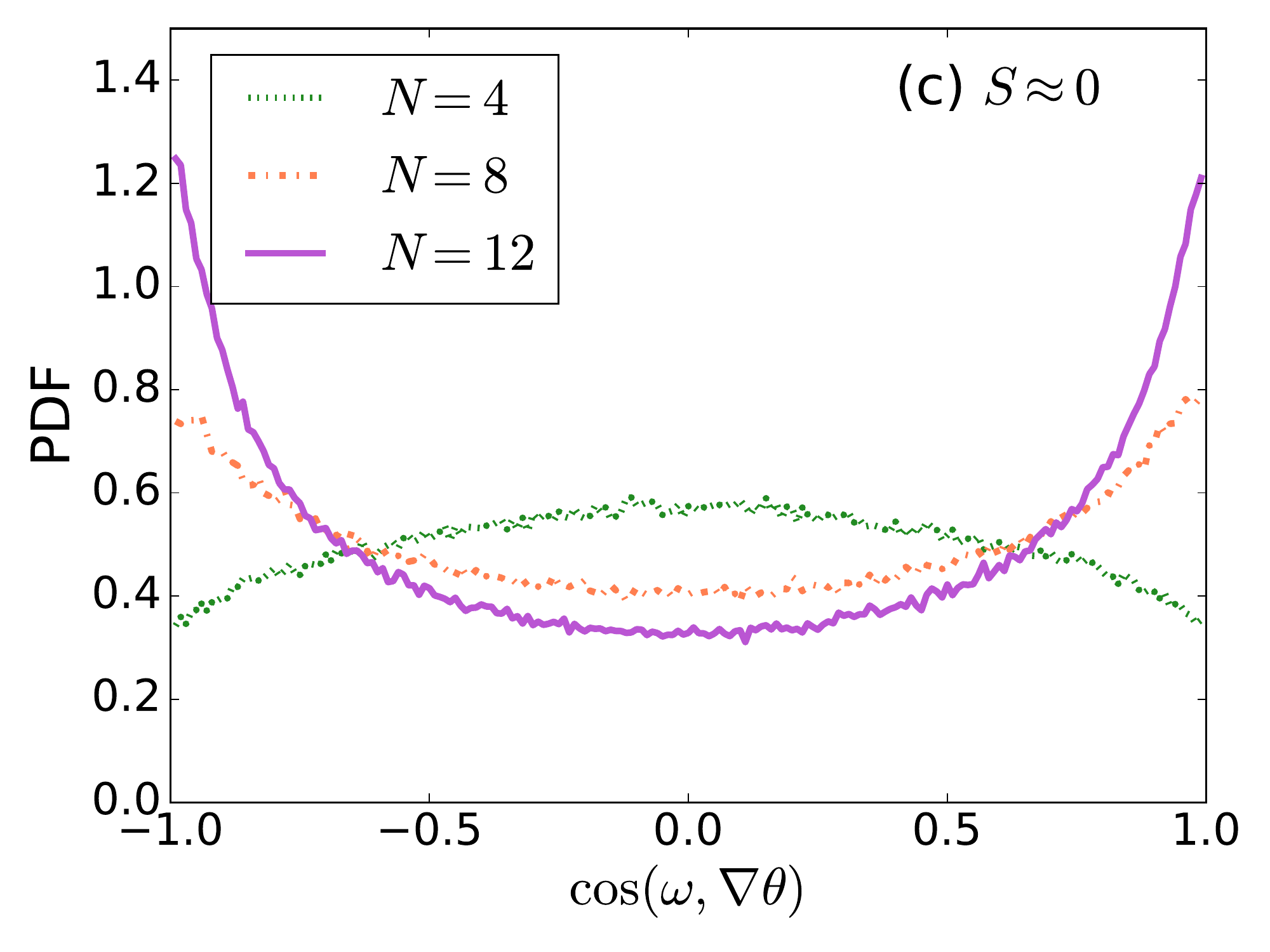}   
\caption{Probability density functions of the cosine of the angle
  between the density gradient ${\boldsymbol \nabla} \theta$ and 
  the vorticity ${\boldsymbol \omega}$ for all simulations in table
  \ref{tab:param}, for (a) all fluid elements, (b) fluid elements with
  $S \approx 0$, and (c) fluid elements with $S \approx N$.} 
\label{f:w_th}
\end{figure}
%%%%%%%%%%%%%%%%%%%%%%%%%%%%%%%%%%%%%%%%%%

%%%%%%%%%%%%%%%%%%%%%%%%%%%%%%%%%%%%%%%%%% 
\begin{figure}
\includegraphics[width=8.5cm]{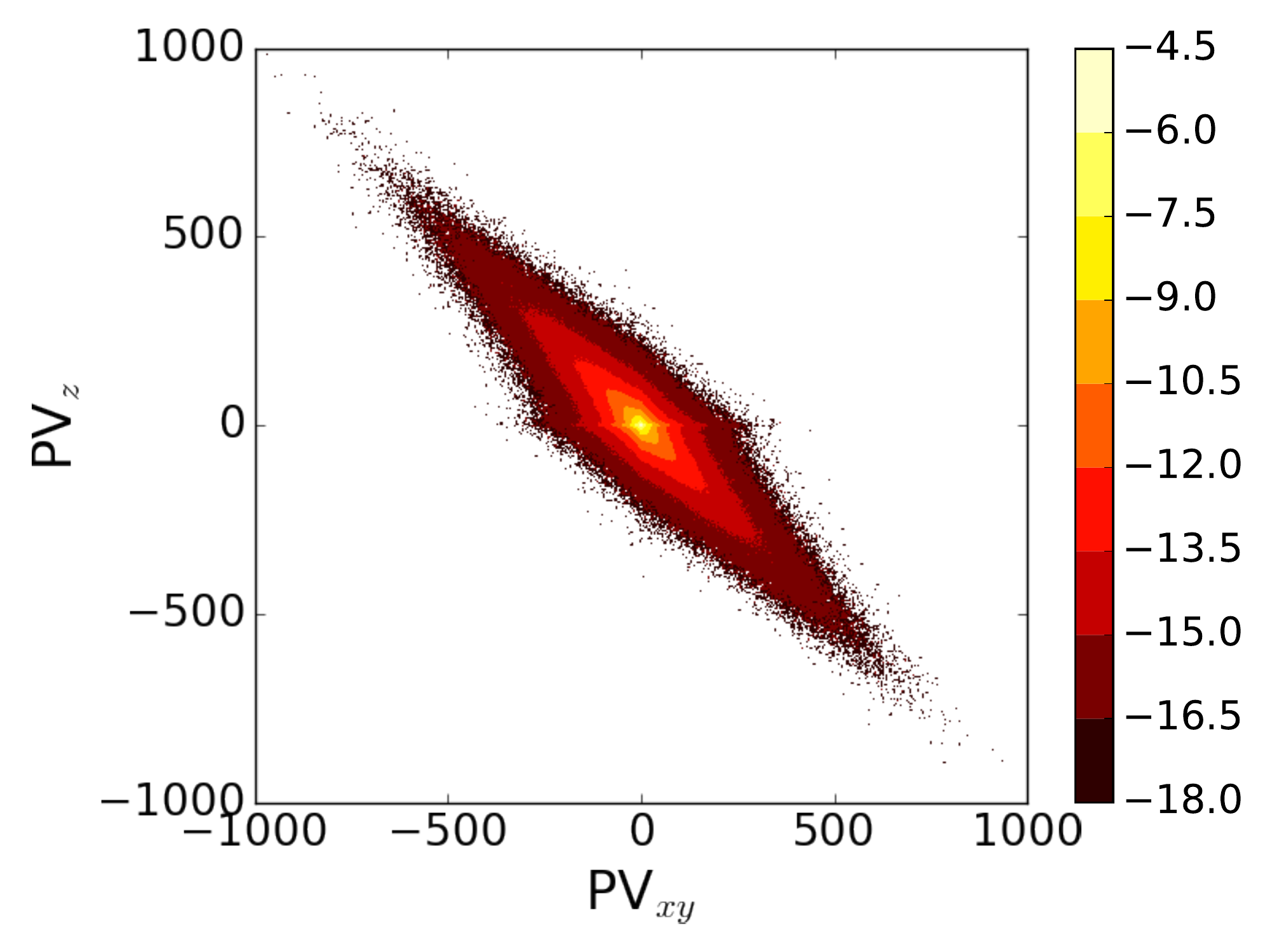}
\includegraphics[width=8.5cm]{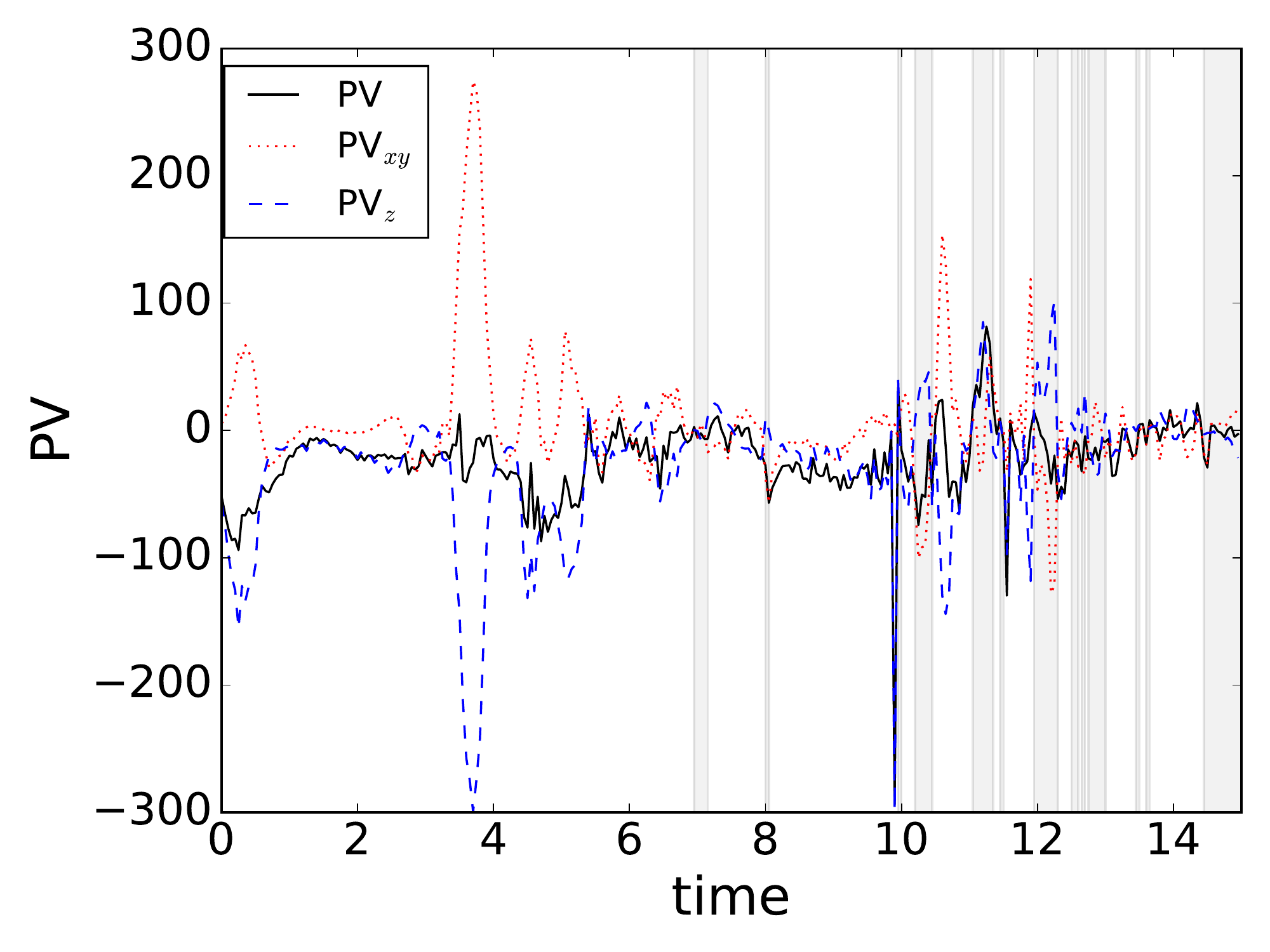}
\caption{{\it Left:} Joint probability density function of $PV_{xy}$
  and $PV_{z}$ for all fluid elements in the simulation with $N=4$. 
  {\it Right:} Values of $PV$, $PV_{xy}$ and $PV_{z}$ following a
  single particle trajectory in the simulation with $N=4$. The value
  of PV fluctuates around zero, with sudden bursts of $PV_{xy}$ and
  $PV_{z}$ cancelling each other. The grey areas indicate changes in
  sign of $PV_{xy}$ (and thus also of $PV_{z}$) as a result of $S$
  becoming larger than $N$.}
\label{f:PV}
\end{figure}
%%%%%%%%%%%%%%%%%%%%%%%%%%%%%%%%%%%%%%%%%%

This could suggest that vortex stretching mechanisms in stably
stratified turbulence are similar to those reported in homogeneous and
isotropic turbulence. However, the alignment observed in
Fig.~\ref{f:w_abg} is, in this case, also related with the anisotropy
of the flow. Indeed, the intermediate eigenvector
${\boldsymbol \beta}$ of the strain-rate tensor is preferentially
perpendicular to gravity (which points in the $z$ direction), and as a
result also preferentially in the $x$-$y$ plane (while the other two
eigenvectors, ${\boldsymbol \alpha}$ and ${\boldsymbol \gamma}$, 
show a larger projection in $z$). This is shown in
Fig.~\ref{f:abg_xyz}, which considers the statistics of the alignment
between the eigenvectors of the strain-rate tensor and the $x$, $y$,
and $z$ axes for the simulation with $N=8$. Note the differences
between probability densities peaking near $\pm 1$ (indicating
the corresponding eigenvalue is mostly aligned with the axis
considered), probability densities peaking near $0$ (indicating the
eigenvalue is mostly perpendicular to the axis considered), and flat
probability densities denoting the absence of a clear correlation.
The alignment of ${\boldsymbol \beta}$ with $x$ or $y$ happens on the
average for the entire flow, and also in fluid elements in the
$\Sigma_\RN{1}$ (``wavy'') manifold, but less so in fluid elements in
the $\Sigma_0$ (``convective'') manifold (see
Fig.~\ref{f:abg_xyz}). This can be expected: fluid elements in the
``wavy'' state have the eigenvectors with the largest and smallest
strains pointing in the vertical direction (and thus, resulting in
anisotropic structures), while interestingly, fluid elements at the
brink of local convection display less alignment of 
${\boldsymbol \alpha}$, ${\boldsymbol \beta}$, and
${\boldsymbol \gamma}$ with the Cartesian axes (and as a result, of
${\boldsymbol \omega}$ with these axes). While this speaks of a state
in which the three Cartesian directions are more similar, note this
does not result in a perfect isotropization (as peaks in the
probability densities in Fig.~\ref{f:abg_xyz} for fluid elements with 
$S\approx N$ are smaller than for all fluid elements or for the fluid
elements with $S\approx 0$, but still visible). In spite of this, we
can associate the second invariant manifold $\Sigma_0$ with flow
regions that are more efficient at mixing.

Besides ${\boldsymbol \omega}$, in stably stratified turbulence we can
also study the alignment of the vector defined by the buoyancy
gradient, ${\boldsymbol \nabla}\theta$, with ${\boldsymbol \alpha}$,
${\boldsymbol \beta}$, and ${\boldsymbol \gamma}$. The vector
${\boldsymbol \nabla}\theta$ provides us with information on how
buoyancy changes in the vicinity of the fluid elements. In homogeneous
and isotropic turbulence, when $\theta$ is a passive scalar,
${\boldsymbol \nabla}\theta$ aligns preferentially with
${\boldsymbol \gamma}$ (the compressive strain direction) 
\cite{ashurst1987alignment}, which can be expected as compression
increases the gradients of a scalar that is advected by the
flow. Figure \ref{f:th_abg} shows the probability density functions of
the cosine of the angle between ${\boldsymbol \nabla}\theta$ and the
eigenvectors of the strain-rate tensor in our simulations. On the
average, for all fluid elements, ${\boldsymbol \nabla}\theta$ is
perpendicular to ${\boldsymbol \beta}$ and becomes more so as
stratification increases. When all fluid elements are considered,
${\boldsymbol \nabla}\theta$ also seems to align almost equally with
either ${\boldsymbol \gamma}$ or ${\boldsymbol \alpha}$ for strong
stratification, which a small preference for aligning with
${\boldsymbol \gamma}$ (albeit this preference decreases with
$N$). This can be related to the flow anisotropy, with gradients in
the density fluctuations being mostly vertical and associated to the
formation of strata or of pancake-like structures. But interestingly,
we see again a strong difference in the alignment when fluid elements
in the two invariant manifolds are considered separately. Now, fluid
elements in the global invariant manifold $\Sigma_0$ ($S\approx N$)
display a similar behavior as the average over all fluid elements (but
with a stronger alignment with ${\boldsymbol \gamma}$, as is observed
in the case of passive scalars in homogeneous and isotropic turbulence
\cite{ashurst1987alignment}), while fluid elements in $\Sigma_\RN{1}$
($S\approx 0$) display the opposite trend and present a weak alignment
of ${\boldsymbol \nabla}\theta$ with ${\boldsymbol \beta}$ for
sufficiently large $N$ (and thus ${\boldsymbol \nabla}\theta$ lies
partially in the $x$-$y$ plane). The latter behavior is to be expected
for propagating internal gravity waves.

From these results, it can be expected that
${\boldsymbol \nabla} \theta$ and ${\boldsymbol \omega}$ should be
perpendicular on the average, and display different behaviors in the
two slow manifolds of the system. This is indeed seen in
Fig.~\ref{f:w_th}, which shows the probability density functions of
the cosine of the angle between these two vectors in all numerical
simulations. The  probability densities peak at $0$ for all fluid
elements and when restricted to fluid elements in $\Sigma_0$, while
for fluid elements in $\Sigma_\RN{1}$ (i.e., for wave-like behavior)
these two vectors go from weakly perpendicular to strongly parallel as
$N$ is increased.

Besides giving information on the local geometry of the flow, and
confirming that fluid elements are in different states when exploring
the two invariant manifolds $\Sigma_0$ and $\Sigma_\RN{1}$, the
results in this section also confirm a requirement for the validity of
the reduced model in Eq.~(\ref{eq:ODEs}). When deriving this model, we
reduced the information in the velocity gradient tensor $A_{ij}$ and
in the buoyancy gradient $\theta_i$ to seven scalars by using the fact 
that the flow has a preferred direction (given by the direction of
stratification, or of gravity). Thus, scalar quantities defined in
Eq.~(\ref{eq:variables}) treat differently $A_{zz}$, $A_{iz}$,
$A_{zi}$, and $\theta_z$, but not other components of these tensors 
and vectors involving the $x$ or $y$ Cartesian directions (in other
words, we reduce the information in the tensors and vectors by using
the axisymmetry of the equations). Probability density functions
presented in this section confirm that there is a preference of the
eigenvectors of the strain-rate tensor, and of other vectors, to align
parallel or perpendicular to $z$, but with no clear preferences in the
$x$ or $y$ direction.

\section{Potential vorticity \label{sec:PV}}

The relations reported above for field gradients in each of the
invariant manifolds of the system also have implications for the
evolution of the potential vorticity. In a stably stratified flow
under the Boussinesq approximation, the potential vorticity is
\begin{equation}
\textrm{PV} = \omega \cdot {\boldsymbol \nabla}\theta - N \omega_{z}
  = (S-N) (A_{yx}-A_{xy}) + \theta_{x} (A_{zy} - A_{yz})  + \theta_{y}
  (A_{xz} - A_{zx}) ,
\end{equation}
which is a conserved quantity following fluid trajectories in the
ideal (and unforced) case. The potential vorticity is an important
quantity in geophysical flows as $\textrm{PV}$ conservation, unlike
circulation conservation or Kelvin’s theorem, holds even for
baroclinic flows. This makes the quantity of relevance for the
atmosphere and the oceans. Indeed, $\textrm{PV}$ is a scalar quantity
that is advected by the flow, and can be used as a means to track
fluid elements, as well as to reconstruct the flow properties in the
vicinity of the fluid elements \cite{vallis2017atmospheric,
  Waite_2013}.

From the previous section it is clear that
$\omega \cdot {\boldsymbol \nabla}\theta \approx 0$ in most of the
phase space. The only exception being the $\Sigma_{1}$ stable
manifold, where all gradients are small, and so $\textrm{PV} \approx
0$. We can further decompose the potential vorticity as
\begin{equation}
\textrm{PV} =\textrm{PV}_{z} + \textrm{PV}_{xy} ,
\end{equation}
where $\textrm{PV}_{z}=(S-N) (A_{yx}-A_{xy})$ and
$\textrm{PV}_{xy}=\theta_{x}(A_{zy}-A_{yz})+\theta_{y}
  (A_{xz}-A_{zx})$. Note that $\textrm{PV}_{z}=0$ in $\Sigma_0$. But
if $\textrm{PV}$ is conserved, this implies that as fluid elements
explore phase space, $\textrm{PV}_{xy}$ must be anti-correlated with
$\textrm{PV}_{z}$ and must compensate for its changes. Moreover, even
when $\textrm{PV}$ is not perfectly conserved (e.g., in the presence
of dissipation), we can expect variations in $\textrm{PV}$ to be slow
compared with the fast evolution when fluid elements escape from one
of the invariant manifolds to the other.

Indeed, Fig.~\ref{f:PV} confirms this anti-correlation from data
obtained from the direct numerical simulations with $N=4$. On the
average, $\textrm{PV}_{z} \approx -\textrm{PV}_{xy}$ for most fluid
elements, and the most probable values in the joint probability
density function of $\textrm{PV}_{z}$ and $\textrm{PV}_{xy}$
correspond to relatively small values of these quantities. Figure
\ref{f:PV} also shows the time evolution of $\textrm{PV}$ for a single
fluid element in the same simulation, as a function of time, as well
as the two components $\textrm{PV}_{z}$ and $\textrm{PV}_{xy}$. The
potential vorticity fluctuates around zero. However, sudden bursts of
$S$ (as those seen in Fig.~\ref{f:scheme}) affect the evolution of
$\textrm{PV}_{z}$, and as a result of $\textrm{PV}_{xy}$. Note how
each sudden burst of $\textrm{PV}_{z}$ has an associated burst 
of opposite sign of $\textrm{PV}_{xy}$, so that $\textrm{PV}$ displays
less fluctuations than the two components separately. In
Fig.~\ref{f:PV} we also shade some time intervals in which the change
of sign of $\textrm{PV}_{z}$ (and, as a consequence, also in
$\textrm{PV}_{xy}$) is not due to $S$ displaying a burst, but becoming
instead larger than $N$, which can occur when fluid elements are in
the vicinity or escape from the ``convective'' manifold
$\Sigma_{0}$. The dynamic of these quantities is reminiscent of that
observed in Fig.~\ref{f:scheme}, with fluid elements displaying a slow
evolution, with fast bursts that change the state from one regime to
another.

\section{Conclusions \label{sec:conclusions}}

Just like Bilbo Baggins, fluid elements in stably stratified
turbulence explore a complicated phase space by moving between two
places (associated with slow invariant manifolds). And these two
places, just like the Shire and Khazad-d\^um, put fluid elements in
two very different states. In the first manifold, fluid elements are
in a quiet ``wavy'' state. When energy becomes sufficiently large,
fluid elements escape from this manifold and evolve rapidly towards
another manifold in which evolution is also slow. This second manifold
is at the brink of the local convective instability, and thus can be
expected to correspond to a disordered state characterized by
efficient mixing and dissipation. Once the strength of gradients
decreases again, fluid elements return rapidly to the first
manifold. But as stratification is increased, the stability of the
first manifold increases, and the journeys to the second manifold
become less and less frequent.

The reduced system presented here, as well as the results from direct
numerical simulations, indicate that the Boussinesq equations have
only these two invariant manifolds, and as a result evolution outside
these manifolds is fast. Such a picture, in which fluid elements
explore slowly two solutions, and travel fast through the rest of
phase space, is in good agreement with recent developments in wave
turbulence \cite{newell2013wave, dyachenko2016whitecapping}, and
provide, at least for the case of stably stratified turbulence and in
the full Boussinesq system, a much needed route for dissipation. While
wave-like motions can bring energy to smaller scales, eventually the
amplitude of the nonlinearity becomes such that the fluid elements
must search for other surfaces of solutions in phase space. These
solutions, which do not correspond to waves, dissipate energy
efficiently, and fluid elements can then return to their previous
state.

The existence of the second invariant manifold, which is at the brink
of a convective instability, explains recent observations that a
significant fraction of fluid elements in stratified turbulence is
always at the threshold of a linear instability
\cite{pouquet2019linking} or in a critical state \cite{Smyth_2013,
  Smyth_2019}, that only an intermittent fraction of fluid elements in
the ocean are responsible for dissipation \cite{Pearson_2018}, and
that extreme vertical drafts develop sporadically in these flows
resulting in non-Gaussian statistics \cite{rorai_stably_2015, 
  feraco_vertical_2018}.

Besides, the invariant manifolds of the reduced system for the
Lagrangian evolution of velocity and density gradients indicate that:
(1) Certain balance relations must hold as fluid elements are advected
in the flow. These balance relations impose conditions, e.g., in the
turbulent production of gradients of density fluctuations, with
implications for subgrid models of turbulence. (2) The change of  
the invariant manifolds with respect to the homogeneous and isotropic
case (and in particular, the destruction of the Vieillefosse tail),
modifies the geometry of vortex stretching, thus providing a
different explanation for the change in vortical structures and the
development of pancake-like structures as stratification is
increased. Moreover, these changes are different depending on what
region of phase space the fluid elements are exploring. And (3),
correlations exist in the evolution of different terms in the
potential vorticity, as fluid elements travel from one manifold to the
other.

The model also has clear limitations. The first is that it neglects
the effect of forcing and dissipation. The second is that pressure
gradient effects are partially neglected (as the pressure Hessian is
dropped). This second limitation is shared with restricted 
Euler models of homogeneous and isotropic turbulence
\cite{cantwell_exact_1992, chevillard2006lagrangian,
  meneveau2011lagrangian}, although in the stably stratified
case the role of the pressure Hessian is smaller as stratification is
increased. In spite of these limitations, it is interesting that
direct numerical simulations of the full Boussinesq equations
including forcing and dissipation are in agreement with predictions
from the model. This suggests that some of the manifolds and balance
relations that follow from the model could be used for the development
of subgrid models (as an example, to derive new estimations for
turbulent production of vertical density fluctuations), as predictors
of the occurrence of extreme events such as those observed in clean
air turbulence or in extreme vertical updrafts and downdrafts, or to
explain the origin of the anomalous dissipation reported in recent
oceanic models.

\begin{acknowledgments}
The authors acknowledge support from PICT Grant No.~2015-3530.
\end{acknowledgments}

\bibliography{ms}

\end{document}